\documentclass[format=acmsmall,screen,authorversion=true, nonacm=true]{acmart}
\pdfoutput=1 
\makeatletter
\if@ACM@authorversion
\usepackage{fancyhdr}
\AtBeginDocument{%
    \addtolength{\footskip}{1.0\baselineskip}%
    \fancyfoot[L]{\small\textit{Pre-print}}%
}
\fi
\makeatother

\hypersetup{
    pdfauthor={Avishek Anand, Lijun Lyu, Maximilian Idahl, Yumeng Wang, Jonas Wallat, Zijian Zhang},
    pdftitle={Explainable Information Retrieval: A Survey},
    pdfkeywords={Survey, Overview, Information Retrieval, Explainable AI, Interpretability, Explainability, Explainable, IR, Ranking, Search, Explainable IR, ExIR, XIR, Explainable Search}
}

\setcopyright{rightsretained}

\usepackage[capitalize, noabbrev]{cleveref}
\usepackage[]{forest}
\usepackage[acronym]{glossaries}
\glsdisablehyper

\newacronym{ir}{IR}{information retrieval}
\newacronym{exir}{ExIR}{explainable information retrieval}
\newacronym{ml}{ML}{machine learning}
\newacronym{nlp}{NLP}{natural language processing}
\newacronym{rs}{RS}{recommender systems}
\newacronym{ltr}{LTR}{learning-to-rank}
\newacronym{ibd}{IBD}{interpretable-by-design}

\newcommand{\pref}[1]{\hyperref[#1]{§\ref*{#1}}} 

\begin{document}
\title{Explainable Information Retrieval: A Survey}

\author{Avishek Anand}
\email{avishek.anand@tudelft.nl}
\orcid{0000-0002-0163-0739}

\author{Lijun Lyu}
\email{L.Lyu@tudelft.nl}
\orcid{0000-0002-7268-4902} 
\affiliation{%
  \institution{Delft University of Technology}
  \streetaddress{P.O. Box 1212}
  \city{Delft}
  \country{The Netherlands}
}

\author{Maximilian Idahl}
\email{idahl@l3s.de}
\orcid{0000-0003-1866-6436}

\author{Yumeng Wang}
\email{wang@l3s.de}
\orcid{0000-0002-2105-8477}

\author{Jonas Wallat}
\email{wallat@l3s.de}
\orcid{0000-0003-1239-2067}

\author{Zijian Zhang}
\email{zzhang@l3s.de}
\orcid{0000-0001-9000-4678}

\affiliation{%
    \institution{L3S Research Center, Leibniz University Hannover}
    \streetaddress{Appelstr. 9a}
    \city{Hannover}
    \state{Lower Saxony}
    \country{Germany}
}

\begin{abstract}
Explainable information retrieval is an emerging research area aiming to make transparent and trustworthy information retrieval systems. Given the increasing use of complex machine learning models in search systems, explainability is essential in building and auditing responsible information retrieval models. This survey fills a vital gap in the otherwise topically diverse literature of explainable information retrieval. It categorizes and discusses recent explainability methods developed for different application domains in information retrieval, providing a common framework and unifying perspectives. In addition, it reflects on the common concern of evaluating explanations and highlights open challenges and opportunities.
\end{abstract}



\maketitle


\section{Introduction}
\label{sec:intro}

\Gls{ir} systems are one of the most user-centric systems on the Web, in digital libraries, and enterprises.
Search engines can be general-purpose (e.g., Web search) to specialized expert systems that are geared towards expert consumption or support, including legal and patent retrieval \gls{ir}~\citep{chakraborty:excir:2021:patent}, historical search~\citep{holzmann:www:2016:tempas,holzmann:webscience:2017:exploring}, and scholarly search~\citep{hanauer:2006:emerse:medicalsearch,singh:sigir:2016:expedition}. 
On the one hand, riding on the recent advances of complex \gls{ml} models trained on large amounts of data, \gls{ir} has seen impressive performance gains over classical models~\cite{lin:sigir:2019:recantation}.
On the other hand, complex models also tend to be opaque and less transparent than their classical and arguably simpler counterparts. 
Therefore, towards an important goal of ensuring a \textit{reliable} and \textit{trustworthy} \gls{ir} systems, recent years have seen increased interest in the area of \gls{exir}.

\subsection{Motivation}

Firstly, in \gls{ir}, there has been sufficient evidence of how user interaction data from search engines can be a source of biases, especially associated with gender and ethnicity~\cite{mowshowitz:ipm:2005:measuring-ir-bias,rekabsaz:sigir:2020:ir-gender-bias, bigdeli:sigir:2022:bias:irsystems}.
When undetected and unidentified, the users of an \gls{ir} system too are exposed to stereotypical biases that reinforce known yet unfair prejudices.
Secondly, model retrieval models based on transformer-style over-parameterized models can be brittle and sensitive to small adversarial errors~\cite{wang:2022:ictir:bert}.
Recently developed inductive biases,  pre-training procedures, and transfer learning practices might lead these statistical over-parameterized models to learn shortcuts~\cite{geirhos:2020:natmachintell:shortcutlearningDNNs}.
Consequently, shortcuts that do not align with human understanding results in learning patterns that are \textit{right for the wrong reasons}.
Finally, expert users using specialized search systems -- in legal search, medicine, journalism, and patent search -- need control, agency, and lineage of the search results.
For all the above \gls{ir}-centric reasons, among many other general reasons -- like utility for legal compliance, scientific investigation, and model debugging -- the field of \gls{exir} provides the \textit{tools/primitives} to examine learning models and the capability to build transparent \gls{ir} systems.

\subsection{The Landscape of Explainable Information Retrieval}
\label{sec:intro-landscape}

Although interpretability in \gls{ir} is a fairly recent phenomenon, there has been a large amount of growing yet unorganized work that covers many tasks and aspects of data-driven models in \gls{ir}.
This survey aims to collect, organize and synthesize the progress in \gls{exir}  in the last few years.
\Gls{exir} has quite a diverse landscape owing to the continued and sustained interest in the last few years. The initial approaches in \gls{exir} were adaptations of widely popular feature-attribution approaches (e.g., LIME~\cite{ribeiro:2016:kdd:lime} and SHAP's~\cite{lundberg:2018:treeshap}). 
However, in the following years, there has been a multitude of approaches that tackle specific problems in \gls{ir}.
We cover a wide range of approaches, from \textit{post-hoc approaches} (cf.~\cref{sec:feature-attribution,,sec:adversarial-examples,,sec:free-text}), grounding to \textit{axiomatic approaches} (cf.~\cref{sec:axioms}), to \textit{interpretable-by-design} methods (cf.~\cref{sec:explainable-by-architecture} and  \cref{sec:rationale-based-methods}). 

\subsection{Methodology and Scope}
\label{sec:scope}

Before we started our literature review, we needed to collect a corpus of relevant papers for \gls{exir} and delineate the boundaries of the review. 

\subsubsection{Corpus Creation.}
We started with very first works in \gls{exir} (e.g.,~\cite{singh:2020:fat:intentmodel,singh:2019:wsdm:exs,cummins:2007:ai:axiomaticComparison}), to build up an initial pool of papers. 
We did then forward search from this initial set of papers that mention terms ``(\textit{explain*} OR \textit{interpretab*} OR \textit{explanation*} OR \textit{transparen*})'' AND ``(\textit{retriev*} OR \textit{rank*}''. 
Secondly, we limited our search to articles published in the past five years (2018 – 2022) to provide a representative window into current best practices that have emerged since the inception of the earliest works in \gls{exir} in the following \gls{ir} venues -- ACM Special Interest Group on Information Retrieval \textit{(SIGIR)}, International Conference on the Theory of Information Retrieval \textit{(ICTIR)}, International Conference on Web Search and Data Mining \textit{(WSDM)}, Conference on Information and Knowledge Management \textit{(CIKM)}, the ACM Web Conference \textit{(TheWebConf)}.
In total, after filtering, we ended up with $68$ papers that we consider in this review that are partially relevant.
A subset of $32$ papers of those partially relevant papers find more detailed treatment in this survey.

\subsubsection{Scope.}
We note that many of the methods in \gls{exir} have methodological overlap with those invented in \gls{ml}, \gls{nlp}, and \gls{rs} communities. 
In fact, most of the approaches in \gls{exir} are based on seminal papers in these communities.
We only focus on core-\gls{ir} issues in this survey and, wherever possible, clearly spell out the distinctions from similar approaches in NLP, RS and ML in general. 
Rationale-based models have been heavily investigated in NLP. We cover only the methods popularized in \gls{ir}-centric or venues. 
Our survey focuses on rationale-based models, i.e., document-ranking tasks, in \gls{ltr}, and tasks that rely on a retrieval component. Also, \gls{rs} have a lot of work and even surveys in explainability~\citep{zhang:2020:explainable-recommendation-book}. We only survey those approaches that are useful for query modeling in query-based systems.
The papers on the topics of personalization search or explainable \gls{rs}, although they can be considered as user modeling applications of \gls{exir}, were not selected due to either lack of specific interpretability methods or being more suitable to be classified into a relatively independent field of study.
We also exclude \gls{ir} approaches dealing with image or multi-modal data.

\section{Notions and Categorization}
\label{sec:categorization}

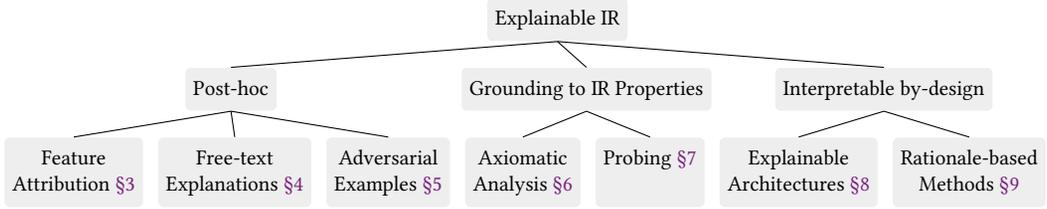
\begin{figure}
\centering
{\footnotesize
\definecolor{nodefill}{RGB}{238, 238, 238}
\forestset{qtree/.style={for tree={parent anchor=south, 
           child anchor=north,align=center,inner sep=3pt, fill=nodefill, rounded corners=2pt}}}
\begin{forest} , baseline, qtree
    [Explainable IR
        [Post-hoc
            [Feature\\Attribution~\pref{sec:feature-attribution}]
            [Free-text\\Explanations~\pref{sec:free-text}]
            [Adversarial\\Examples~\pref{sec:adversarial-examples}]
        ]
        [Grounding to IR Properties
            [Axiomatic\\Analysis~\pref{sec:axioms}]
            [Probing ~\pref{sec:probing}\\~]
        ]
        [Interpretable by-design
            [Explainable\\Architectures~\pref{sec:explainable-by-architecture}]
            [Rationale-based\\Methods~\pref{sec:rationale-based-methods}]
        ]
    ]
\end{forest}
}
\caption{Categorization of explainable IR approaches, where § indicates the section the approach is discussed.}
\label{fig:categorization_tree}
\end{figure}

We start the survey by first introducing the notions and terminologies that are commonly used in \gls{exir}.
Note that most of the terminologies in \gls{exir} are adapted from the general area of interpretable machine learning~\cite{molnar:2022:interpretable-ML-book}, explainable vision~\cite{wojciech:2019:explainable-ai-book}, natural language processing~\cite{Sogaard:2021:explainable-nlp}, and recommendation systems~\cite{zhang:2020:explainable-recommendation-book}.
We harmonize the differences in the categorizations used in these areas to distill a specific method-centric classification of all approaches used in \gls{exir} in \cref{fig:categorization_tree}.
Our classification permeates the binary divides of post-hoc and interpretable-by-design approaches by covering \gls{ir}-specific dimensions of axiomatic characterization and free-text explanations.

\subsection{Notions in Explainable Information Retrieval}
\label{sec:notions}

Explanations are the outputs of an \textit{interpretable machine learning} procedure or an \textit{interpretability method}. 
In general machine learning, explanations vary in \textit{scope} and \textit{type}.
The scope of an explanation can be a single instance or the entire dataset.
The type of explanation refers to the style or form of the explanation.
Notions in \gls{exir} share commonalities for the most part with general XAI.
However, there are some variations due to different tasks, inputs, and output types in \gls{ir}.
In the following, we describe these IR-specific notions pertaining to explainability.

\subsubsection{Local vs global interpretability}
Local interpretability refers to \textit{per-instance} interpretability. 
For the task of document ranking, an individual query is usually considered as a single instance even though multiple decisions might be involved (e.g., multiple query-document pairs and multiple preference pairs). 
Specifically, local interpretability aims to explain the model decisions in the locality of a specific query. 
On the other hand, global interpretability refers to the case when there is no distinction across instances/queries in terms of model parameters, input spaces, etc. 

\subsubsection{Pointwise, Pairwise, Listwise.} 
Ranking models output a ranked candidate list for a given query.
Therefore, the explanation of pointwise methods can only explain the models' decision of a single element in the list; while pairwise methods intend to explain the model's preference of a candidate pair.
The explanation of listwise methods, however, aims to cover all individual decisions in the entire ranking list.

\subsubsection{Type of Explanations}
A model decision can be explained differently in terms of input features, training data, model parameters, or human-understandable decision structures. 
When an explanation method measures the contribution of each feature in the input instance leading to a specific decision, the generated explanation can be a \textit{feature attribution}. 
On the one hand, feature attributions can be soft masks, i.e., real numbers denoting feature importance.
On the other hand, they can also be presented as boolean or hard masks where a feature is either present or absent in the explanation.
An explanation is understandable to humans or users based not only if the feature space is understandable but also if the explanation is small. 
An attribution over a feature space of hundreds of dimensions is hard to interpret, even if it is over words and phrases that are themselves understandable.
In IR, we typically deal with long text documents, and using feature attributions and sparsity is a key design criterion.
Explanation procedures can enforce sparsity constraints to have short extractive attributions or generate a small set of words or terms called \textit{free-text} explanation.
 Unlike feature-based explanations, explanations can be in terms of input instances. 
\textit{Contrastive} explanations are such types of explanations where the objective is to generate example instances with minor differences from the input example but with contrasting predictions.
The value of contrastive examples as explanations is grounded in social sciences~\cite{Miller:2019:insights-from-social-science}.
Therefore, using contrastive explanations to understand model behavior is one crucial aspect of gaining more transparency into the model's decision-making process.
 Finally, rules are also one of the prevalent explanations.
We denote the explicit decision-making rules as \textit{hard-rule}, such as a decision-tree path and the well-established IR principles (axioms).
On the other hand, a \textit{soft-rule} refers to those that partially impact the model decision.

\subsection{Post-hoc Interpretability}  
\label{sec:posthoc-prelims}

Post-hoc interpretability methods explain the decisions of already trained machine learning models.
Post-hoc approaches are either \emph{model-agnostic} (\textit{black-box}) where the interpretability approach has no access to the trained model parameters~\cite{ribeiro:2016:kdd:lime,lundberg:2017:neurips:shap}, 
or \emph{model introspective} (\textit{white-box}) which have full access to the parameters of the underlying model~\cite{sundararajan:2017:integratedgradients,shrikumar:2017:icml:deeplift} or data~\cite{Koh:2017:icml:influencefunction}.
In this survey, we will review approaches for both white-box and black-box settings. 
Moreover, specifically in \gls{ir}, we make a distinction between a strongly- and weakly-agnostic setting depending on if we are provided only access to a ranking of documents or also the score of a document given a query. Most of the work in the existing literature only considers our definition of a weakly agnostic model.

\subsubsection{Methods of post-hoc interpretability.}
A dominant class of post-hoc explanation approaches output what is known as \textit{feature attributions} or saliency maps. 
Most of the white-box approaches adapt gradient-based attribution approaches with task-specific calibrations.
For black-box approaches,  explanation methods use words/sentences/passages in text retrieval and ranking, and numeric and categorical features in \gls{ltr} for modeling the feature space.
We discuss methods in detail about feature attribution in \cref{sec:feature-attribution}, free-text explanations in \cref{sec:free-text}, and adversarial examples in \cref{sec:adversarial-examples}.

\subsection{Interpretability by Design}
\label{sec:intrinsic-prelims}
A common problem with post-hoc approaches is that it is often unclear how much the model behavior is indeed understood.
In fact, \citet{rudin:2019:nature:stopexplain} advocates using \Gls{ibd} models as much as possible specifically for high-stakes decision-making. 
However, building an \gls{ibd} model that is indeed fully transparent and meanwhile maintaining competitive performance is challenging, especially for complex non-linear and over-parameterized neural models. 
We note that most proposals in literature are partially interpretable, instead of exhibiting full transparency.

\subsubsection{Explainable by Architecture vs Rationales}
Many approaches brand themselves as \gls{ibd} methods, when in fact they are partially interpretable.
On one hand, some methods have only interpretable feature interactions and score compositions~\cite{khattab:2020:sigir:colbert,formal:2021:ecir:WhiteBoxColBERT}.
On the other hand, methods choose extractive input sequences as explanations while the models themselves are non-interpretable~\cite{zhang:2021:wsdm:expred,leonhardt:2021:arxiv:hard:kuma:ranking}.
In this survey, we firstly subdivide the family of \gls{ibd} approaches by \textit{explainable by architecture} (cf. \cref{sec:explainable-by-architecture}) where components of the model architecture are partially or fully interpretable. 
Secondly, \gls{ibd} methods that enforce input feature sparsity are detailed in \cref{sec:rationale-based-methods} as \textit{rationale-based} methods.

\begin{figure}
    \centering
    \includegraphics[]{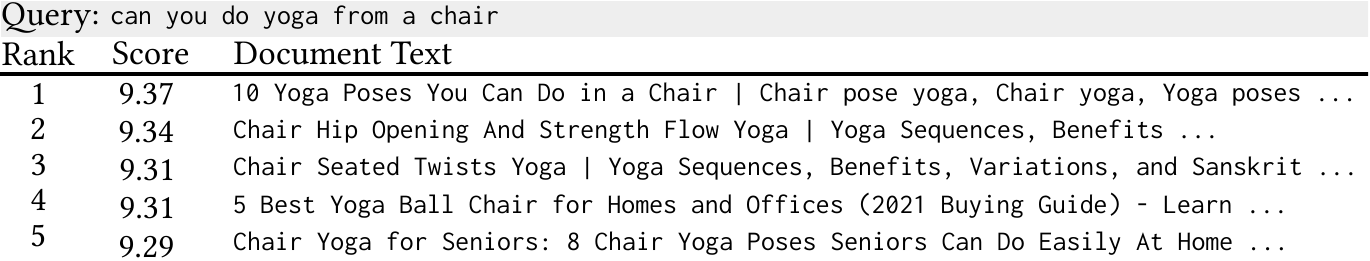}
    \caption{Example ranking result showing top-5 ranked documents with predicted relevance scores for the query~``\texttt{can~you~do~yoga~from~a~chair}''. Query and Documents are selected from TREC-DL (2021)~\citep{craswell:2021:trec-dl} and MS MARCO~\citep{nguyen:2016:msmarco}, respectively.}
    \label{fig:mainExample}
\end{figure}

\subsection{Grounding to Information Retrieval Principles}
There is a long-standing history of building text ranking models in \gls{ir}.
Most of the well-known and robust approaches for understanding relevance are based on establishing closed-formed relevance equations based on probabilistic~\cite{pontecroft:1998:sigir:language} or axiomatic foundations~\cite{bruza:1994:sigir:axiomsIR}.
A possible improve way to improve the transparency of data-driven complex \gls{ml} models is to determine if the learned models adhere to well-understood \gls{ir} principles.
Towards this, there are two streams of research efforts that attempt to ground the predictions of learned ranking models into axioms or probing models for known relevance factors of matching, term proximity, and semantic similarity. We review these approaches in \cref{sec:axioms,sec:probing}. Note that the methods utilizing \gls{ir} principles can be implemented in both post-hoc and IBD manner.

\subsection{Evaluation of Explanations}
\label{sec:notions-eval}

Evaluation of interpretability or explainability approaches has long been an arduous and challenging task.
There is no agreed-upon set of experimental protocols leading to various design decisions due to a lack of ground truths and differences in the perceived utility, stakeholders, and forms.
~\citet{doshi:2017:arxiv:towards} classify evaluation approaches as application-grounded, human-grounded, and functionality-grounded. 
The difference between application- and human-grounded evaluations is using experts and non-experts as evaluation subjects. 
Functionality-grounded evaluation does not involve humans and relies on a closed-form definition of interpretability that serves as a proxy to evaluate the explanation quality.
We introduce the following three classes of evaluation strategies employed in \gls{exir}. 

\subsubsection{Human evaluation} 
Most current papers in \gls{exir} involve human evaluation, but primarily do not differentiate between expert- and non-expert users. 
Evaluations can be simply \textit{anecdotal}. 
In this case, example explanations are shown to users, and typically binary judgments regarding the goodness of the explanations are elicited. 
A surprising number of \gls{exir} papers claim interpretability of their approaches but conduct simple anecdotal experiments.
A more fine-grained human evaluation is to ask users to solve specific tasks with the assistance of explanations. Such an approach evaluates the \textit{utility} of the explanations or answers the question -- \textit{how helpful are the explanations in the context of a given application?}

\subsubsection{Fidelity-based Evaluation}
Fidelity measures to which degree the explanations can replicate the underlying model decisions. Fidelity is measured by generating a second prediction and computing the agreement between the actual and the generated prediction.
The second prediction could be derived from either 1) using a part of the input, 2) using a surrogate model, or 3) generating a counterfactual or adversarial example. A more fine-grained category of fidelity can include evaluating the \textit{comprehensiveness}, \textit{sufficiency}, etc. We will further discuss the detailed metrics when we come to specific methods.

\subsubsection{Reference-based Evaluation}
The lack of ground truths for explanations is a central problem in explainable AI.
Whenever the ground-truth explanations are available, we can use them as the \textit{reference} to compare with the generated explanations. 
In case of a lack of ground truth explanations, some methods choose a well-understood and fully explainable/transparent model as a \textit{reference} model. 
In such cases,  we can evaluate the truthfulness of the explanation methods by comparing the explanations generated by the reference model and the explanation method.

\section{Feature Attribution}
\label{sec:feature-attribution}

\begin{figure}[h]
    \centering
    \includegraphics[]{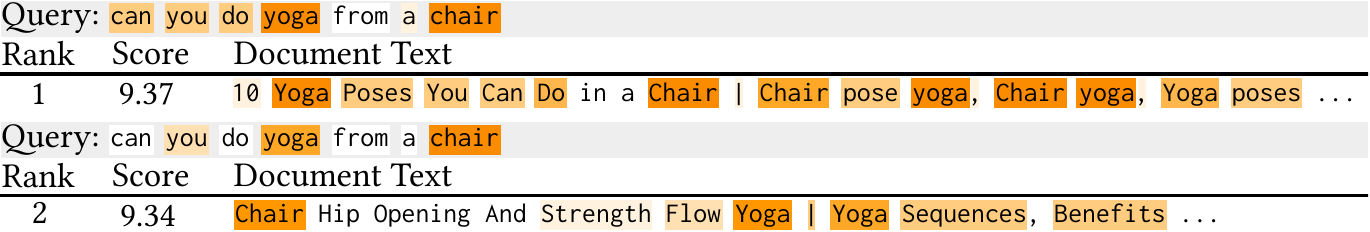}
    \caption{A fictive example using a heatmap to visualize feature attributions for the top-2 ranked documents for the query~``\texttt{can~you~do~yoga~from~a~chair}''. Feature importance is highlighted in orange.}
    \label{fig:exampleFeatureAttr1}
\end{figure}

Feature attribution methods, also known as feature importance or saliency methods, typically generate explanations for individual predictions by attributing the model output to the input features.
A scalar representing the importance is assigned to each input feature or groups of input features.
These scores are then commonly visualized using a heatmap or a bar chart, informing the user about which features the model's prediction is most sensitive to.
\cref{fig:exampleFeatureAttr1} demonstrates example feature attributions for the top-2 ranked documents, following the example from \cref{fig:mainExample}.
Feature attribution methods have been found to be the most popular explanation technique and are used in many domains~\citep{bhatt:2020:xaideployment}.
However, as is common for interpretation techniques, most feature attribution methods were originally designed to explain the predictions of classification models.

Recent work explores how such methods can be applied or adapted to explain the output of ranking models, where feature importance scores can be computed for the query or document input individually, or for both, as shown in \cref{fig:exampleFeatureAttr1}.
Following our categorization (\cref{sec:categorization}), we differentiate between model-agnostic and model-introspective feature attribution methods.

\subsection{Model-agnostic Feature Attribution}
\label{sec:feature_attribution_model-agnostic}
A variety of feature attribution methods generate explanations in a model-agnostic way by perturbing input features and observing the change in the model output.
The underlying model is treated as a black box.

\subsubsection{Feature Ablation}
\label{sec:feature-attribution-model-agnostic-ablation}
Feature ablation is a simple perturbation-based approach to computing importance scores.
Individual (or groups of) input features are removed one at a time, and an importance score is assigned based on the observed difference between the model predictions.

To interpret a BERT-based ranking model, \citet{qiao:2019:understanding-behavior-bert-ranking} compute the importance of tokens through feature ablation.
To produce feature importance scores, they compare the ranking score of an unmodified document with the ranking score for the same document when removing a randomly chosen input token.
Specifically, they only remove input tokens corresponding to regular words and keep all tokens that are special tokens or correspond to stopwords.
They find that the ranking score produced by a BERT model depends on only a few tokens in each document.
The ranking score often decreases significantly when these tokens are removed.
When manually examining the important tokens, the authors find that they often correspond to exact match terms, i.e., terms that also appear in the input query, and terms in close semantic context.
In contrast, when examining token importance scores for a neural ranker based on convolutions and interactions~\citep{dai:2018:convolutional} that soft-matches n-grams for ad-hoc search, the most important terms appear to be rather loosely related to the input query.

\subsubsection{Surrogate Models}
\label{sec:feature-attribution-model-agnostic-surrogate}
Local Interpretable Model-agnostic Explanations~(LIME)~\citep{ribeiro:2016:kdd:lime} is an interpretability method that generates explanations by training a surrogate model on a dataset of perturbed samples to locally approximate the behavior of the underlying black-box model.
Typically, a linear model, preferably sparse, is chosen as the interpretable surrogate model since the weights directly specify the importance of each feature.
Using LIME to generate feature attributions, \citet{singh:2019:wsdm:exs} propose EXS, an explainable search system that provides explanations to users through feature attribution.
Specifically, EXS aims to provide information on three questions: 1) Why is a document relevant to the query, 2) Why is a document ranked higher than another document, and 3) What the intent of the query is according to the ranker?
LIME is designed to explain the output of a classifier, and EXS casts the output of a pointwise ranker into a classification problem by transforming query-document scores into class probabilities.
A binary classification problem is created by considering the top-$k$ documents in an input ranking as relevant and the rest as irrelevant, essentially considering document ranking as a classification problem where the black box ranker is considered as a classifier. 
\citet{polley:2021:exdocs} compare EXS with their evidence-based explainable document search system, ExDocS, which performs reranking using interpretable features.
In a user study, they found that EXS is on par with the ExDocS system in completeness and transparency metrics, although users rated ExDocS as more interpretable compared to EXS.
At the same time, the use of ExDocS resulted in a drop in ranking performance, whereas the use of EXS does not affect performance at all.

Similarly, \citet{verma:2019:sigir:lirme} adapt LIME to create locally interpretable ranking model explanations (LIRME).
In contrast to EXS, LIRME trains the local surrogate model directly on the query-document scores and does not transform them into class probabilities.
Instead, they experiment with different strategies to sample documents in the neighborhood of the document to be explained.
In their experiments, they create explanations for the output of a Jelinek-Mercer smoothed language model on the TREC-8 dataset and find that uniform or TF-IDF-biased term replacement strategies produce better explanations than replacement strategies that use term position information.

Instead of training a local surrogate model to generate explanations for individual examples, \citet{singh:2018:ears:posthocSecondary} distill an already trained black-box \gls{ltr} model into an interpretable global surrogate model that is used to generate explanations.
This global surrogate model only operates on the interpretable subset of features and is trained to mimic the predictions of the black-box ranker.
For training, they create numerous artificial training examples
In their experiments, they validate whether it is possible to train an interpretable model that approximates a complex model.
On the \gls{ltr} datasets~\cite{qin:2013:corr:letor4} they find that a faithful interpretable ranker can only be learned for certain query localities.
This showcases the limitation that simple models, even when trained with a much larger quantity of training data, are not able to faithfully explain all localities of the decision boundary of a complex model and that using local surrogate models can be advantageous.

\subsubsection{Searching for Explanations}
\label{sec:feature-attribution-model-agnostic-search}
An alternative to the above approaches is to search the space of all possible explanations, optimizing for a metric of choice.
For \gls{ltr} models, \citet{singh:2021:ValidSetLtR} propose a simple, yet effective greedy search-based approach to find explanations.
Their approach aims to find a subset of explanatory features that maximizes two measures, validity and completeness.
The validity of an explanation is defined as the amount of predictive capacity contained in a subset of explanatory features.
The idea is that the explanatory features should be sufficient to produce the original output ranking.
In fact, this measure aligns with the sufficiency metric introduced by~\citet{deyoung:2019:acl:eraser}.
The completeness metric measures whether removing explanatory features from the input significantly changes the output.
When all explanatory features are removed, it should not be possible to produce the original output ranking.
Kendall's tau rank correlation measures differences in output rankings; the underlying model is treated as a black-box.

\subsection{Model-introspective Feature Attribution}
\label{sec:feature-attribution-model-introspective}
In contrast to model-agnostic methods, model-introspective feature attribution methods require white-box access to the model being explained.
Model-introspective methods typically rely on gradients or other properties of the model to compute feature importance scores.

\subsubsection{Gradient-based Methods}
\label{sec:feature-attribution-model-introspective-gradient}
Many feature attribution methods generate an explanation by computing the gradient with respect to the input features.
This gradient reflects how a small change in the input features affects the prediction.
The vanilla gradient method can produce noisy explanations and suffers from a saturation problem.
A variety of methods aim to remedy these issues.
For example, Integrated Gradients~\citep{sundararajan:2017:integratedgradients} accumulates gradients on a path between a baseline input and the actual input.
While this resolves the saturation problem, the baseline input is a hyperparameter to be chosen carefully.
It is unclear what baseline is best, and each baseline makes assumptions about the distribution of the data and the concept of missingness in the feature space~\citep{sturmfels:2020:baselines}.
Other gradient-based feature attribution methods, such as Layer-wise Relevance Propagation~\citep{bach:2015:lrp}, Guided Backpropagation~\citep{springenberg:2014:guidedBP}, or DeepLIFT~\citep{shrikumar:2017:icml:deeplift} back-propagate custom relevance scores using modified, sometimes layer-specific, rules.

\citet{fernando:2019:sigir:studyshaponretrieval} apply DeepSHAP~\citep{lundberg:2017:neurips:shap}, a combination of SHAP~\citep{lundberg:2017:neurips:shap} and DeepLIFT~\citep{shrikumar:2017:icml:deeplift}, to neural retrieval models.
Specifically, they investigate the sensitivity of the explanations to different choices for constructing a baseline input document.
Generating explanations for a subset of queries from the TREC Robust04 test collection and the corresponding top-3 ranked documents, they find that the explanations are indeed sensitive to the baseline input.
The DeepSHAP explanations are also compared to explanations produced by LIME, and while for some baseline inputs there is high overlap in the most important features, there is a lack of overlap for others.

\citet{purpura:2021:ecir:neuralfs} use simple gradient-based feature attribution to find the most important features used by \gls{ltr} models.
They generate a saliency map for each instance in a training dataset and select feature groups by thresholding the normalized importance values.
Feature selection is then performed by counting how often each feature group is considered important across all extracted saliency maps.

\begin{figure}[h]
    \centering
    \includegraphics[]{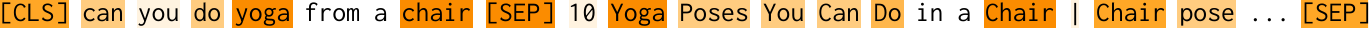}
    \caption{Example visualization of feature attributions for a single query-document pair using the BERT-style input format, which is ``[CLS] query [SEP] document [SEP]''. Important tokens are highlighted in orange.}
    \label{fig:exampleFeatureAttr2}
\end{figure}

\citet{zhan:2020:sigir:AnalysisBERTDocRanking} use Integrated Gradients~\citep{sundararajan:2017:integratedgradients} to obtain feature attributions for a BERT-based ranking model.
As a baseline input, they create an empty query and an empty document input by replacing the corresponding tokens with the special padding token ``\texttt{[PAD]}''.
An example of feature attributions for BERT-style input is visualized in \cref{fig:exampleFeatureAttr2}.

\subsubsection{Attention-based Methods}
\label{sec:feature-attribution-model-introspective-attention}
Instead of using gradients, attention-based feature attribution methods use the attention weights contained in attention layers, which are a core building block of transformer models.
The attention weights can be used to explain what part of the input a model attends to when making a prediction, for example, by visualizing the attention weights at certain layers~\citep{vig:2019:bertviz}.
However, whether attention weights actually provide explanations is subject to an ongoing debate~\citep{bibal:2022:attentionisexplanation?,bastings:2020:elephant}.

\begin{figure}[h]
    \centering
    \includegraphics[]{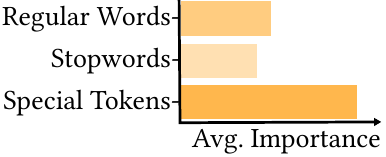}
    \caption{Example bar chart visualization of feature attributions for different groups of tokens.}
    \label{fig:exampleFeatureAttr3}
\end{figure}

\citet{qiao:2019:understanding-behavior-bert-ranking} analyze the learned attentions of BERT-based ranking models, using attention weights to measure the importance of features.
They group input tokens into three categories, as visualized in \cref{fig:exampleFeatureAttr3}:
Regular Words, Stopwords, and Markers, which are the special tokens ``[CLS]'' and ``[SEP]''.
In their experiments on the MS MARCO passage reranking dataset~\citep{nguyen:2016:msmarco}, they find that marker tokens receive the highest attention.
The importance of the marker tokens is confirmed by observing a strong decrease in model performance when they are removed from the inputs.
Stopwords appear to be as important as regular words; however, removing them does not appear to affect the ranking performance.
Additionally, they observe that the attention scores spread more uniformly across the input sequence in deeper layers of BERT, as the embeddings become more contextualized.

In addition to Integrated Gradients, \citet{zhan:2020:sigir:AnalysisBERTDocRanking} also use attention weights to obtain feature attributions for a BERT-based ranking model.
With an experimental setup similar to \citet{qiao:2019:understanding-behavior-bert-ranking}, they compute attribution scores for different groups of input tokens: The special ``[CLS]'' and ``[SEP]'' tokens, the query tokens, the document tokens, and the period token.
While confirming that a significant amount of attention weight is distributed to the special tokens and the period token, the authors also find that the attributions produced using attention weights are negatively correlated with the attributions produced by Integrated Gradients.
Based on their results, the authors speculate that these tokens receive high attention weights due to their high document frequency.
They argue that the model dumps redundant attention on these tokens, while these actually carry little relevance information.

\subsection{Evaluating Feature Attributions}
Input feature attributions can be evaluated in many ways. However, there is little agreement on which evaluation strategy is best.
Sanity-checks~\citep{adebayo:2018:sanity-checks, tomsett:2020:saliency-metrics} test functionally grounded assumptions behind feature attributions.
Whether feature attributions are faithful to the model that is explained can also be evaluated by removing important features and re-evaluating model performance, either with or without retraining~\citep{hooker:2019:roar-bechnmark, madsen:2021:eval-faithfulness, Rong:2022:eval-attribution-information-theory}.
However, if the model is not retrained, removing or replacing features can result in out-of-distribution inputs.
Other works propose shortcut, artifact, or spurious correlation detection tasks to evaluate feature attributions~\citep{yang:2019:bam-benchmark, bastings:2021:arxiv:shortcut, idahl:2021:utility, adebayo:2022:ineffective}, where bugs are added to a model on purpose and then used as ground-truth for explanation evaluation.
Feature attribution methods that rely on surrogate models need to evaluate their fidelity, that is, how well the surrogate model approximates the black box model being explained.
Unfortunately, the evaluation of feature attributions in \gls{ir} is often limited to anecdotal examples.
\citet{singh:2019:wsdm:exs} neither evaluate the explanation quality of EXS nor the fidelity of the local surrogate models used to generate explanations.
\citet{verma:2019:sigir:lirme} evaluate LIRME by comparing the explanations to a reference of important terms obtained from relevance judgments but also do not explicitly evaluate faithfulness.
\citet{fernando:2019:sigir:studyshaponretrieval} include an analysis of the faithfulness of LIME explanations for neural ranking models by measuring accuracy and mean-squared error of the local surrogate model.
To evaluate explanations produced by DeepSHAP, they use LIME explanations as a reference.
Directly optimizing explanations based on evaluation metrics, as done by \citet{singh:2021:ValidSetLtR}, seems advantageous, but does not provide any guarantees of finding a good explanation.
Based on the limited work on evaluating feature attributions in \gls{ir}, we argue that claims and hypotheses based on insights from feature attribution explanations should be handled with caution unless the explanation methodology has been evaluated rigorously.

\begin{table}[H]
	\centering
	\small
    {\renewcommand{\arraystretch}{1.2}
	\begin{tabular}{llcc}
		\hline
    		\bf Approach & \bf Task & \bf Explanation & \bf Evaluation \\
		\hline
			EXS~\citep{singh:2019:wsdm:exs} & Text Ranking & Feature Attribution & Anecdotal\\
            LIRME~\citep{verma:2019:sigir:lirme} & Text Ranking & Feature Attribution & Anecdotal/Reference\\
            DeepSHAP~\citep{fernando:2019:sigir:studyshaponretrieval} & Text Ranking & Feature Attribution & Reference\\
            Attention~\citep{qiao:2019:understanding-behavior-bert-ranking, zhan:2020:sigir:AnalysisBERTDocRanking} & Text Ranking & Feature Attribution & Visualization\\
            Global Surrogate Model~\citep{singh:2018:ears:posthocSecondary} & LTR & Global Feature Attribution & Faithfulness \\
            Greedy Search~\citep{singh:2021:ValidSetLtR} & LTR & Feature Attribution & Sufficiency/Completeness\\
            Gradient Saliency~\citep{purpura:2021:ecir:neuralfs} & LTR & Feature Attribution & Faithfulness\\
            Intent Modeling~\citep{singh:2020:fat:intentmodel} & Text Ranking & Terms/Words & Faithfulness/Reference \\
            CtrsGen~\citep{zhang:2020:cikm:intentGeneration} & Text Ranking & Free-text & Reference \\
            GenEx~\citep{rahimi:2021:arxiv:explainQuery} & Text Ranking & Free-text & Reference/Human \\
            LiEGe~\citep{yu:2022:sigir:generateListExplanation} & Text Ranking & Topic Words & Reference \\
            Universal Adv. Triggers~\citep{wang:2022:ictir:bert} & Text Ranking & Trigger & Anecdotal/Visualization \\
        \hline
    \end{tabular}
    }
	\caption{\small{Overview of post-hoc explanation methods. The evaluation of post-hoc methods can be \textit{anecdotal, visualized}, or can be intrinsically measured by a corresponding \textit{faithfulness} measure.
	\textit{``Reference''} refers to comparison with ground-truth explanations, an interpretable model, or another attribution method.}}
\end{table}

\section{Free-text Explanations}
\label{sec:free-text}
Free-text explanations methods aim to generate explanations using natural language and are thus also called natural language explanations.
Compared to feature attributions, the explanations can be more expressive, as they are not limited to words that already contain the input.
Typical free-text explanations are not more than a few sentences long, and sometimes even limited to a few words.
This form of explanation is popular for both textual and visual-textual tasks, for which a variety of datasets have been collected or expanded to include explanations~\citep{wiegreffe:2021:teachmetoexplain}.
However, apart from a few question-answering datasets, none of them are closely related to \gls{ir}.
Instead, this explanation style is commonly used for tasks that involve reasoning.
Since for such tasks, the information contained in the inputs is often insufficient to achieve good task performance, the explanations must also contain external information apart from what is contained in the inputs.
In fact, many datasets that include free-text explanations are used to improve the task performance of the model.
The idea is that a model will generalize better if it can also explain its predictions~\citep{camburu:2018:e-snli, kumar:2020:nile,liu:2019:generative-explanation-framework, rajani:2019:explain-yourself}.

Approaches to generating free-text explanations for text ranking models focus either on interpreting the query intent as understood by a ranking model or on producing a short text summary to explain why an individual document or a list of documents is relevant.  

\subsection{Explaining Query Intent}
\label{sec:free-text-query-intent}
Satisfying the information need of a user that issues a search query is a key concept in \gls{ir}.
Explaining the intent as understood by black box ranking models can be useful to examine whether complex ranking models perform in accordance with a user's intent.

\subsubsection{Query Expansion}
\label{sec:free-text-query-intent-expansion}
\begin{figure}[h]
    \centering
    \includegraphics[]{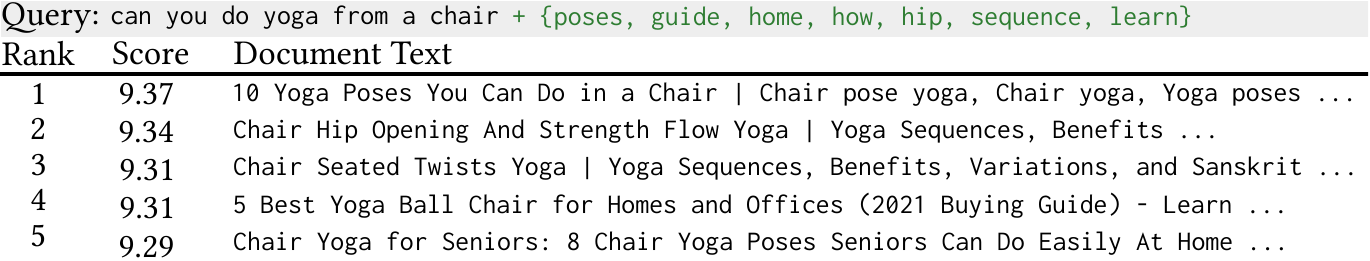}
    \caption{Example of query expansion terms (green) as explanations. The expansion terms are chosen such that an interpretable term-based ranker best approximates the ranking of documents produced by a black-box ranking model.}
    \label{fig:exampleIntentModelling}
\end{figure}

\citet{singh:2020:fat:intentmodel} propose a model-agnostic approach to interpret a query intent as understood by a black-box ranker.
Given a single query and a set of expansion terms as input, they fit an interpretable term-based ranking model to mimic the complex model to be interpreted.
The goal is to identify a set of query expansion terms such that most of the pairwise preferences in the output ranking are preserved.
Query expansion terms are selected by optimizing the preference pair coverage using greedy search.
The expanded query terms act as an explanation for the intent perceived by the black-box ranking model, as \cref{fig:exampleIntentModelling} demonstrates.
In experiments with a variety of ranking models, including RM3~\citep{lavrenko:2001:rm3}, DESM~\citep{nalisnick:2016:DESM}, DRMM~\citep{guo:2016:cikm:drmm}, P-DRMM~\citep{mcdonald:2018:P-DRMM}, trained on the Robust04 collection~\citep{voorhees:2006:trec:robust}, they show that this approach can produce explanations with high fidelity.

\subsubsection{Generating Query Descriptions}
\label{sec:free-text-query-intent-descriptions}
\citet{zhang:2020:cikm:intentGeneration} introduce a Query-to-Intent-Description task for query understanding.
Given a query and a set of both relevant and irrelevant documents, the goal is to generate a natural language intent description.
To solve this task, they propose CtrsGen, a contrastive generation model that generates a query intent description by contrasting the relevant and irrelevant documents.
The training data for CtrsGen consists of multiple TREC and SemEval~\citep{chatterjee:2019:semeval} collections that already include query descriptions.
Although not explicitly discussed by~\citet{zhang:2020:cikm:intentGeneration}, CtrsGen can be used to explain query intents as understood by a black-box ranker by selecting relevant and irrelevant documents based on the output of the ranking model.
However, it has not yet been examined whether the generations of CtrsGen explain the underlying ranking model faithfully.

\subsection{Explaining Document Relevance}
\label{sec:free-text-document-relevance}
A recent line of work in explainable document retrieval aims to explain why a document or a set of documents is considered relevant to a query by generating free-text explanations.
Compared to other model-agnostic explanation methods, free-text explanations are not limited to explaining document relevance using features that are already contained in the input.
A user study by~\citep{rahimi:2021:arxiv:explainQuery} suggests that adding free-text document relevance explanations to search engine result pages can help users identify relevant documents faster and more accurately.

\subsubsection{Pointwise Explanations}
\citet{rahimi:2021:arxiv:explainQuery} generate document relevance explanations for individual query-document pairs.
They propose GenEx, a transformer-based model that outputs free-text document relevance explanations.
Given a query-document pair, GenEx learns to generate a text sequence that explains why the document is relevant to the query.
The explanations consist of only a few words instead of whole snippets, and explicitly avoid reusing the terms already contained in the query.
The model uses an encoder-decoder architecture, with the decoder being extended by a query-masking mechanism to decrease the probability of
generating tokens that are already contained in the query.
The training data consists of query-document-explanation triplets and is automatically constructed from Wikipedia articles and the ClueWeb09 dataset~\citep{callan:2009:clueweb09}.

\subsubsection{Listwise Explanations}
\citet{yu:2022:sigir:generateListExplanation} argue that explaining documents independently is inherently limited.
Per-document explanations do not explain differences between documents, and a single document can potentially cover multiple query aspects at the same time.
As a solution, they propose a listwise explanation generator~(LiEGe) that for a given query jointly explains all the documents contained in a ranked result list.
LiEGe is based on an encoder-decoder transformer architecture and uses pre-trained weights from BART~\citep{lewis:2020:BART}.
The authors introduce two settings for search result explanations: 1) comprehensive explanation generation, where the explanation contains all query aspects covered by each document, and 2) novelty explanation generation, where the explanation contains a description of the relevant information of a document that is novel, considering all the preceding documents in the ranked list.
Two weakly labeled datasets are constructed from Wikipedia to train LiEGe for these two settings, the evaluation dataset is constructed using query logs from the MIMICS dataset~\citep{Zamani:2020:MIMICS}.

\subsection{Evaluation of Free-text Explanations}
The evaluation of free-text explanations is generally based on the availability of ground-truth explanations.
Although explanations are not included in most \gls{ir} datasets, proxy explanations can be created from query descriptions, query aspect annotations, topic annotations, or click logs~\citep{zhang:2020:cikm:intentGeneration, rahimi:2021:arxiv:explainQuery, yu:2022:sigir:generateListExplanation}.
BLEU~\citep{Papineni:2002:BLEU} and ROUGE~\citep{lin:2004:ROUGE}, two metrics commonly used to evaluate text summarization and machine translation tasks, can be used to compare generated free-text explanations with reference explanations.
Furthermore, \citet{rahimi:2021:arxiv:explainQuery} and \citet{yu:2022:sigir:generateListExplanation} use BERTScore~\citep{Zhang:2020:BERTScore} to measure semantic coherence.
However, human-annotated but model-independent ground-truth explanations can only be used to evaluate the plausibility of generated explanations.
Whether the generated explanations are faithful to the ranking model being explained remains an open question.
Only \citet{singh:2020:fat:intentmodel} evaluate the faithfulness of their query intent explanations since they have to ensure that the interpretable ranker used during optimization closely mimics the black-box ranking model being explained.
To examine whether GenEx explanations actually help users, \citet{rahimi:2021:arxiv:explainQuery} conduct a user study.
Specifically, they collect explanation preferences, linguistic quality ratings, and relevance judgments from crowd-workers, comparing GenEx explanations with different baseline explanations.

\section{Adversarial Examples}
\label{sec:adversarial-examples}

Adversarial examples are commonly used to demonstrate the fragility or robustness of machine learning models.
However, they can also serve as explanations and provide valuable insight.
In fact, adversarial examples are closely related to counterfactual examples, but instead of providing actionable recourse, the goal is to fool machine learning models.
Given an individual input to a model, a corresponding adversarial example is crafted by applying small deliberate perturbations to deceive a model into making a wrong prediction.
The resulting adversarial examples inform about the minimal input changes required to change a prediction and thus provide insight into the decision behavior of the model.
Specifically, the adversarial perturbations indicate which input features have to change by how much to alter a predicted outcome.
Compared to feature attributions~(\cref{sec:feature-attribution}), adversarial explanations are contrastive explanations, since the adversarial example is always compared to the unmodified input example.
From the perspective of social science, \citet{Miller:2019:insights-from-social-science} argues that such contrastive explanations can be considered more human-grounded.

\subsection{Adversarial Examples in Ranking}
\label{sec:adversarial-examples-ranking}
Most of the work on adversarial examples is concerned with classification tasks, where a wrong prediction is defined by comparing the predicted label with a target label.
For ranking tasks, the main objective of an adversarial perturbation is to cause a relatively large rank promotion or rank demotion of a document.
For example, a company aiming to optimize search engines could leverage adversarial attacks to promote a specific web page to the top of a search result page with minor changes in the page content itself.

\citet{raval:2020:adv-attack-retrieval-models} generate adversarial examples for black-box retrieval models that lower the position of a top-ranked document with minimal changes to the document text.
Given the non-differentiability of replacing discrete tokens, they optimize adversarial examples using a stochastic evolutionary algorithm with a one-token-at-a-time replacement strategy.
\citet{wu:2022:arxiv:prada} take a different approach by training a surrogate model based on pseudo-relevance feedback, which is used to approximate the gradient of the underlying black box ranking model.
This approximated gradient is then used to find adversarial perturbations that promote a target document.
Additionally, the adversarial perturbations are restricted by semantic similarity to the original document.
The authors argue that the perturbations are imperceptible and evade spam detection when constraining the perturbations to semantic synonyms.
\citet{goren:2020:sigir:ranking} craft adversarial examples for the LambdaMART \gls{ltr} model.
For a given query, they use past rankings to create perturbations by replacing passages in the target document with passages from other high-ranked documents. 
\citet{wang:2022:ictir:bert} use gradient-based optimization to generate adversarial examples for BERT-based ranking models.
They add or replace a few tokens in documents that cause significant rank promotions and demotions.

\subsection{Universal Adversarial Triggers}
\label{sec:adversarial-triggers}
While adversarial examples focus on input perturbations that change the prediction of individual inputs, universal adversarial triggers~\citep{wallace:2019:emnlp:universal} are input-agnostic perturbations that lead to a model making a specific prediction whenever the trigger is concatenated to any input.
Starting from an initial sequence of tokens, a trigger is optimized via a gradient-based search algorithm that iteratively replaces tokens.
The effect of replacing a discrete token is usually approximated using HotFlip~\citep{ebrahimi:2018:acl:hotflip}.
Since the resulting triggers transfer across input examples, they can be used to explain the global behavior of a model and can reveal global patterns.

\subsubsection{Universal Triggers for Text Ranking}
\citet{wang:2022:ictir:bert} adapt universal adversarial triggers for text-based ranking models.
They propose a global ranking attack to find trigger tokens that are adversarial to all queries contained in a dataset.
Specifically, they optimize a fixed-length trigger so that any document to which it is concatenated will be demoted (or promoted) as much as possible for any given query.
In their experiments with BERT-based ranking models fine-tuned on ClueWeb09~\citep{callan:2009:clueweb09} and MS MARCO~\citep{nguyen:2016:msmarco}, they discover topical patterns within and between datasets and expose potential dataset and model biases.
For example, the trigger
\begin{center}
    \texttt{hinduism earthquakes childbirth tornadoes Wikipedia}
\end{center} promotes a document by 63 ranks on average, and the trigger 
\begin{center}
    \texttt{acceptable competition rayon favour \#\#kei} 
\end{center}
demotes a document by 84 ranks on average across all queries.
In general, finding triggers for which highly relevant documents get demoted appears easier than finding triggers for which low-ranked documents are promoted.

\section{Axiomatic Analysis of Text Ranking Models}
\label{sec:axioms}

Unlike current data-driven, parameterized models for relevance estimation, traditional IR approaches to ranking involve probabilistic models of relevance such as BM25 \citep{amati:2002:tois:bm25} and axiomatic approaches.
Both approaches have a top-down defined notion of relevance, allowing for some sort of interpretability.
Yet, the probabilistic models are currently dominant and axiomatic approaches less popular.
In contrast to the recent development of neural, and therefore less interpretable, rankers, \emph{Axiomatic IR} postulates and formalizes the properties of principled rankers.
The term \emph{axiom} in IR was first coined by \citet{bruza:1994:sigir:axiomsIR}, who proposed to describe retrieval mechanisms using axioms expressed through concepts in the field of \gls{ir}.

\begin{figure}[h]
    \centering
    \includegraphics[]{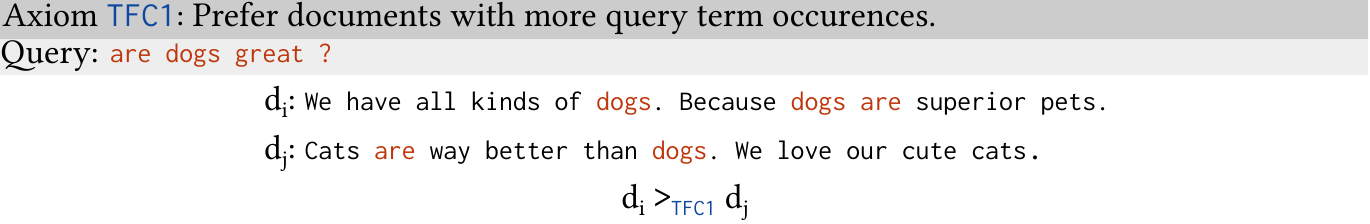}
    \caption{Example of applying the TFC1~\citep{fang:2004:sigir:studyIR} axiom to rank two documents. Query terms are highlighted. $d_i$ is ranked higher than $d_j$ because it contains more query terms.}
    \label{fig:axiom_example}
\end{figure}

An example axiom is TFC1~\citep{fang:2004:sigir:studyIR} which proposes to prefer documents having more query terms occurrences (\cref{fig:axiom_example}). Formally, given a query $q = {t}$ and two documents $d_1, d_2$ with $|d_1| = |d_2|$, TFC1 is defined as

\begin{equation}
    tf(t, d_1) > tf(t, d_2) \Rightarrow d_1 >_{TFC1} d_2.
\end{equation}

Similarly, a large set of axioms has been proposed in recent decades, ranging over different aspects of relevance such as term frequency \citep{fang:2004:sigir:studyIR, fang:2011:tis:diagnosticIR}, document length \citep{fang:2004:sigir:studyIR}, semantic similarity \citep{fang:2006:sigir:semanticTermMatch}, or term proximity \citep{hagen:2016:cikm:aximaticReranking} among others (see \cref{table:axioms}).
For a more detailed description of the various axioms, we refer to an overview by \citet{bondarenko:2022:sigir:AxiomRetrievalIRAxioms}. 

\begin{table}[tb]
\centering
\small
{\renewcommand{\arraystretch}{1.2}
\begin{tabular}{@{}lll@{}}
\toprule
\bfseries Property             & \bfseries Axiom     & \bfseries Details \\
\midrule
Term Frequency      & TFC1~\cite{fang:2004:sigir:studyIR}  & \begin{tabular}[c]{@{}l@{}}Prefer documents with more query term occurrences\\ 
                                Given $Q={q}, |D_1|=|D_2|$, 
                                $tf(q,D_1) > tf(q,D_2)$ 
                                $\Rightarrow D_1 >_{TFC1} D_2$
                                \end{tabular} \\ \hline
Document Length     & LNC1~\cite{fang:2004:sigir:studyIR}  & \begin{tabular}[c]{@{}l@{}}Penalize long documents for non-relevant terms\\ 
                                Given $t \notin Q$, arbitrary term $w$, 
                                $tf(t,D_1) > tf(t,D_2) \land$ \\ $\forall_{w\neq t}tf(w,D_1)=tf(w,D_2)$ 
                                $\Rightarrow D_1 >_{LNC1} D_2$
                                \end{tabular} \\ \hline
Semantic Similarity & STMC1~\cite{fang:2006:sigir:semanticTermMatch} & \begin{tabular}[c]{@{}l@{}}Prefer terms more similar to query terms\\ 
                                Given $Q={q},D_1={t_1}, D_2={t_2},q\neq t_1, q\neq t_2$, 
                                $sim(q,t_1) > sim(q,t_2)$ \\
                                $\Rightarrow D_1 >_{STMC1} D_2$
                                \end{tabular} \\ \hline
Query Aspect        & AND~\cite{zheng:2010:ecir:queryTermWeight}   & \begin{tabular}[c]{@{}l@{}}Prefer documents containing all query terms\\ 
                                Given $Q={q_1,q_2}, td(q_1) \geq td(q\_2)$, 
                                $tf(q_1,D_1)=1 \land tf(q_2,D_1) = 1$ \\ $\land tf(q_1,D_2) > 1 \land tf(q_2,D_2) = 0$ 
                                $\Rightarrow D_1 >_{AND} D_2$ \\
                                \end{tabular} \\ 
\bottomrule
\end{tabular}
}
\caption{Selection of proposed retrieval axioms. Adapted from a more complete list of axioms available in \citep{bondarenko:2022:sigir:AxiomRetrievalIRAxioms}.}
\label{table:axioms}
\end{table}

Axioms are human-understandable concepts. This is in stark contrast to neural networks, which have been shown time and time again to learn spurious correlations \citep{geirhos:2020:natmachintell:shortcutlearningDNNs} and to be susceptible to adversarial attacks \citep{wang:2022:ictir:bert}.
Although not yet achieved, a long-term goal of axiomatic \gls{ir} could be a concept of relevance built on axioms.
This conceptualization of relevance would then be robust to attacks, generalize to novel distributions, and be interpretable for humans.

Although there is no general model of relevance yet, previous work aggregated axioms to build axiomatic rankers (\cref{sec:axioms-axiomatic-rankers}), analyze and explain existing neural ranking approaches by aligning them to known axioms (\cref{sec:axiom-diagnostics}), and use axioms to regularize the training of neural rankers (\cref{sec:axiom-regularization}).
An overview of this classification and papers in this section can be found in \cref{table:axiom_papers}.

\begin{table}[tb]
\centering
\small
{\renewcommand{\arraystretch}{1.2}
\begin{tabular}{@{}llccc@{}}
\toprule
\bfseries Paper             & \bfseries Task              &  \bfseries Approach         & \bfseries Dataset          & \bfseries Evaluation \\
\midrule
\citet{hagen:2016:cikm:aximaticReranking}                                                        & LTR          & IBD & TREC Web tracks 2009-2014                           & -               \\
\citet{rennings:2019:ecir:axiomDiagnoseNeural}                            & Text Ranking & Post-hoc                     & WikiPassageQA                                       & -               \\
\citet{camara:2020:ecir:diagnoseBert}                                     & Text Ranking & Post-hoc                     & TREC 2019 DL                                        & -               \\
\citet{voelske:2021:ictir:axiomaticExplanations4NeuralRanking}                       & Text Ranking & Post-hoc & Robust04, MS MARCO                                  & Fidelity        \\
\citet{rosset:2019:sigir:axiomaticNeural}               & Text Ranking              & Regularization   & MS MARCO                                            & -               \\
\citet{cheng:2020:ictir:AxiomPerturbNeuralRanking} & Text Ranking & Regularization   & WikiQA, MS MARCO                                    & -               \\
\citet{chen:2022:sigir:axiomResularizationPreTraining}     & Text Ranking & Regularization   & MS MARCO, TREC 2019 DL & Anecdotal      \\
\bottomrule
\end{tabular}
}
\caption{Classification of axiomatic methods. The evaluation w.r.t. interpretability can be \textit{anecdotal} or intrinsically measured by a corresponding \textit{faithfulness} measure.}
\label{table:axiom_papers}
\end{table}

\subsection{Interpretable Axiomatic Rankers}
\label{sec:axioms-axiomatic-rankers}

\citet{hagen:2016:cikm:aximaticReranking} is one of the first to operationalize retrieval axioms to perform axiomatic re-ranking. By learning the importance of individual axioms, they aggregate the axioms' partial orderings.
Despite being inherently more interpretable, they evaluate their axiomatic re-ranking step with a selection of retrieval models, showing that for most of them the performance significantly increases.
Given that the axioms and the aggregation method are fully interpretable, the resulting re-ranking is also fully interpretable.
\citet{bondarenko:2022:sigir:AxiomRetrievalIRAxioms} proposed a utility library called \emph{ir\_axioms} that allows experimenting with a collection of 25 different axioms and allows one to add new axioms.
This library can be used for axiomatic result re-ranking and diagnostic experiments to explain neural ranking models.

\subsection{Axioms for Model Diagnostics}\label{sec:axiom-diagnostics}
More directly related to the classical post-hoc interpretability work is a line of recent works diagnosing and explaining ranking models using axioms. 
\citet{rennings:2019:ecir:axiomDiagnoseNeural} constructed \textit{diagnostic datasets} based on existing axioms and checked whether classical neural ranking models are in agreement with the axiomatic rules.
They find that out-of-the-box neural rankers conform with the axiomatic rankings to only a limited extent.
However, they hypothesize that including diagnostic datasets in the training process could boost this conformity. 
\citet{camara:2020:ecir:diagnoseBert} extend this work and apply diagnostic datasets similarly to ad-hoc retrieval with BERT.
They find that \textit{BERT does not align with most of the ranking axioms} but significantly outperforms other neural and classical approaches.
The authors conclude that the current set of axioms is insufficient to understand BERT's notion of relevance. 
Last in this line of work is an approach to produce \textit{axiomatic explanations} for neural ranking models by~\citet{voelske:2021:ictir:axiomaticExplanations4NeuralRanking}.
Similar to existing work on axiomatic re-ranking~\citep{hagen:2016:cikm:aximaticReranking} and diagnosing neural rankers~\citep{rennings:2019:ecir:axiomDiagnoseNeural, camara:2020:ecir:diagnoseBert}, this study investigates whether neural rankings can be explained by the combination of existing axioms.
To do so, they train a small random forest explanation model on the axioms' partial orderings to reconstruct the ranking list produced by the neural ranking model.
They find that axiomatic explanations work well in cases where the ranking models are confident in their relevance estimation.
However, these explanations fail for pairs with similar retrieval scores and conclude that more axioms are needed to close this gap. 

\subsection{Axioms for Regularizing Neural Rankers}
\label{sec:axiom-regularization}
Recently, a variety of approaches for \textit{axiomatic regularization} of neural ranking models has been proposed~\citep{rosset:2019:sigir:axiomaticNeural, cheng:2020:ictir:AxiomPerturbNeuralRanking, chen:2022:sigir:axiomResularizationPreTraining}.
These approaches aim to regularize opaque neural rankers to incentivize learning of the principled, axiomatic notions of relevance.
This has the benefits of faster convergence ~\citep{rosset:2019:sigir:axiomaticNeural}, improved performance~\citep{cheng:2020:ictir:AxiomPerturbNeuralRanking} or generalization ability~\citep{rosset:2019:sigir:axiomaticNeural, chen:2022:sigir:axiomResularizationPreTraining}, and improved interpretability~\citep{chen:2022:sigir:axiomResularizationPreTraining}.
The method by which the ranking models are regularized varies from adding a regularization term to the loss function~\citep{rosset:2019:sigir:axiomaticNeural, chen:2022:sigir:axiomResularizationPreTraining} to axiomatically perturbing the training data to amplify desirable properties~\citep{cheng:2020:ictir:AxiomPerturbNeuralRanking}.
An example of such a regularization term is applied by~\citet{chen:2022:sigir:axiomResularizationPreTraining} who add a relevance loss to their final loss function that checks how well the model's relevance judgments coincides with the axioms'.
\citet{cheng:2020:ictir:AxiomPerturbNeuralRanking} extend the training dataset by randomly sampling instances and perturbing them according to three document length normalization axioms, such as by adding noise terms.
Then, these more noisy documents are assigned a lower relevancy value.
From such perturbed data examples, the model is expected to understand the corresponding normalization axiom based on document length.
While current regularization methods offer only limited (perceived) interpretability, the approach similar to the neuro-symbolic approaches~\cite{sarker:2021:AIC:neurosymbolic} marry the benefits of both axioms and data-driven models.

\subsection{Evaluation}
\gls{ir} axioms have been applied in various works over the past decades, and many revolve around interpretability.
However, little formal evaluation of the insights gained through the axioms has been done from an interpretability perspective.
One exception is \citet{chen:2022:sigir:axiomResularizationPreTraining}, who give anecdotal examples of their axiomatically regularized model's input attribution being more sparse and focused on relevant tokens.
In addition, only \citet{voelske:2021:ictir:axiomaticExplanations4NeuralRanking} use established interpretability evaluation metrics and measure the fidelity of their generated (post-hoc) explanations.
From the interpretability perspective, two steps are needed for upcoming work: 1), proposing new axioms or methods to better explain neural ranking models and 2), rigorously evaluating the produced explanations with established metrics and eventually human acceptance studies.

\section{Probing and Parametric Analysis of Text Ranking Models }
\label{sec:probing}

Probing is a method to analyze the content of latent embeddings.
It allows us to understand the information encoded in the model's representations.
Usually, probing includes training a small classifier to predict the property of interest (e.g., part-of-speech tags or question types) directly from the embeddings~\citep{tenney:2019:iclr:you, tenney:2019:acl:bert, voita:2020:emnlp:ProbingMDL, belinkov:2022:coli:probingcls}. 
\subsection{The Probing Methodology}
\begin{figure}[h]
    \centering
    \includegraphics[]{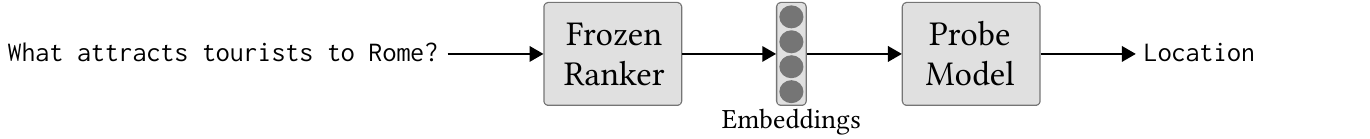}
    \caption{Example of the probing paradigm. A small classifier (the probe model) is used to predict properties (in this case the question type) from a ranker's frozen representations.}
    \label{fig:probing_example}
\end{figure}

\cref{fig:probing_example} shows an example in which we test whether a ranking model encodes information on different question types. 

To do so, we need a small, labeled dataset of questions and their respective question types.
We then train the probing classifier to recover the question type information from the ranker's frozen embeddings.
Originally, the model would be considered to encode the property of interest if the classifier can better predict it than a majority classifier.
However, depending on the task's difficulty, dataset size, and classifier complexity, large portions of the resulting performance must be attributed to the classifier.
Therefore, a large set of \textit{improvements to the probing paradigm} have been proposed -- from introducing suitable baselines~\citep{zhang:2018:bbnlp:ProbingBaselines} and control tasks~\citep{hewitt:2019:emnlp:probingControlTasks}, over varying the classifier complexity~\citep{pimentel:2020:emnlp:paretoprobing}, to measuring the target property's ease of extraction from the embeddings~\citep{voita:2020:emnlp:ProbingMDL}.
For a more comprehensive overview of the initial probing paradigm and the proposed improvements, we refer to the paper by \citet{belinkov:2022:coli:probingcls}.  

\begin{table}[tb]
\centering
\small
{\renewcommand{\arraystretch}{1.2}
\begin{tabular}{@{}llcc@{}}
\toprule
\bfseries Paper             & \bfseries Task          & \bfseries Concept under Investigation          & \bfseries Architectural component \\
\midrule
\citet{choi:2022:arxiv:IDFProbing}                                      & Text Ranking                       & IDF                                 & Attention                     \\
\citet{zhan:2020:sigir:AnalysisBERTDocRanking}                                         & Text Ranking                       & Attention, <Q, D> Interactions                                    & Attention, Embeddings                     \\
\citet{formal:2022:ecir:lexicalMatchingNeuralIR} & Text Ranking                       & Lexical Matching                    & Behavioral                         \\
\citet{formal:2021:ecir:WhiteBoxColBERT}                                                       & Text Ranking                       & Matching, Term Importance        & Behavioral                         \\
\citet{sen:2020:explain-document-scores-within-and-across-ranking-models}                                                       & Text Ranking                       & TF, IDF, Document Length        & Behavioral                         \\
\citet{macavaney:2020:arxiv:abnirml}                                       & Text Ranking                                   & Matching, Manipulation, Style                                    & Embeddings                                    \\
\citet{fan:2021:www:RelevanceModelingIR}                     & Various IR Tasks                                   & Relevance Modeling                                    & Embeddings                                   \\
\citet{vanaken:2019:cikm:howdoesbertQA}                     & QA                                 & QA Subtasks & Embeddings                         \\
\citet{cai:2020:sigir:ProbeFinetunedMRC}  & RC                                 & MRC Subtasks                                    &    --                                \\
\citet{DBLP:conf/blackboxnlp/bertnesia}               & Various NLP Tasks                           & Factual Knowledge                  & Embeddings                         \\
\citet{petroni:2021:naacl:kilt}                                       & \textless{}Benchmark\textgreater{} & Factual Knowledge                 & \textless{}Benchmark\textgreater{} \\

\bottomrule
\end{tabular}
}
\caption{Classification of the probing literature (Section~\ref{sec:probing}). These papers usually investigate whether models trained on a downstream (\gls{ir}) task encode a concept (such as lexical matching) in different architectural components (e.g., the attention maps). Behavioral studies do not probe a specific model component but investigate the model's general behavior.}
\label{table:probing_papers}
\end{table}

\subsection{Probing Ranking Models}
\label{sec:probing-rankingmodels}
Several variations of the probing paradigm have also been applied to various \gls{ir} tasks and models.
An overview of the papers, together with a classification, can be found in \cref{table:probing_papers}.
As a \textit{text-ranking model}, the approach of \citet{zhan:2020:sigir:AnalysisBERTDocRanking} investigates the attention patterns of BERT after fine-tuning on the document ranking task.
Their experiments show that large parts of the attention are off-loaded to low information tokens such as punctuation, which might lead to increased susceptibility to adversarial attacks.
Similarly, a recent study by \citet{choi:2022:arxiv:IDFProbing} probes the attention maps of a BERT ranker, finding that inverse document frequency is captured. 
As discussed in \cref{sec:axioms}, the existing ranking axioms are insufficient to explain rankings produced by BERT-based models.
Therefore, \citet{formal:2021:ecir:WhiteBoxColBERT} investigate the ColBERT regarding its term-matching mechanism.
By stratifying on IDF bins, they show that ColBERT indeed captures a notion of term importance, which is enhanced by fine-tuning.
However, the results suggest that estimating term importance is limited when no exact matches are available.  
Given the limited ability of current neural retrieval models to generalize to new datasets, \citet{formal:2022:ecir:lexicalMatchingNeuralIR} question whether this is caused by their inability to perform lexical matching in the out-of-domain scenario.
While general lexical matching ability is present in neural retrievers (such as TAS-B or ColBERT), the understanding of which terms are important to match seems to be missing in the out-of-domain setting.
\citet{sen:2020:explain-document-scores-within-and-across-ranking-models} aim to attribute relevance prediction performance to term frequency, document frequency, or document length.
To do so, they train a linear model using these aspects to approximate the ranking model.
The resulting coefficients are then used to understand the importance of the corresponding aspects.
The resulting explanations confirm that the model behavior follows certain constraints used in axiomatic \gls{ir} (\cref{sec:axioms}). 
\citet{macavaney:2020:arxiv:abnirml} also further investigate the hidden abilities of neural rankers that lead to their good ranking performance.
They attribute the model's matching ability to three properties (concepts), relevance, document length, and term frequency.
They devise a \textit{behavioral-probing setup} that verifies to what extent the model could capture these concepts.
For \textit{manipulation-sensitivity analysis}, they test the effect of shuffled words, sentences, or typos on the model performance.
Lastly,~\citet{macavaney:2020:arxiv:abnirml} create probing sets for \textit{writing style concepts} such as fluency, formality, or factuality.
Their results suggest that neural rankers are biased toward factually correct articles and that appending irrelevant text can improve the relevance scores.  
Similarly, the work by \citet{fan:2021:www:RelevanceModelingIR} strives to understand the relevance-modeling of \gls{ir} models. They also propose to probe for a large set of lexical, syntactic, and semantic concepts such as named entities or coreference resolution ability.
By comparing the performance of their fine-tuned models to a pre-trained BERT, they find that these \gls{ir} models generally seem to sacrifice small parts of their ability to perform lexical and syntactic tasks and improve especially in semantic matching (e.g., identifying synonyms).
Furthermore, causal intervention analysis is applied to the model parameters, input features, and training objectives, resulting in suggesting that a careful intervention on linguistic properties can improve the performance of downstream \gls{ir} models.  
\subsection{Probing other Information Retrieval Models}
\label{sec:probing-irmodels}
In addition to the core ranking objective, models for \textbf{other \gls{ir}-related tasks} have been probed.
\citet{vanaken:2019:cikm:howdoesbertQA} investigate BERT embeddings of a QA model and how do they interact over the layers when answering questions. 
Specifically, they probed a pre-trained BERT and a QA model, finding that training the model for QA improves the performance on related tasks such as question type classification or identification of supporting facts. 
The question of how BERT reacts to fine-tuning has also been investigated in several studies~\citep{DBLP:conf/blackboxnlp/bertnesia, fan:2021:www:RelevanceModelingIR, vanaken:2019:cikm:howdoesbertQA}. \citet{cai:2020:sigir:ProbeFinetunedMRC} probe MRC (machine reading comprehension) models for relevant subtasks (synonyms, abbreviations, coreference, as well as question type classification).
They find that only for core MRC subtasks, the token representation varies in the later layers of the MRC model.
The core MRC subtasks include tasks such as coreference, question type classification, and answer boundary detection.
However, for tasks like synonym and abbreviation detection, the representations are only moderately different from the pre-trained BERT representations.
\citet{DBLP:conf/blackboxnlp/bertnesia} probe models fine-tuned for various tasks to assess the effect of fine-tuning on (factual) knowledge retention.
In their layer-wise experiments, they find the ranking model to be specifically knowledgeable, dropping the least amount of knowledge compared to the question-answering and named entity recognition models.
Additionally, large parts, though not all, of the factual knowledge seem to be captured in the latter layers. 
\citet{petroni:2021:naacl:kilt} identify the requirement of world knowledge for many \gls{ir} tasks such as open-domain question-answering, slot filling, entity linking, or fact-checking.
To understand to what extent do current models capture real-world knowledge, \citet{petroni:2021:naacl:kilt} propose a benchmark containing knowledge-intensive tasks (QA, slot filling, entity linking, fact-checking, among others) all derived from a single Wikipedia corpus. 

\subsection{Evaluation}
In the past, probing results have been evaluated differently by the interpretability community than other post hoc methods.
Whereas other methods such as feature attributions have been rigorously evaluated concerning metrics such as fidelity or faithfulness, this has not been the case in the probing literature.
As suggested by~\citet{belinkov:2022:coli:probingcls}, a standard probing setting can answer the question: \emph{What information can be decoded from the model's embeddings?}
It does not offer a human-centered explanation for a specific data instance, but rather provides general information about the model.
Thereafter, it does not offer interpretability for users but for model developers, although the probing methodology has been scrutinized and extended in various works~\citep{sturmfels:2020:baselines, hewitt:2019:emnlp:probingControlTasks, voita:2020:emnlp:ProbingMDL}.
Given the correct baselines and a tightly controlled setup, it might be able to shed light on the question of \emph{What information is learned by training on a specific task?} or \emph{How easily extractable is information about a concept from the model?} \citep{voita:2020:emnlp:ProbingMDL}.
However, it is unclear whether this information is actually being used by the model at inference time \citep{belinkov:2022:coli:probingcls}.
To resolve this, recent studies borrow ideas from causality research to understand whether a specific concept is utilized during the inference using counterfactual representations, where the concept is voided~\citep{elazar:2021:tacl:amnesicprobing, lasri:2022:acl:probingusagegramnumber}.
The model is proven to have used the concept if the counterfactual representations result in worse task performance.
In conclusion, while there has been an in-depth evaluation of the probing paradigm by the NLP and interpretability community and many improvements have been proposed, little of that found its way into \gls{ir}-related probing studies.
Future probing studies in \gls{ir} will need to include learnings and best practices from established research and use them to evaluate and validate the findings for \gls{ir} models.

\begin{figure}
    \centering
    \includegraphics[]{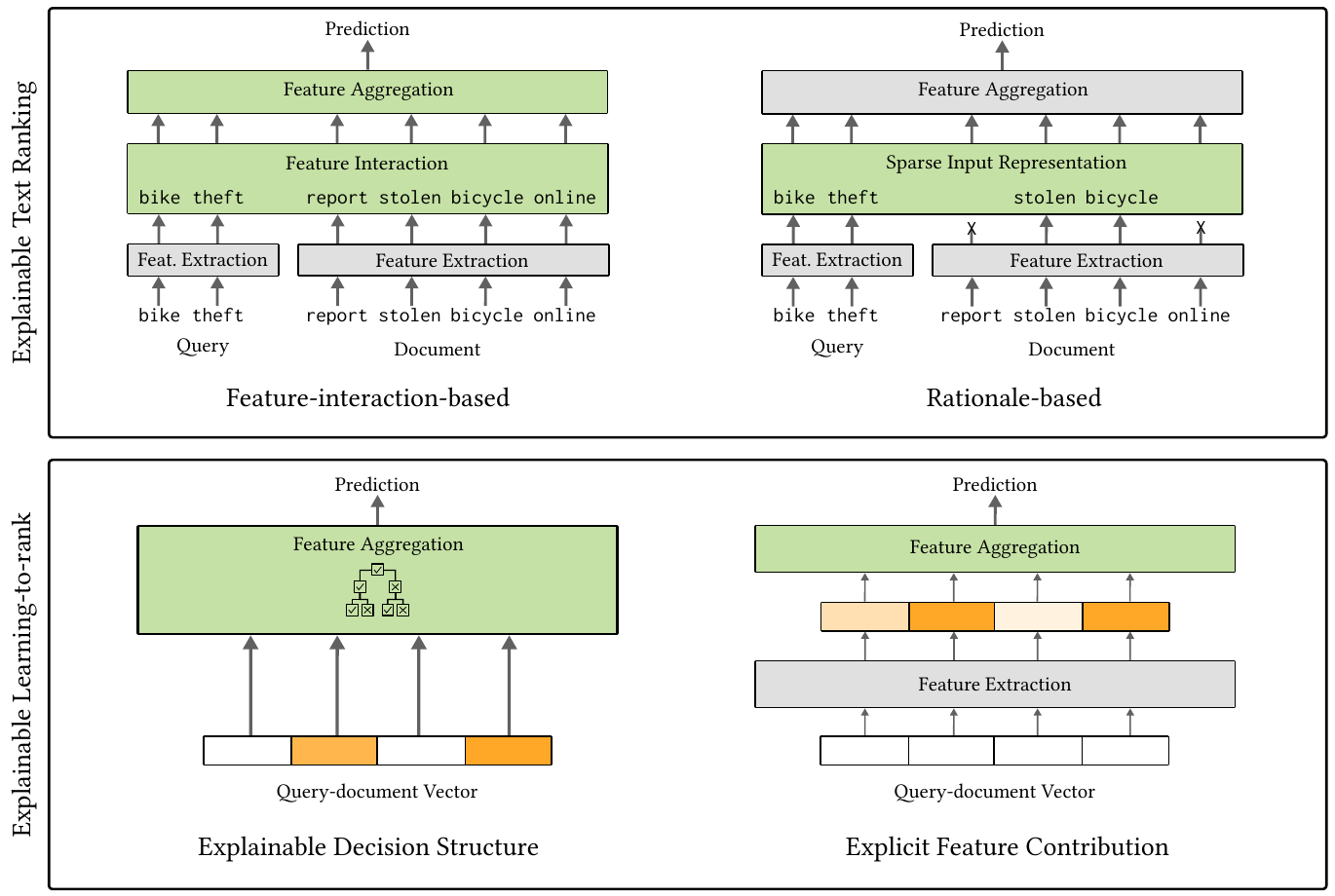}
    \caption{Types of \Gls{ibd} models. Green and gray color refers to \textit{mostly} interpretable/non-interpretable components, respectively.}
    \label{fig:exir-ibd}
\end{figure}

\section{Explainable-by-Architecture Models}
\label{sec:explainable-by-architecture}

We refer to the first family of \gls{ibd} models as explainable-by-architecture models.
Those models can be viewed as a modular framework of multiple components (see \cref{fig:exir-ibd}).
The general architecture of these models involves intermediate feature extraction (that might involve feature attributions), and a task-specific decision structure (that might involve feature interactions).
Pragmatically speaking, not all components are fully interpretable to ensure competitive task performance. 
Therefore, most of the \gls{ibd} resort to making only specific components interpretable or transparent.
In the following, we look at two major use cases of such models in text ranking and \gls{ltr} tasks.

\subsection{Explainable Text Rankers}
\label{sec:eba-interaction}

In text ranking, the need for interpretability is based on large input sizes and complex feature interactions. 
Since documents can be long, it is hard to ascertain what sections of text the query terms interact with within a complex model.
This problem is particularly acute in the case of contextual models with transformers, where the self-attention mechanism essentially considers all pairs of interactions between the query and the document terms.
Therefore, one strategy of the IBD models in the text ranking family focuses on building \textit{interpretable query-document interaction functions} and, in turn, leading to a more transparent decision-making path.
In this setup, the query and the document are encoded separately by two individual models and each token (or word) is represented by a fixed-size embedding vector.
Note that this encoding process remains opaque for both context-free and contextualized embeddings.
A (partially) explainable model employs human-understandable functions to measure the degree of query-document interactions, which essentially indicates the similarity of the query and the document. The final relevance judgment can then be made based on the interactions. 
Another line of IBD text rankers is focusing on reducing the large input space, which we refer to as \textit{rationale-based} methods. The idea is to use a small set of explicit words or sentences as input leading to the final prediction, whereas how the input is selected, and how the prediction is made, remains agnostic. There are extensive works in building such sorts of models, to highlight the popularity, we will further discuss this method family in Section~\ref{sec:rationale-based-methods}.

\subsubsection{Feature Interaction}
We summarize \textit{three} ranking models, which utilize two BERT/Transformer-style encoders to generate the vectorized representations for query and document individually.
In the following paragraphs, we emphasize on their interaction and decision-making processes, showing how the relevance decision can be explained. 

Colbert~\cite{khattab:2020:sigir:colbert} follows the conventional term-matching strategy. For each query token, it computes the cosine similarity scores with each token from the document and keeps the maximum similarity score. The final document relevance is computed by simply summing up the maximum scores of all query tokens. Essentially, Colbert measures the semantic similarity between the query and the document, and a document is deemed more relevant if it contains more terms that are semantically closer to the query.
\citet{boytsov:2021:ecir:neuralLexical} propose NeuralModel1, which
adds an explainable layer, namely Model1~\cite{berger:1999:sigir:model1} on top of the input embedding.  
Specifically, the non-parametric Model1 layer maintains pairwise similarity statistics between query-document tokens, which are learned/computed from parallel datasets beforehand. The final document relevance is combined from all query-document similarity scores by the product-of-sum formula. This approach is very similar to Colbert, where the cosine similarity computation can also be viewed as an explainable layer. NeuralModel1 experimented with slightly more comprehensive similarity learning, resulting in lower interpretability.
Nevertheless, with a more complex interaction mechanism, NeuralModel1 achieves better balance in terms of ranking performance and efficiency. 

Transformer-Kernel~\cite{hofstatter:2020:ecai:interpretablererank} maintains a matching matrix, where each row represents the cosine similarity scores between a particular query token and all document tokens. In contrast to Colbert, which simply takes the maximum similarity from each row and sums them up to obtain the query-document relevance, Transformer-Kernel transforms the matching matrix to a set of isomorphic matrices with RBF-kernels~\cite{xiong:2017:sigir:kernelpooling} and each kernel focuses on a specific similarity range. This interaction shares a similar idea as the similarity histogram in DRMM~\cite{guo:2016:cikm:drmm} model but employs the kernel-pooling technique to solve the non-differentiation of the hard histogram. The final relevance score is learned by a feed-forward layer, given the semantic matrices as the input. Therefore, the interaction in Transformer-Kernel can be viewed as smoothed semantic similarity, and the relevance decision is learned via a neural layer, making the Transformer-Kernel less interpretable in comparison to Colbert and NeuralModel1.

\subsection{Explainable Learning-to-Rank}
\label{sec:eba-ltr}

For \gls{ltr} task dealing with smaller numerical input features, there are works relying on explicitly \textit{aggregating feature contribution} for relevance measurement, or a \textit{fully transparent tree model}.
The goal of \Gls{ltr} is to predict a relevance score for a fixed-size input vector. Because of the smaller and structured input features, it is more practical to build a transparent model in input space or the entire decision path.
In the following paragraphs, we introduce one \gls{ltr} model with \textit{explicit feature contribution} and one transparent \textit{decision-tree} model incorporated with reduced input space. 

\subsubsection{Explicit Feature Contribution.}
Different from the previously discussed feature-attribution methods, explicit feature contribution indicates a simple and transparent correlation between each input feature and the relevance prediction, in addition to showing importance heatmaps. 

NeuralGAM~\cite{zhuang:2021:wsdm:gam} is based on Generalized Additive Models (GAMs). For each individual feature, NeuralGAM employs an isolated black-box (e.g., neural networks) model to generate a score indicating the contribution (or importance) of the feature.
The relevance of the input is aggregated by simply summing up all contribution scores. 
NeuralGAM is explainable in terms of feature contribution, as the relevance is aggregated from the feature importance score directly by a simple \textit{sum} operation.
Nevertheless, it remains opaque how each feature importance score is generated by the black-box model. 

\subsubsection{Explainable Tree Models}

The main challenge of interpreting tree models is the over-complex decision path caused by the massive number of features and their interactions. Thus, an explainable tree model should have a limited number of features and interactions and, in turn, be able to provide a simple and understandable decision-making path.

ILMART~\cite{lucchese:2022:sigir:ilmart} shares a similar structure as GAM, while using LambdaMART as the sub-model. ILMART starts from learning a set of trees, with each dealing with one single distinct feature only. This step enables ILMART to identify a small yet crucial set of features and exclude the rest. Then, ILMART enforces a new ensemble of trees to explore the interactions between every two remaining features only. This design can effectively reduce the model's complexity. Finally, ILMART combines trees from the previous two steps and learns a much smaller and simpler ensemble-tree model with the input space hugely reduced.

\subsection{Evaluation}
A key attribute of interpretable models is, it does not just highlight the importance of input snippets/dimensions (e.g., tokens in a query or document), but also suggest why those snippets lead to the decision. Namely, a set of rules can be implicitly inferred from the explanations, even when only the input features are presented. 
This is the usual case when the audience group of explanation is system developers or domain experts. One explanation example for Colbert can be \textit{a small set of tokens} in the query and document, together with their \textit{cosine similarity} degree. We denote this type of explanation as \textit{soft-rule}, to distinguish from the \textit{hard-rule} of an explicit path in a tree model. NeuralGAM presents feature attribution scores (similar to \cref{sec:feature-attribution}) as explanations and moreover, the relevance decision can be explicitly induced from the scores. 

Except for Colbert, all methods evaluate the goodness of explanations by showing anecdotal examples. Additionally, NeuralGAM compares the features to a referenced tree-model, and justifies the faithfulness of explanations by a similar trend. A summary of methods can be found in \cref{table:interpretable-by-archi}.

\begin{table}[tb]
\centering
\small
{\renewcommand{\arraystretch}{1.2}
\begin{tabular}{@{}llcccc@{}}
\toprule
\bfseries Method             & \bfseries Task              &         \bfseries  Components & \bfseries Explanation & \bfseries Dataset & \bfseries Evaluation\\
\midrule
Colbert~\cite{khattab:2020:sigir:colbert}   & Text Ranking   & Interaction & Soft-rule  & MS MARCO & -\\
Transformer-Kernel~\cite{hofstatter:2020:ecai:interpretablererank} & Text Ranking & Interaction & Soft-rule & MS MARCO & Anecdotal\\
NeuralModel1~\cite{boytsov:2021:ecir:neuralLexical}  & Text Ranking & Interaction & Soft-rule &  MS MARCO& Anecdotal \\
 
NeuralGAM~\cite{zhuang:2021:wsdm:gam} & LTR & Input  & Feature Attr. &  Yahoo & Reference\\
 
ILMART~\cite{lucchese:2022:sigir:ilmart} & LTR & Fully & Hard-rule & Yahoo & Anecdotal\\

\bottomrule
\end{tabular}
}
\caption{Explainable-by-architecture Methods. Components indicate which component of the model architecture is explainable. Note that Colbert did not discuss or evaluate explainability. More similar datasets are used in each paper, and we choose one as representative.}
\label{table:interpretable-by-archi}
\end{table}

\section{Rationale-based Methods}
\label{sec:rationale-based-methods}

\begin{figure}[h]
    \centering
    \includegraphics[]{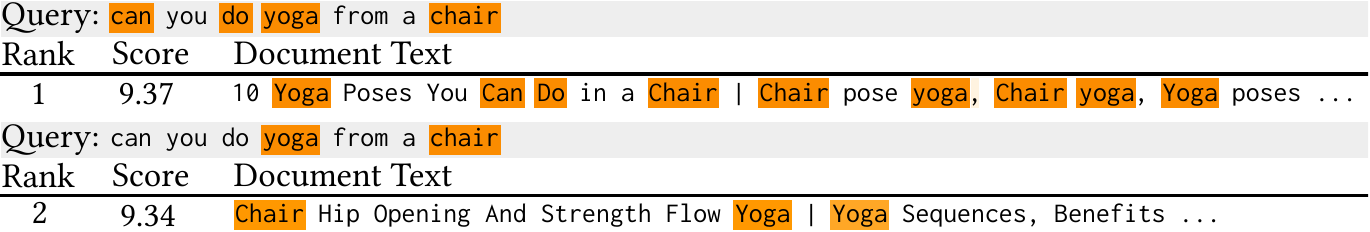}
    \caption{Example of a binary rationale selection. First, a subset of tokens is selected. Then, a prediction is made based on the selected rationale tokens. Selected rationales are highlighted in orange.}
    \label{fig:exampleRationales}
\end{figure}

The second class of \gls{ibd} methods deals to enhance the interpretability of \gls{ir} models by generating \textit{rationales} as an intermediate sparse input representation (see \cref{fig:exir-ibd}).
A \textit{rationale} is defined as an extractive piece of the input text that is responsible for the decision of the model.
A rationale-based method performs the task prediction two-stage. In the first feature-extraction phase, a model learns to extract the rationale from the input text.
In the subsequent prediction phase, another independent task model predicts the task output solely based on the extractive explanation.
Note that in such a setup, each prediction can be unambiguously attributed to the generated rationale that is both human-understandable and acts as an explanation.
Examples of rationales are provided below in \cref{fig:exampleRationales}.

We summarize the approaches in this section in \cref{tab:intp-by-sparsity}.
The feature extraction stage in rationale-based models is sometimes called the selection or rationale-generation stage~\citep{zhang:2021:wsdm:expred,leonhardt:2021:arxiv:hard:kuma:ranking}.
The major challenge in rationale-based methods is training the rationale-extraction module due to the discrete rationale output of this stage.
There are essentially two types of rationale-based methods based on the optimization styles -- \textit{pipeline} or \textit{end-to-end}.

\subsection{Pipeline Approaches}
\label{sec:rationale-pipeline}

A rationale-based model is a pipeline model if the rationale-extraction module is trained separately from the task prediction module.
Most pipeline methods require the presence of human-annotated, extractive rationale data to train the rationale-extraction network.
The first pipeline model was proposed by~\citet{lehman:2019:naacl:inferringdataset}.
Their approach was proposed for the analysis of clinical trial reports, where the aim is to predict whether the clinical trial causes a significant effect compared with the control group.
The reports are themselves annotated by human experts, where experts do annotate not only the significance of the trial but also the snippet of the reports as the rationale supporting such prediction.
This constitutes the training data for the rationale-extraction module.
During the inference, the prediction model takes the output of the rationale-extraction module as its input.

\begin{table}[tb]
\centering
\small
\resizebox{\linewidth}{!}{
\begin{tabular}{llccc}
\toprule
\bfseries Method & \bfseries Task & \bfseries Training  & \bfseries Dataset & \bfseries Evaluation\\
\midrule
Expred~\citep{zhang:2021:wsdm:expred} & Knowledge Intensive & Pipeline & ERASER Benchmark\cite{deyoung:2019:acl:eraser} & PRF Score, C/S\\
\citet{wojtas:2020:nips:featureranking} & Medical Search & Pipeline & GM12878 & Feature Ranking \\
Jiang's IB~\cite{jiang:2021:blackboxnlp:bertbottleneck} & Text Ranking & End-to-End & MS MARCO & -- \\
Select and Rank\cite{leonhardt:2021:arxiv:hard:kuma:ranking} & Text Ranking & End-to-End & \textsc{TrecDL}, \textsc{Core17}, CW09 & Anecdotal\\
IDCM\cite{hoftaetter:2021:sigir:esm:etm} & Text Ranking & Pipeline & MS MARCO & -- \\
IOT-Match\cite{Yu:2022:sigir:iot-match} & Legal Case Matching & Pipeline & ELAM, eCAIL & PRF Scores\\
\bottomrule
\end{tabular}
}\\
\caption{Rationale-based approaches. The C/S score in the evaluation refers to comprehensive and sufficiency.}
\label{tab:intp-by-sparsity}
\end{table}

Later in the same year,~\citet{deyoung:2019:acl:eraser} released a benchmark called ERASER to evaluate rationale-based interpretability.
The ERASER benchmark consists of a large variety of knowledge-intensive tasks that presupposes an \gls{ir} system, like question answering (QA) and fact-checking.
Despite the reasonable performance benefits of such select-and-predict approaches, they suffer from a crucial deficiency.
That is, the rationale-extraction module could ``cheat'' to overfit the pattern of the rationale sub-sequences instead of selecting the rationales based on their semantic meaning~\citep{jacovi:2021:acl:socialattribution, bastings:2021:arxiv:shortcut}.
To this end, another pipeline approach \textsc{ExPred}~\citep{zhang:2021:wsdm:expred} was proposed.
The main idea of \textsc{ExPred} was to make the rationale-extraction phase task-aware by training it in a multitask fashion with the downstream task.
By doing so, they use an auxiliary output to force the rationale selector to learn the semantics of the inputs with respect to the classification task.

The pipeline models introduced above contain only one extractor-predictor training cycle.
~\citet{wojtas:2020:nips:featureranking}, however, propose to train the rationale extractor and the task predictor alternatively for multiple rounds and select the masks using a genetic algorithm.
The alternative training cycle is initialized by training the classifier on multiple randomly sampled rationales and keeping the best rationale mask, resulting in the best classification performance.
Then they optimize the rationale-extractor and the task-predictor alternatively.

For document ranking tasks,~\citet{hoftaetter:2021:sigir:esm:etm} propose the IDCM (Intra-Document Cascading Model) approach to overcome the input length limitations of modern transformer-based rankers.
IDCM is a pipeline approach whose rationale extractor is an effectively simple model (student model) trained to mimic the passage-selection behavior of a more complex model (teacher model).
The student extractor model selects important passages as rationales from the huge amount of documents before calculating the ranking score of selected passages with respect to the current query using another complex model.
Evaluated on the MS MARCO dataset~\citep{nguyen:2016:msmarco}, IDCM turns to outperform traditional ranking models.

Finally, another pipeline model called IOT-match~\citep{Yu:2022:sigir:iot-match} focuses on the \textit{case-matching problem}.
The case-matching problem is defined as: given two different legal cases, the model should predict whether the two cases are related.
They try to solve the problem using optimal transport theory.
The intuition behind their algorithm is that the predicted sentence matching matrix is also an optimal transport matrix that minimizes the transport distance given the sentence discrepancy matrix.
The sentence matching matrix is a selection matrix that selects sentence pairs from both cases, where the sum of their discrepancies is minimized (similarity maximized).
After selecting the most similar sentences from both cases, they utilize the task prediction model to predict whether the two cases are related based on matched sentence pairs as the rationales.

\subsection{End-to-End Approaches}
\label{sec:rationale-based-end2end}
As its name suggests, we can train both the task and the rationale-generation module jointly using the gradients from the task supervision signal.
The major technical challenge in this setting is that of gradient estimation for the rationale generation parameters.
\citet{lei:2016:emnlp:rationalizing} is the first work that proposes an end-to-end approach for the problem of rationale-based models in the context of vanilla text classification.
The rationale extraction module parameterizes a per-token distribution of relevance.
The output of this layer is a hard binary mask as rationales.
The parameters of this module are optimized by estimated gradients, using a REINFORCE-style algorithm.
Additionally, they also applied constraints like the continuity of the selected rationales and the sparsity of the rationales to further enhance the sparsity.
Extensions of this approach include ~\citep{lehman:2019:naacl:inferringdataset} and~\citep{bastings:2019:acl:interpretablebinary} that also focus on text classification, albeit using reparameterization trick for better numerical stability and convergence rate when training the rationale-extractor.

The first work to propose end-to-end methods for ranking tasks is~\cite{leonhardt:2021:arxiv:hard:kuma:ranking}.
Their approach is called \textit{select and rank} and is based on the observation that only a few sentences in a related document are relevant given a query.
In the rationale-selection phase, they first select relevant sentences from a document with respect to the query input.
The selected rationales act as an extractive summary of the document. 
After that, only these rationales are used in the re-ranking phase with the query in a cross-encoder ranker.
Both the selector and the rankers are trained end-to-end using a combination of the \textit{Gumbel-Softmax} and \textit{reservoir sampling} to ensure a user-specified $k$-sentences to be sampled.

Unlike previous rationale-based models,~\citet{chen:2018:icml:l2x} use a mutual-information-based formulation.
Their theory is to select the rationales containing the most mutual information with the final prediction.
\citet{jiang:2021:blackboxnlp:bertbottleneck} and ~\citet{bang:2021:aaai:bottleneck} further extend this information-theory-based approach by considering the \textit{information bottleneck (IB) as the rationale}.
Specifically, the information bottleneck $\mathbf{T}$ is an intermediate (usually latent) representation that maximizes the mutual information between $\mathbf{T}$ and the prediction $\mathbf{Y}$, while its mutual information with the input $\mathbf{X}$ is minimized, i.e., $\mathbf{I}(\mathbf{Y};\mathbf{T}) - \beta I(\mathbf{X};\mathbf{T})\text{,}$ where $\beta$ is a hyper-parameter that balances both terms.
Specifically, the information bottleneck can be seen as a rationale mask applied to the input, i.e. $T = \mathbf{m}$.
The approach from~\citet{bang:2021:aaai:bottleneck} leverages the Gumbel-Softmax trick to sub-sample $k$ tokens as the rationale, while ~\citet{jiang:2021:blackboxnlp:bertbottleneck}'s approach predicts the probability of being the rationale for each feature individually and obtain the rationale mask by rounding the probability.

\subsection{Evaluation}
\label{ref:explanable-by-sparsity-evaluation}
Evaluation regimes to evaluate rationale-based models typically trade off task performance and the interpretability achieved.
The desirable objective for these approaches is: a good \gls{ibd} approach should provide a task-prediction model that performs at least no worse, if not better, than its non-explainable comparators, and delivers valuable rationales.

The quality of the rationales can be measured by the degree of their agreement with the ground-truth rationales. 
Essentially, they try to answer the interpretability question: \textit{To what degree do the rationales agree to what humans consider as true reasons?}
Benchmarks like~\citep{deyoung:2019:acl:eraser} collect multiple human-annotated datasets in \gls{ir} ranging from sentiment analysis, and fact-checking to entailment prediction.
Therefore, given the human-annotated rationale data, one can also evaluate the rationales output by the rationale-extractor by calculating their similarity to the human annotations.
The similarity metrics include but are not restricted to the \textit{accuracy}, \textit{precision}, \textit{recall}, and \textit{F1 score} of the rationale selection.

Apart from correspondence with human reasoning,
\citet{deyoung:2019:acl:eraser} also introduces C/S scores, two evaluation metrics that evaluate semantic attribution of selected rationales: ``comprehensiveness'' and ``sufficiency''.
For an arbitrary input $\mathbf{x}$ with its corresponding label $l$ on a fine-tuned model $M$, the comprehensiveness of a rationale-selection mask $\mathbf{m}$ is defined as the difference between the model prediction made based on the whole input and on all-but-rationale tokens, i.e.:
\begin{equation}
    \mathtt{comprehensiveness}(M, \mathbf{m}, (\mathbf{x}, l)) := P(y=l|\mathbf{x};M) - P(y=l|\mathbf{x} \odot \bar{\mathbf{m}};M)\text{,}
\end{equation}
where $\bar{\mathbf{m}}$ is the inverse mask and $P(\cdot)$ indicates the predicted probability.
The sufficiency is defined as the difference between the full-input prediction and the prediction based on the rationale-only tokens, i.e.:
\begin{equation}
    \mathtt{sufficiency}(M, \mathbf{m}, (\mathbf{x}, l)) := P(y=l|\mathbf{x}; M) - P(M(\mathbf{x} \odot \mathbf{m}) = l)\text{.}
\end{equation}
Furthermore,~\citet{bang:2021:aaai:bottleneck} evaluate their information-bottleneck model with the fidelity of the rationales.
They define the fidelity similar to the sufficiency score introduced above, i.e., to answer \textit{how well does the rationale-based prediction match the prediction on the full input?}

\section{Limitations and Open Questions}

In this section, we will discuss the limitations, challenges, and open questions in the area of explainable information retrieval. We have reviewed many interpretability methods and approaches that cover various aspects and tasks in \gls{ir}.
However, there are many unanswered questions, use cases, and scenarios that need further research. We feel that most interpretability approaches have focussed on the functional aspect of the central \gls{ir} tasks of ranking items. 
There are, however, many more \gls{ir} tasks that employ learning systems. Similarly, an \gls{ir} system has different stakeholders – most prominently, the benefactor of the \gls{ir} system is the user, but much of the work has focused on the system developer as the most likely stakeholder.
Finally, most of the explanation methods have relied on feature attributions as the dominant type of explanations.
However, explanations can be in terms of training instances, adversarial examples, rules, etc.

\subsection{Limitations}
\label{sec:limitations}

There are multiple limitations and challenges in facilitating and developing interpretable approaches for information retrieval tasks.
For the common task of document retrieval, we discussed early heard that we require listwise or pairwise explanations instead of pointwise explanations.  

\subsubsection{Limiting Assumptions.} The underlying assumption for surrogate models is that a simple model can locally approximate the behavior of a complex black-box ranker. However, the ranked output from a complex retrieval model can involve multiple relevance factors. While one document in the ranking might rely on term matching with the query, another document in the same ranking might be deemed relevant by the same ranking model due to the proximity of query terms in the document. Therefore, rankings with multiple and sometimes conflicting relevance factors for a single simple surrogate model might not be able to provide high fidelity. 

\subsubsection{Disentangling Explanations.} Many of the feature attribution methods provide one explanation, but complex machine learning models learn multiple features for the same behavior, which are also difficult to disentangle.
This problem is exaggerated when it is coupled with the problem of correlated reasons.
Specifically, many relevance factors are known to be correlated. A document that exhibits high semantic similarity with the query might also have a high term-level matching score. In these cases, it is likely that the methods covered in this survey (for example, probing approaches) will not be able to disentangle the effects of the underlying relevant factors from each other.

\subsection{Open Questions}

Now we turn to some of the open questions in the area of explainable information retrieval. We divide the questions into three main categories - types of explanations, explaining user models, evaluation of explanations, causal explanations and the utility of explanations.

\subsubsection{Explanation by Example}
As discussed before, most of the explanation methods have been limited to explaining the feature space – words, sentences, or numerical features in learning to rank tasks.  Prominent among these are attribution methods and hard masking techniques. However, data attribution or instance attributes and methods have not been explored in the context of information retrieval tasks. Current papers that deal with explanation by examples are limited to finding adversarial examples of generated text documents that result in errors of contextual rankers. Instance attribution methods attribute the rationale or reason behind the prediction back to the input instances in the training data. Examples of instance attribution methods include influence functions and data poisoning methods.
The interpretability question that instance attribution method methods answer is which of the input instances in the training data is responsible for training the model in a certain way to cause the following decision. 

For a document retrieval task, the interpretability question could be ``which of the queries in the training set affect a certain test query ?''. The output of instance attribution tasks can result in isolating mislabelled training instances, identifying dataset biases, and providing insights into query representation of the complex encoders. Other types of explanations can be generated explanations for the stakeholders who are end users. These generative explanations can take the form of fully-fledged natural language that is composed of extractive explanations, feature attributions, or even instance attribution methods.

\subsubsection{Explaining User Models}
Personalized ranking models tailor the search result list according to a user's profile as inferred by the search engine. While it is useful, modern personalization techniques cause growing anxiety in their users – ``Why am I seeing these search results? What does the search engine believe my interests are?''

Search engines have recently come under increased scrutiny due to their influence on people's decision-making in critical scenarios such as health and elections. Search personalization typically involves tailoring the ranking of results for individual users based on models of their past preferences and interests. 
Consequently, there is a growing concern in users due to the possible negative effects of personalization that include bias~\cite{hannak2013measuring,mustafaraj2020case}, filter bubbles~\cite{pariser2011filter,haim2018burst,flaxman2016filter} and increased opacity of the ranking mechanism. Modern personalization techniques are based on learning an effective representation of a user by mining sensitive behavioral data like click-throughs~\cite{joachims2002optimizingclick}, query logs~\cite{carman2010towards} and topical interests~\cite{harvey2013building} from social media. 
Given today's landscape of partisan news coupled with the fact that commercial search engines do not highlight personalized results, there is a real need to help us better understand what a search engine infers about its users.
Specifically, an interesting interpretability question to ask is \emph{what does the search engine perceive the user as when they issue a certain query?}  This manner of post-hoc auditing of retrieval models can be useful in a variety of downstream bias detection and validation applications.

\subsubsection{Evaluation of Explanations}
Evaluation of explanations is a general problem in the area of interpretable machine learning. There is a large spectrum of evaluation methods, starting from functionally grounded evaluations to Human-centered evaluation in the wider domain of machine learning and natural language processing. 
However, in information retrieval, most of the explanation evaluation techniques have focused on functionally grounded evaluation.
Approaches that we reviewed in this paper propose and evaluate explanation methods by their fidelity, validity, completeness, and human congruence. We refer to these methods as intrinsic methods. 

A deeper problem lies in the absence of ground truth for evaluating or validating the accuracy of output explanations of post-hoc methods. Unfortunately, this leads to a chicken-and-egg problem that is hard to fix -- \emph{to evaluate an explanation, one needs a mechanism to generate or collect ground truth, which in the first place is the objective of the interpretability task}.  If we indeed have a procedure to create ground-truth explanations from a black-box model, that is, to determine what exactly the model pays attention to, then we would have solved the problem.  Note that this is in stark contrast to standard ML tasks, where the ground-truth are indeed the observed variables that are explicitly specified in the data.
While intrinsic methods in the absence of ground-truth explanations are reasonable proxies, they still do not answer the utility question of explanations -- that is, to what extent do the explanations assist the end-user in performing a given task. Examples of tasks depend upon the stakeholder. 
For a machine learning expert, the task can be explanation-based model debugging, while for an end-user the question would be why the machine learning model ranks an obviously relevant document lower than an irrelevant document.
Apart from these open questions, we believe that there is ample opportunity for explainable IR methods to many vertical search applications like medical search~\cite{hanauer:2006:emerse:medicalsearch}, high-recall search~\cite{chakraborty:excir:2021:patent}, scholarly and historical search~\cite{holzmann:2017:websci:temporalanchor,holzmann:webscience:2017:exploring,holzmann:www:2016:tempas}. 
Apart from specialized search application, explainable IR has direct applications in knowledge intensive tasks that use an information retrieval component like fact checking~\cite{pratapa:emnlp:2020:factchecking,zhang:cikm:2021:faxplainac,fetahu:cikm:2016:finding:wiki}, question answering~\cite{roy:2021:book:question}, entity addition~\cite{singh:cikm:2016:entity}.

\section{Conclusion}
\label{sec:conclusion}

We provided an extensive investigation into the state of \gls{exir} research. 
We fill a distinct gap in the \gls{ir} literature to curate, organize, and synthesize works relating to explainability of learning systems.
Our analysis reveals that while post-hoc interpretability was heavily researched in the initial years, current efforts are trying to propose approaches that are interpretable by design (\gls{ibd}).
Due to a variety of design choices in \gls{ibd} models, we find that authors are often vague about the extent and style of interpretability in their \gls{ibd} approaches.
We explored the feature-attribution, free-text generation, and adversarial examples for post-hoc interpretability. Moreover, we summarize methods that make use of well-established \gls{ir} principles to explain and probe ranking models. 
Finally, we explored the two major subtypes of \gls{ibd} methods for \gls{ir} tasks. Based on our findings, we reflect on the design trade-offs and experimental protocols that are used in evaluating \gls{exir} approaches.
In the end, we present some limitations and open questions that we foresee as the next steps toward building transparent, trustworthy search systems.

\begin{acks}
We acknowledge all the helpful comments from the anonymous reviewers, and funding from DFG AN 996/1-1.
\end{acks}

\bibliographystyle{ACM-Reference-Format}
\bibliography{references}


\begin{thebibliography}{149}


\ifx \showCODEN    \undefined \def \showCODEN     #1{\unskip}     \fi
\ifx \showDOI      \undefined \def \showDOI       #1{#1}\fi
\ifx \showISBNx    \undefined \def \showISBNx     #1{\unskip}     \fi
\ifx \showISBNxiii \undefined \def \showISBNxiii  #1{\unskip}     \fi
\ifx \showISSN     \undefined \def \showISSN      #1{\unskip}     \fi
\ifx \showLCCN     \undefined \def \showLCCN      #1{\unskip}     \fi
\ifx \shownote     \undefined \def \shownote      #1{#1}          \fi
\ifx \showarticletitle \undefined \def \showarticletitle #1{#1}   \fi
\ifx \showURL      \undefined \def \showURL       {\relax}        \fi
\providecommand\bibfield[2]{#2}
\providecommand\bibinfo[2]{#2}
\providecommand\natexlab[1]{#1}
\providecommand\showeprint[2][]{arXiv:#2}

\bibitem[Adebayo et~al\mbox{.}(2018)]%
        {adebayo:2018:sanity-checks}
\bibfield{author}{\bibinfo{person}{Julius Adebayo}, \bibinfo{person}{Justin
  Gilmer}, \bibinfo{person}{Michael Muelly}, \bibinfo{person}{Ian~J.
  Goodfellow}, \bibinfo{person}{Moritz Hardt}, {and} \bibinfo{person}{Been
  Kim}.} \bibinfo{year}{2018}\natexlab{}.
\newblock \showarticletitle{Sanity Checks for Saliency Maps}. In
  \bibinfo{booktitle}{\emph{Advances in Neural Information Processing Systems
  31: Annual Conference on Neural Information Processing Systems 2018, NeurIPS
  2018, December 3-8, 2018, Montr{\'{e}}al, Canada}},
  \bibfield{editor}{\bibinfo{person}{Samy Bengio}, \bibinfo{person}{Hanna~M.
  Wallach}, \bibinfo{person}{Hugo Larochelle}, \bibinfo{person}{Kristen
  Grauman}, \bibinfo{person}{Nicol{\`{o}} Cesa{-}Bianchi}, {and}
  \bibinfo{person}{Roman Garnett}} (Eds.). \bibinfo{pages}{9525--9536}.
\newblock
\urldef\tempurl%
\url{https://proceedings.neurips.cc/paper/2018/hash/294a8ed24b1ad22ec2e7efea049b8737-Abstract.html}
\showURL{%
\tempurl}


\bibitem[Adebayo et~al\mbox{.}(2022)]%
        {adebayo:2022:ineffective}
\bibfield{author}{\bibinfo{person}{Julius Adebayo}, \bibinfo{person}{Michael
  Muelly}, \bibinfo{person}{Harold Abelson}, {and} \bibinfo{person}{Been Kim}.}
  \bibinfo{year}{2022}\natexlab{}.
\newblock \showarticletitle{Post hoc Explanations may be Ineffective for
  Detecting Unknown Spurious Correlation}. In \bibinfo{booktitle}{\emph{The
  Tenth International Conference on Learning Representations, {ICLR} 2022,
  Virtual Event, April 25-29, 2022}}. \bibinfo{publisher}{OpenReview.net}.
\newblock
\urldef\tempurl%
\url{https://openreview.net/forum?id=xNOVfCCvDpM}
\showURL{%
\tempurl}


\bibitem[Amati and Van~Rijsbergen(2002)]%
        {amati:2002:tois:bm25}
\bibfield{author}{\bibinfo{person}{Gianni Amati} {and}
  \bibinfo{person}{Cornelis~Joost Van~Rijsbergen}.}
  \bibinfo{year}{2002}\natexlab{}.
\newblock \showarticletitle{Probabilistic Models of Information Retrieval Based
  on Measuring the Divergence from Randomness}.
\newblock \bibinfo{journal}{\emph{ACM Trans. Inf. Syst.}} \bibinfo{volume}{20},
  \bibinfo{number}{4} (\bibinfo{year}{2002}), \bibinfo{pages}{357--389}.
\newblock
\showISSN{1046-8188}
\urldef\tempurl%
\url{https://doi.org/10.1145/582415.582416}
\showDOI{\tempurl}


\bibitem[Bach et~al\mbox{.}(2015)]%
        {bach:2015:lrp}
\bibfield{author}{\bibinfo{person}{Sebastian Bach}, \bibinfo{person}{Alexander
  Binder}, \bibinfo{person}{Gr{\'e}goire Montavon}, \bibinfo{person}{Frederick
  Klauschen}, \bibinfo{person}{Klaus-Robert M{\"u}ller}, {and}
  \bibinfo{person}{Wojciech Samek}.} \bibinfo{year}{2015}\natexlab{}.
\newblock \showarticletitle{On pixel-wise explanations for non-linear
  classifier decisions by layer-wise relevance propagation}.
\newblock \bibinfo{journal}{\emph{PloS one}} \bibinfo{volume}{10},
  \bibinfo{number}{7} (\bibinfo{year}{2015}), \bibinfo{pages}{e0130140}.
\newblock


\bibitem[Bang et~al\mbox{.}(2021)]%
        {bang:2021:aaai:bottleneck}
\bibfield{author}{\bibinfo{person}{Seojin Bang}, \bibinfo{person}{Pengtao Xie},
  \bibinfo{person}{Heewook Lee}, \bibinfo{person}{Wei Wu}, {and}
  \bibinfo{person}{Eric Xing}.} \bibinfo{year}{2021}\natexlab{}.
\newblock \showarticletitle{Explaining A Black-box By Using A Deep Variational
  Information Bottleneck Approach}.
\newblock \bibinfo{journal}{\emph{Proceedings of the AAAI Conference on
  Artificial Intelligence}} \bibinfo{volume}{35}, \bibinfo{number}{13}
  (\bibinfo{year}{2021}), \bibinfo{pages}{11396–11404}.
\newblock
\urldef\tempurl%
\url{https://doi.org/10.1609/aaai.v35i13.17358}
\showDOI{\tempurl}


\bibitem[Bastings et~al\mbox{.}(2019)]%
        {bastings:2019:acl:interpretablebinary}
\bibfield{author}{\bibinfo{person}{Jasmijn Bastings}, \bibinfo{person}{Wilker
  Aziz}, {and} \bibinfo{person}{Ivan Titov}.} \bibinfo{year}{2019}\natexlab{}.
\newblock \showarticletitle{Interpretable Neural Predictions with
  Differentiable Binary Variables}. In \bibinfo{booktitle}{\emph{Proceedings of
  the 57th Annual Meeting of the Association for Computational Linguistics}}.
  \bibinfo{publisher}{Association for Computational Linguistics},
  \bibinfo{address}{Florence, Italy}, \bibinfo{pages}{2963--2977}.
\newblock
\urldef\tempurl%
\url{https://doi.org/10.18653/v1/P19-1284}
\showDOI{\tempurl}


\bibitem[Bastings et~al\mbox{.}(2021)]%
        {bastings:2021:arxiv:shortcut}
\bibfield{author}{\bibinfo{person}{Jasmijn Bastings},
  \bibinfo{person}{Sebastian Ebert}, \bibinfo{person}{Polina Zablotskaia},
  \bibinfo{person}{Anders Sandholm}, {and} \bibinfo{person}{Katja Filippova}.}
  \bibinfo{year}{2021}\natexlab{}.
\newblock \showarticletitle{"Will You Find These Shortcuts?" {A} Protocol for
  Evaluating the Faithfulness of Input Salience Methods for Text
  Classification}.
\newblock \bibinfo{journal}{\emph{ArXiv preprint}}
  \bibinfo{volume}{abs/2111.07367} (\bibinfo{year}{2021}).
\newblock
\urldef\tempurl%
\url{https://arxiv.org/abs/2111.07367}
\showURL{%
\tempurl}


\bibitem[Bastings and Filippova(2020)]%
        {bastings:2020:elephant}
\bibfield{author}{\bibinfo{person}{Jasmijn Bastings} {and}
  \bibinfo{person}{Katja Filippova}.} \bibinfo{year}{2020}\natexlab{}.
\newblock \showarticletitle{The elephant in the interpretability room: Why use
  attention as explanation when we have saliency methods?}. In
  \bibinfo{booktitle}{\emph{Proceedings of the Third BlackboxNLP Workshop on
  Analyzing and Interpreting Neural Networks for NLP}}.
  \bibinfo{publisher}{Association for Computational Linguistics},
  \bibinfo{address}{Online}, \bibinfo{pages}{149--155}.
\newblock
\urldef\tempurl%
\url{https://doi.org/10.18653/v1/2020.blackboxnlp-1.14}
\showDOI{\tempurl}


\bibitem[Belinkov(2022)]%
        {belinkov:2022:coli:probingcls}
\bibfield{author}{\bibinfo{person}{Yonatan Belinkov}.}
  \bibinfo{year}{2022}\natexlab{}.
\newblock \showarticletitle{Probing Classifiers: Promises, Shortcomings, and
  Advances}.
\newblock \bibinfo{journal}{\emph{Comput. Linguistics}} \bibinfo{volume}{48},
  \bibinfo{number}{1} (\bibinfo{year}{2022}), \bibinfo{pages}{207--219}.
\newblock
\urldef\tempurl%
\url{https://doi.org/10.1162/coli\_a\_00422}
\showDOI{\tempurl}


\bibitem[Berger and Lafferty(1999)]%
        {berger:1999:sigir:model1}
\bibfield{author}{\bibinfo{person}{Adam Berger} {and} \bibinfo{person}{John
  Lafferty}.} \bibinfo{year}{1999}\natexlab{}.
\newblock \showarticletitle{Information retrieval as statistical translation}.
  In \bibinfo{booktitle}{\emph{Proceedings of the 22nd annual international ACM
  SIGIR conference on Research and development in information retrieval}}.
  \bibinfo{pages}{222--229}.
\newblock


\bibitem[Bhatt et~al\mbox{.}(2020)]%
        {bhatt:2020:xaideployment}
\bibfield{author}{\bibinfo{person}{Umang Bhatt}, \bibinfo{person}{Alice Xiang},
  \bibinfo{person}{Shubham Sharma}, \bibinfo{person}{Adrian Weller},
  \bibinfo{person}{Ankur Taly}, \bibinfo{person}{Yunhan Jia},
  \bibinfo{person}{Joydeep Ghosh}, \bibinfo{person}{Ruchir Puri},
  \bibinfo{person}{Jos\'{e} M.~F. Moura}, {and} \bibinfo{person}{Peter
  Eckersley}.} \bibinfo{year}{2020}\natexlab{}.
\newblock \showarticletitle{Explainable Machine Learning in Deployment}. In
  \bibinfo{booktitle}{\emph{Proceedings of the 2020 Conference on Fairness,
  Accountability, and Transparency}} (Barcelona, Spain)
  \emph{(\bibinfo{series}{FAT* '20})}. \bibinfo{publisher}{Association for
  Computing Machinery}, \bibinfo{address}{New York, NY, USA},
  \bibinfo{pages}{648–657}.
\newblock
\showISBNx{9781450369367}
\urldef\tempurl%
\url{https://doi.org/10.1145/3351095.3375624}
\showDOI{\tempurl}


\bibitem[Bibal et~al\mbox{.}(2022)]%
        {bibal:2022:attentionisexplanation?}
\bibfield{author}{\bibinfo{person}{Adrien Bibal}, \bibinfo{person}{R{\'e}mi
  Cardon}, \bibinfo{person}{David Alfter}, \bibinfo{person}{Rodrigo Wilkens},
  \bibinfo{person}{Xiaoou Wang}, \bibinfo{person}{Thomas Fran{\c{c}}ois}, {and}
  \bibinfo{person}{Patrick Watrin}.} \bibinfo{year}{2022}\natexlab{}.
\newblock \showarticletitle{Is Attention Explanation? An Introduction to the
  Debate}. In \bibinfo{booktitle}{\emph{Proceedings of the 60th Annual Meeting
  of the Association for Computational Linguistics (Volume 1: Long Papers)}}.
  \bibinfo{publisher}{Association for Computational Linguistics},
  \bibinfo{address}{Dublin, Ireland}, \bibinfo{pages}{3889--3900}.
\newblock
\urldef\tempurl%
\url{https://doi.org/10.18653/v1/2022.acl-long.269}
\showDOI{\tempurl}


\bibitem[Bigdeli et~al\mbox{.}(2022)]%
        {bigdeli:sigir:2022:bias:irsystems}
\bibfield{author}{\bibinfo{person}{Amin Bigdeli}, \bibinfo{person}{Negar
  Arabzadeh}, \bibinfo{person}{Shirin Seyedsalehi}, \bibinfo{person}{Morteza
  Zihayat}, {and} \bibinfo{person}{Ebrahim Bagheri}.}
  \bibinfo{year}{2022}\natexlab{}.
\newblock \showarticletitle{Gender Fairness in Information Retrieval Systems}.
  In \bibinfo{booktitle}{\emph{{SIGIR} '22: The 45th International {ACM}
  {SIGIR} Conference on Research and Development in Information Retrieval,
  Madrid, Spain, July 11 - 15, 2022}}. \bibinfo{publisher}{{ACM}},
  \bibinfo{pages}{3436--3439}.
\newblock
\urldef\tempurl%
\url{https://doi.org/10.1145/3477495.3532680}
\showDOI{\tempurl}


\bibitem[Bondarenko et~al\mbox{.}(2022)]%
        {bondarenko:2022:sigir:AxiomRetrievalIRAxioms}
\bibfield{author}{\bibinfo{person}{Alexander Bondarenko}, \bibinfo{person}{Maik
  Fr{\"{o}}be}, \bibinfo{person}{Jan~Heinrich Reimer}, \bibinfo{person}{Benno
  Stein}, \bibinfo{person}{Michael V{\"{o}}lske}, {and}
  \bibinfo{person}{Matthias Hagen}.} \bibinfo{year}{2022}\natexlab{}.
\newblock \showarticletitle{Axiomatic Retrieval Experimentation with
  ir{\_}axioms}. In \bibinfo{booktitle}{\emph{{SIGIR} '22: The 45th
  International {ACM} {SIGIR} Conference on Research and Development in
  Information Retrieval, Madrid, Spain, July 11 - 15, 2022}}.
  \bibinfo{publisher}{{ACM}}, \bibinfo{pages}{3131--3140}.
\newblock
\urldef\tempurl%
\url{https://doi.org/10.1145/3477495.3531743}
\showDOI{\tempurl}


\bibitem[Boytsov and Kolter(2021)]%
        {boytsov:2021:ecir:neuralLexical}
\bibfield{author}{\bibinfo{person}{Leonid Boytsov} {and} \bibinfo{person}{Zico
  Kolter}.} \bibinfo{year}{2021}\natexlab{}.
\newblock \showarticletitle{Exploring classic and neural lexical translation
  models for information retrieval: Interpretability, effectiveness, and
  efficiency benefits}. In \bibinfo{booktitle}{\emph{European Conference on
  Information Retrieval}}. Springer, \bibinfo{pages}{63--78}.
\newblock


\bibitem[Bruza and Huibers(1994)]%
        {bruza:1994:sigir:axiomsIR}
\bibfield{author}{\bibinfo{person}{Peter Bruza} {and} \bibinfo{person}{Theo
  W.~C. Huibers}.} \bibinfo{year}{1994}\natexlab{}.
\newblock \showarticletitle{Investigating Aboutness Axioms using Information
  Fields}. In \bibinfo{booktitle}{\emph{Proceedings of {SIGIR} Forum 1994}}.
  \bibinfo{pages}{112--121}.
\newblock


\bibitem[Cai et~al\mbox{.}(2020)]%
        {cai:2020:sigir:ProbeFinetunedMRC}
\bibfield{author}{\bibinfo{person}{Jie Cai}, \bibinfo{person}{Zhengzhou Zhu},
  \bibinfo{person}{Ping Nie}, {and} \bibinfo{person}{Qian Liu}.}
  \bibinfo{year}{2020}\natexlab{}.
\newblock \showarticletitle{A Pairwise Probe for Understanding {BERT}
  Fine-Tuning on Machine Reading Comprehension}. In
  \bibinfo{booktitle}{\emph{Proceedings of the 43rd International {ACM} {SIGIR}
  conference on research and development in Information Retrieval, {SIGIR}
  2020, Virtual Event, China, July 25-30, 2020}},
  \bibfield{editor}{\bibinfo{person}{Jimmy Huang}, \bibinfo{person}{Yi~Chang},
  \bibinfo{person}{Xueqi Cheng}, \bibinfo{person}{Jaap Kamps},
  \bibinfo{person}{Vanessa Murdock}, \bibinfo{person}{Ji{-}Rong Wen}, {and}
  \bibinfo{person}{Yiqun Liu}} (Eds.). \bibinfo{publisher}{{ACM}},
  \bibinfo{pages}{1665--1668}.
\newblock
\urldef\tempurl%
\url{https://doi.org/10.1145/3397271.3401195}
\showDOI{\tempurl}


\bibitem[Callan et~al\mbox{.}(2009)]%
        {callan:2009:clueweb09}
\bibfield{author}{\bibinfo{person}{Jamie Callan}, \bibinfo{person}{Mark Hoy},
  \bibinfo{person}{Changkuk Yoo}, {and} \bibinfo{person}{Le Zhao}.}
  \bibinfo{year}{2009}\natexlab{}.
\newblock \bibinfo{title}{Clueweb09 data set}.
\newblock
\newblock
\urldef\tempurl%
\url{http://lemurproject.org/clueweb09/}
\showURL{%
\tempurl}


\bibitem[C{\^{a}}mara and Hauff(2020)]%
        {camara:2020:ecir:diagnoseBert}
\bibfield{author}{\bibinfo{person}{Arthur C{\^{a}}mara} {and}
  \bibinfo{person}{Claudia Hauff}.} \bibinfo{year}{2020}\natexlab{}.
\newblock \showarticletitle{Diagnosing {BERT} with Retrieval Heuristics}. In
  \bibinfo{booktitle}{\emph{Proceedings of {ECIR} 2020}},
  Vol.~\bibinfo{volume}{12035}. \bibinfo{publisher}{Springer},
  \bibinfo{pages}{605--618}.
\newblock


\bibitem[Camburu et~al\mbox{.}(2018)]%
        {camburu:2018:e-snli}
\bibfield{author}{\bibinfo{person}{Oana{-}Maria Camburu}, \bibinfo{person}{Tim
  Rockt{\"{a}}schel}, \bibinfo{person}{Thomas Lukasiewicz}, {and}
  \bibinfo{person}{Phil Blunsom}.} \bibinfo{year}{2018}\natexlab{}.
\newblock \showarticletitle{e-SNLI: Natural Language Inference with Natural
  Language Explanations}. In \bibinfo{booktitle}{\emph{Advances in Neural
  Information Processing Systems 31: Annual Conference on Neural Information
  Processing Systems 2018, NeurIPS 2018, December 3-8, 2018, Montr{\'{e}}al,
  Canada}}. \bibinfo{pages}{9560--9572}.
\newblock
\urldef\tempurl%
\url{https://proceedings.neurips.cc/paper/2018/hash/4c7a167bb329bd92580a99ce422d6fa6-Abstract.html}
\showURL{%
\tempurl}


\bibitem[Carman et~al\mbox{.}(2010)]%
        {carman2010towards}
\bibfield{author}{\bibinfo{person}{Mark~James Carman}, \bibinfo{person}{Fabio
  Crestani}, \bibinfo{person}{Morgan Harvey}, {and} \bibinfo{person}{Mark
  Baillie}.} \bibinfo{year}{2010}\natexlab{}.
\newblock \showarticletitle{Towards query log based personalization using topic
  models}. In \bibinfo{booktitle}{\emph{Proceedings of the 19th {ACM}
  Conference on Information and Knowledge Management, {CIKM} 2010, Toronto,
  Ontario, Canada, October 26-30, 2010}}. \bibinfo{publisher}{{ACM}},
  \bibinfo{pages}{1849--1852}.
\newblock
\urldef\tempurl%
\url{https://doi.org/10.1145/1871437.1871745}
\showDOI{\tempurl}


\bibitem[Chakraborty et~al\mbox{.}(2021)]%
        {chakraborty:excir:2021:patent}
\bibfield{author}{\bibinfo{person}{Manajit Chakraborty}, \bibinfo{person}{David
  Zimmermann}, {and} \bibinfo{person}{Fabio Crestani}.}
  \bibinfo{year}{2021}\natexlab{}.
\newblock \showarticletitle{PatentQuest: {A} User-Oriented Tool for Integrated
  Patent Search}. In \bibinfo{booktitle}{\emph{Proceedings of the 11th
  International Workshop on Bibliometric-enhanced Information Retrieval
  co-located with 43rd European Conference on Information Retrieval {(ECIR}
  2021), Lucca, Italy (online only), April 1st, 2021}}
  \emph{(\bibinfo{series}{{CEUR} Workshop Proceedings},
  Vol.~\bibinfo{volume}{2847})}. \bibinfo{publisher}{CEUR-WS.org},
  \bibinfo{pages}{89--101}.
\newblock
\urldef\tempurl%
\url{http://ceur-ws.org/Vol-2847/paper-09.pdf}
\showURL{%
\tempurl}


\bibitem[Chatterjee et~al\mbox{.}(2019)]%
        {chatterjee:2019:semeval}
\bibfield{author}{\bibinfo{person}{Ankush Chatterjee},
  \bibinfo{person}{Kedhar~Nath Narahari}, \bibinfo{person}{Meghana Joshi},
  {and} \bibinfo{person}{Puneet Agrawal}.} \bibinfo{year}{2019}\natexlab{}.
\newblock \showarticletitle{Semeval-2019 task 3: Emocontext contextual emotion
  detection in text}. In \bibinfo{booktitle}{\emph{Proceedings of the 13th
  international workshop on semantic evaluation}}. \bibinfo{pages}{39--48}.
\newblock


\bibitem[Chen et~al\mbox{.}(2022)]%
        {chen:2022:sigir:axiomResularizationPreTraining}
\bibfield{author}{\bibinfo{person}{Jia Chen}, \bibinfo{person}{Yiqun Liu},
  \bibinfo{person}{Yan Fang}, \bibinfo{person}{Jiaxin Mao},
  \bibinfo{person}{Hui Fang}, \bibinfo{person}{Shenghao Yang},
  \bibinfo{person}{Xiaohui Xie}, \bibinfo{person}{Min Zhang}, {and}
  \bibinfo{person}{Shaoping Ma}.} \bibinfo{year}{2022}\natexlab{}.
\newblock \showarticletitle{Axiomatically Regularized Pre-training for Ad hoc
  Search}. In \bibinfo{booktitle}{\emph{{SIGIR} '22: The 45th International
  {ACM} {SIGIR} Conference on Research and Development in Information
  Retrieval, Madrid, Spain, July 11 - 15, 2022}}. \bibinfo{publisher}{{ACM}},
  \bibinfo{pages}{1524--1534}.
\newblock
\urldef\tempurl%
\url{https://doi.org/10.1145/3477495.3531943}
\showDOI{\tempurl}


\bibitem[Chen et~al\mbox{.}(2018)]%
        {chen:2018:icml:l2x}
\bibfield{author}{\bibinfo{person}{Jianbo Chen}, \bibinfo{person}{Le Song},
  \bibinfo{person}{Martin~J. Wainwright}, {and} \bibinfo{person}{Michael~I.
  Jordan}.} \bibinfo{year}{2018}\natexlab{}.
\newblock \showarticletitle{Learning to Explain: An Information-Theoretic
  Perspective on Model Interpretation}. In
  \bibinfo{booktitle}{\emph{Proceedings of the 35th International Conference on
  Machine Learning, {ICML} 2018, Stockholmsm{\"{a}}ssan, Stockholm, Sweden,
  July 10-15, 2018}}, \bibfield{editor}{\bibinfo{person}{Jennifer~G. Dy} {and}
  \bibinfo{person}{Andreas Krause}} (Eds.), Vol.~\bibinfo{volume}{80}.
  \bibinfo{publisher}{{PMLR}}.
\newblock
\urldef\tempurl%
\url{http://proceedings.mlr.press/v80/chen18j.html}
\showURL{%
\tempurl}


\bibitem[Cheng and Fang(2020)]%
        {cheng:2020:ictir:AxiomPerturbNeuralRanking}
\bibfield{author}{\bibinfo{person}{Zitong Cheng} {and} \bibinfo{person}{Hui
  Fang}.} \bibinfo{year}{2020}\natexlab{}.
\newblock \showarticletitle{Utilizing Axiomatic Perturbations to Guide Neural
  Ranking Models}. In \bibinfo{booktitle}{\emph{{ICTIR} '20: The 2020 {ACM}
  {SIGIR} International Conference on the Theory of Information Retrieval,
  Virtual Event, Norway, September 14-17, 2020}}. \bibinfo{publisher}{{ACM}},
  \bibinfo{pages}{153--156}.
\newblock
\urldef\tempurl%
\url{https://doi.org/10.1145/3409256.3409828}
\showDOI{\tempurl}


\bibitem[Choi et~al\mbox{.}(2022)]%
        {choi:2022:arxiv:IDFProbing}
\bibfield{author}{\bibinfo{person}{Jaekeol Choi}, \bibinfo{person}{Euna Jung},
  \bibinfo{person}{Sungjun Lim}, {and} \bibinfo{person}{Wonjong Rhee}.}
  \bibinfo{year}{2022}\natexlab{}.
\newblock \showarticletitle{Finding Inverse Document Frequency Information in
  {BERT}}.
\newblock \bibinfo{journal}{\emph{ArXiv preprint}}
  \bibinfo{volume}{abs/2202.12191} (\bibinfo{year}{2022}).
\newblock
\urldef\tempurl%
\url{https://arxiv.org/abs/2202.12191}
\showURL{%
\tempurl}


\bibitem[Craswell et~al\mbox{.}(2021)]%
        {craswell:2021:trec-dl}
\bibfield{author}{\bibinfo{person}{Nick Craswell}, \bibinfo{person}{Bhaskar
  Mitra}, \bibinfo{person}{Emine Yilmaz}, {and} \bibinfo{person}{Daniel
  Campos}.} \bibinfo{year}{2021}\natexlab{}.
\newblock \showarticletitle{Overview of the {TREC} 2020 deep learning track}.
\newblock \bibinfo{journal}{\emph{CoRR}}  \bibinfo{volume}{abs/2102.07662}
  (\bibinfo{year}{2021}).
\newblock
\showeprint[arXiv]{2102.07662}
\urldef\tempurl%
\url{https://arxiv.org/abs/2102.07662}
\showURL{%
\tempurl}


\bibitem[Cummins and O'Riordan(2007)]%
        {cummins:2007:ai:axiomaticComparison}
\bibfield{author}{\bibinfo{person}{Ronan Cummins} {and} \bibinfo{person}{Colm
  O'Riordan}.} \bibinfo{year}{2007}\natexlab{}.
\newblock \showarticletitle{An Axiomatic Comparison of Learned Term-Weighting
  Schemes in Information Retrieval: Clarifications and Extensions}.
\newblock \bibinfo{journal}{\emph{Artif. Intell. Rev.}} \bibinfo{volume}{28},
  \bibinfo{number}{1} (\bibinfo{year}{2007}), \bibinfo{pages}{51--68}.
\newblock


\bibitem[Dai et~al\mbox{.}(2018)]%
        {dai:2018:convolutional}
\bibfield{author}{\bibinfo{person}{Zhuyun Dai}, \bibinfo{person}{Chenyan
  Xiong}, \bibinfo{person}{Jamie Callan}, {and} \bibinfo{person}{Zhiyuan Liu}.}
  \bibinfo{year}{2018}\natexlab{}.
\newblock \showarticletitle{Convolutional Neural Networks for Soft-Matching
  N-Grams in Ad-hoc Search}. In \bibinfo{booktitle}{\emph{Proceedings of the
  Eleventh {ACM} International Conference on Web Search and Data Mining, {WSDM}
  2018, Marina Del Rey, CA, USA, February 5-9, 2018}}.
  \bibinfo{publisher}{{ACM}}, \bibinfo{pages}{126--134}.
\newblock
\urldef\tempurl%
\url{https://doi.org/10.1145/3159652.3159659}
\showDOI{\tempurl}


\bibitem[DeYoung et~al\mbox{.}(2020)]%
        {deyoung:2019:acl:eraser}
\bibfield{author}{\bibinfo{person}{Jay DeYoung}, \bibinfo{person}{Sarthak
  Jain}, \bibinfo{person}{Nazneen~Fatema Rajani}, \bibinfo{person}{Eric
  Lehman}, \bibinfo{person}{Caiming Xiong}, \bibinfo{person}{Richard Socher},
  {and} \bibinfo{person}{Byron~C. Wallace}.} \bibinfo{year}{2020}\natexlab{}.
\newblock \showarticletitle{{ERASER}: {A} Benchmark to Evaluate Rationalized
  {NLP} Models}. In \bibinfo{booktitle}{\emph{Proceedings of the 58th Annual
  Meeting of the Association for Computational Linguistics}}.
  \bibinfo{publisher}{Association for Computational Linguistics},
  \bibinfo{address}{Online}, \bibinfo{pages}{4443--4458}.
\newblock
\urldef\tempurl%
\url{https://doi.org/10.18653/v1/2020.acl-main.408}
\showDOI{\tempurl}


\bibitem[Doshi-Velez and Kim(2017)]%
        {doshi:2017:arxiv:towards}
\bibfield{author}{\bibinfo{person}{Finale Doshi-Velez} {and}
  \bibinfo{person}{Been Kim}.} \bibinfo{year}{2017}\natexlab{}.
\newblock \bibinfo{title}{Towards A Rigorous Science of Interpretable Machine
  Learning}.
\newblock
\newblock
\urldef\tempurl%
\url{https://doi.org/10.48550/ARXIV.1702.08608}
\showDOI{\tempurl}


\bibitem[Ebrahimi et~al\mbox{.}(2018)]%
        {ebrahimi:2018:acl:hotflip}
\bibfield{author}{\bibinfo{person}{Javid Ebrahimi}, \bibinfo{person}{Anyi Rao},
  \bibinfo{person}{Daniel Lowd}, {and} \bibinfo{person}{Dejing Dou}.}
  \bibinfo{year}{2018}\natexlab{}.
\newblock \showarticletitle{{H}ot{F}lip: White-Box Adversarial Examples for
  Text Classification}. In \bibinfo{booktitle}{\emph{Proceedings of the 56th
  Annual Meeting of the Association for Computational Linguistics (Volume 2:
  Short Papers)}}. \bibinfo{publisher}{Association for Computational
  Linguistics}, \bibinfo{address}{Melbourne, Australia},
  \bibinfo{pages}{31--36}.
\newblock
\urldef\tempurl%
\url{https://doi.org/10.18653/v1/P18-2006}
\showDOI{\tempurl}


\bibitem[Elazar et~al\mbox{.}(2021)]%
        {elazar:2021:tacl:amnesicprobing}
\bibfield{author}{\bibinfo{person}{Yanai Elazar}, \bibinfo{person}{Shauli
  Ravfogel}, \bibinfo{person}{Alon Jacovi}, {and} \bibinfo{person}{Yoav
  Goldberg}.} \bibinfo{year}{2021}\natexlab{}.
\newblock \showarticletitle{Amnesic Probing: Behavioral Explanation with
  Amnesic Counterfactuals}.
\newblock \bibinfo{journal}{\emph{Transactions of the Association for
  Computational Linguistics}}  \bibinfo{volume}{9} (\bibinfo{year}{2021}),
  \bibinfo{pages}{160--175}.
\newblock
\urldef\tempurl%
\url{https://doi.org/10.1162/tacl_a_00359}
\showDOI{\tempurl}


\bibitem[Fan et~al\mbox{.}(2021)]%
        {fan:2021:www:RelevanceModelingIR}
\bibfield{author}{\bibinfo{person}{Yixing Fan}, \bibinfo{person}{Jiafeng Guo},
  \bibinfo{person}{Xinyu Ma}, \bibinfo{person}{Ruqing Zhang},
  \bibinfo{person}{Yanyan Lan}, {and} \bibinfo{person}{Xueqi Cheng}.}
  \bibinfo{year}{2021}\natexlab{}.
\newblock \showarticletitle{A Linguistic Study on Relevance Modeling in
  Information Retrieval}. In \bibinfo{booktitle}{\emph{{WWW} '21: The Web
  Conference 2021, Virtual Event / Ljubljana, Slovenia, April 19-23, 2021}}.
  \bibinfo{publisher}{{ACM} / {IW3C2}}, \bibinfo{pages}{1053--1064}.
\newblock
\urldef\tempurl%
\url{https://doi.org/10.1145/3442381.3450009}
\showDOI{\tempurl}


\bibitem[Fang et~al\mbox{.}(2004)]%
        {fang:2004:sigir:studyIR}
\bibfield{author}{\bibinfo{person}{Hui Fang}, \bibinfo{person}{Tao Tao}, {and}
  \bibinfo{person}{ChengXiang Zhai}.} \bibinfo{year}{2004}\natexlab{}.
\newblock \showarticletitle{A Formal Study of Information Retrieval
  Heuristics}. In \bibinfo{booktitle}{\emph{Proceedings of {SIGIR} 2004}}.
  \bibinfo{pages}{49--56}.
\newblock


\bibitem[Fang et~al\mbox{.}(2011)]%
        {fang:2011:tis:diagnosticIR}
\bibfield{author}{\bibinfo{person}{Hui Fang}, \bibinfo{person}{Tao Tao}, {and}
  \bibinfo{person}{ChengXiang Zhai}.} \bibinfo{year}{2011}\natexlab{}.
\newblock \showarticletitle{Diagnostic Evaluation of Information Retrieval
  Models}.
\newblock \bibinfo{journal}{\emph{{ACM} Trans. Inf. Syst.}}
  \bibinfo{volume}{29}, \bibinfo{number}{2} (\bibinfo{year}{2011}),
  \bibinfo{pages}{7:1--7:42}.
\newblock


\bibitem[Fang and Zhai(2006)]%
        {fang:2006:sigir:semanticTermMatch}
\bibfield{author}{\bibinfo{person}{Hui Fang} {and} \bibinfo{person}{ChengXiang
  Zhai}.} \bibinfo{year}{2006}\natexlab{}.
\newblock \showarticletitle{Semantic Term Matching in Axiomatic Approaches to
  Information Retrieval}. In \bibinfo{booktitle}{\emph{Proceedings of {SIGIR}
  2006}}. \bibinfo{pages}{115--122}.
\newblock


\bibitem[Fernando et~al\mbox{.}(2019)]%
        {fernando:2019:sigir:studyshaponretrieval}
\bibfield{author}{\bibinfo{person}{Zeon~Trevor Fernando},
  \bibinfo{person}{Jaspreet Singh}, {and} \bibinfo{person}{Avishek Anand}.}
  \bibinfo{year}{2019}\natexlab{}.
\newblock \showarticletitle{A study on the Interpretability of Neural Retrieval
  Models using DeepSHAP}. In \bibinfo{booktitle}{\emph{Proceedings of the 42nd
  International {ACM} {SIGIR} Conference on Research and Development in
  Information Retrieval, {SIGIR} 2019, Paris, France, July 21-25, 2019}}.
  \bibinfo{publisher}{{ACM}}, \bibinfo{pages}{1005--1008}.
\newblock
\urldef\tempurl%
\url{https://doi.org/10.1145/3331184.3331312}
\showDOI{\tempurl}


\bibitem[Fetahu et~al\mbox{.}(2016)]%
        {fetahu:cikm:2016:finding:wiki}
\bibfield{author}{\bibinfo{person}{Besnik Fetahu}, \bibinfo{person}{Katja
  Markert}, \bibinfo{person}{Wolfgang Nejdl}, {and} \bibinfo{person}{Avishek
  Anand}.} \bibinfo{year}{2016}\natexlab{}.
\newblock \showarticletitle{Finding news citations for wikipedia}. In
  \bibinfo{booktitle}{\emph{Proceedings of the 25th ACM International on
  Conference on Information and Knowledge Management}}.
  \bibinfo{pages}{337--346}.
\newblock
\urldef\tempurl%
\url{https://doi.org/10.1145/2983323.2983808}
\showDOI{\tempurl}


\bibitem[Flaxman et~al\mbox{.}(2016)]%
        {flaxman2016filter}
\bibfield{author}{\bibinfo{person}{Seth Flaxman}, \bibinfo{person}{Sharad
  Goel}, {and} \bibinfo{person}{Justin~M Rao}.}
  \bibinfo{year}{2016}\natexlab{}.
\newblock \showarticletitle{Filter bubbles, echo chambers, and online news
  consumption}.
\newblock \bibinfo{journal}{\emph{Public opinion quarterly}}
  \bibinfo{volume}{80}, \bibinfo{number}{S1} (\bibinfo{year}{2016}),
  \bibinfo{pages}{298--320}.
\newblock


\bibitem[Formal et~al\mbox{.}(2021)]%
        {formal:2021:ecir:WhiteBoxColBERT}
\bibfield{author}{\bibinfo{person}{Thibault Formal}, \bibinfo{person}{Benjamin
  Piwowarski}, {and} \bibinfo{person}{St{\'{e}}phane Clinchant}.}
  \bibinfo{year}{2021}\natexlab{}.
\newblock \showarticletitle{A White Box Analysis of ColBERT}. In
  \bibinfo{booktitle}{\emph{Advances in Information Retrieval - 43rd European
  Conference on {IR} Research, {ECIR} 2021, Virtual Event, March 28 - April 1,
  2021, Proceedings, Part {II}}} \emph{(\bibinfo{series}{Lecture Notes in
  Computer Science}, Vol.~\bibinfo{volume}{12657})}.
  \bibinfo{publisher}{Springer}, \bibinfo{pages}{257--263}.
\newblock
\urldef\tempurl%
\url{https://doi.org/10.1007/978-3-030-72240-1\_23}
\showDOI{\tempurl}


\bibitem[Formal et~al\mbox{.}(2022)]%
        {formal:2022:ecir:lexicalMatchingNeuralIR}
\bibfield{author}{\bibinfo{person}{Thibault Formal}, \bibinfo{person}{Benjamin
  Piwowarski}, {and} \bibinfo{person}{St{\'{e}}phane Clinchant}.}
  \bibinfo{year}{2022}\natexlab{}.
\newblock \showarticletitle{Match Your Words! {A} Study of Lexical Matching in
  Neural Information Retrieval}. In \bibinfo{booktitle}{\emph{Advances in
  Information Retrieval - 44th European Conference on {IR} Research, {ECIR}
  2022, Stavanger, Norway, April 10-14, 2022, Proceedings, Part {II}}}
  \emph{(\bibinfo{series}{Lecture Notes in Computer Science},
  Vol.~\bibinfo{volume}{13186})}. \bibinfo{publisher}{Springer},
  \bibinfo{pages}{120--127}.
\newblock
\urldef\tempurl%
\url{https://doi.org/10.1007/978-3-030-99739-7\_14}
\showDOI{\tempurl}


\bibitem[Geirhos et~al\mbox{.}(2020)]%
        {geirhos:2020:natmachintell:shortcutlearningDNNs}
\bibfield{author}{\bibinfo{person}{Robert Geirhos},
  \bibinfo{person}{J{\"{o}}rn{-}Henrik Jacobsen}, \bibinfo{person}{Claudio
  Michaelis}, \bibinfo{person}{Richard~S. Zemel}, \bibinfo{person}{Wieland
  Brendel}, \bibinfo{person}{Matthias Bethge}, {and} \bibinfo{person}{Felix~A.
  Wichmann}.} \bibinfo{year}{2020}\natexlab{}.
\newblock \showarticletitle{Shortcut learning in deep neural networks}.
\newblock \bibinfo{journal}{\emph{Nat. Mach. Intell.}} \bibinfo{volume}{2},
  \bibinfo{number}{11} (\bibinfo{year}{2020}), \bibinfo{pages}{665--673}.
\newblock
\urldef\tempurl%
\url{https://doi.org/10.1038/s42256-020-00257-z}
\showDOI{\tempurl}


\bibitem[Goren et~al\mbox{.}(2020)]%
        {goren:2020:sigir:ranking}
\bibfield{author}{\bibinfo{person}{Gregory Goren}, \bibinfo{person}{Oren
  Kurland}, \bibinfo{person}{Moshe Tennenholtz}, {and} \bibinfo{person}{Fiana
  Raiber}.} \bibinfo{year}{2020}\natexlab{}.
\newblock \showarticletitle{Ranking-Incentivized Quality Preserving Content
  Modification}. In \bibinfo{booktitle}{\emph{Proceedings of the 43rd
  International {ACM} {SIGIR} conference on research and development in
  Information Retrieval, {SIGIR} 2020, Virtual Event, China, July 25-30,
  2020}}. \bibinfo{publisher}{{ACM}}, \bibinfo{pages}{259--268}.
\newblock
\urldef\tempurl%
\url{https://doi.org/10.1145/3397271.3401058}
\showDOI{\tempurl}


\bibitem[Guo et~al\mbox{.}(2016)]%
        {guo:2016:cikm:drmm}
\bibfield{author}{\bibinfo{person}{Jiafeng Guo}, \bibinfo{person}{Yixing Fan},
  \bibinfo{person}{Qingyao Ai}, {and} \bibinfo{person}{W.~Bruce Croft}.}
  \bibinfo{year}{2016}\natexlab{}.
\newblock \showarticletitle{A Deep Relevance Matching Model for Ad-hoc
  Retrieval}. In \bibinfo{booktitle}{\emph{Proceedings of the 25th {ACM}
  International Conference on Information and Knowledge Management, {CIKM}
  2016, Indianapolis, IN, USA, October 24-28, 2016}}.
  \bibinfo{publisher}{{ACM}}, \bibinfo{pages}{55--64}.
\newblock
\urldef\tempurl%
\url{https://doi.org/10.1145/2983323.2983769}
\showDOI{\tempurl}


\bibitem[Hagen et~al\mbox{.}(2016)]%
        {hagen:2016:cikm:aximaticReranking}
\bibfield{author}{\bibinfo{person}{Matthias Hagen}, \bibinfo{person}{Michael
  V{\"{o}}lske}, \bibinfo{person}{Steve G{\"{o}}ring}, {and}
  \bibinfo{person}{Benno Stein}.} \bibinfo{year}{2016}\natexlab{}.
\newblock \showarticletitle{Axiomatic Result Re-Ranking}. In
  \bibinfo{booktitle}{\emph{Proceedings of the 25th {ACM} International
  Conference on Information and Knowledge Management, {CIKM} 2016,
  Indianapolis, IN, USA, October 24-28, 2016}}. \bibinfo{publisher}{{ACM}},
  \bibinfo{pages}{721--730}.
\newblock
\urldef\tempurl%
\url{https://doi.org/10.1145/2983323.2983704}
\showDOI{\tempurl}


\bibitem[Haim et~al\mbox{.}(2018)]%
        {haim2018burst}
\bibfield{author}{\bibinfo{person}{Mario Haim}, \bibinfo{person}{Andreas
  Graefe}, {and} \bibinfo{person}{Hans-Bernd Brosius}.}
  \bibinfo{year}{2018}\natexlab{}.
\newblock \showarticletitle{Burst of the filter bubble? Effects of
  personalization on the diversity of Google News}.
\newblock \bibinfo{journal}{\emph{Digital Journalism}} \bibinfo{volume}{6},
  \bibinfo{number}{3} (\bibinfo{year}{2018}), \bibinfo{pages}{330--343}.
\newblock


\bibitem[Hanauer(2006)]%
        {hanauer:2006:emerse:medicalsearch}
\bibfield{author}{\bibinfo{person}{David~A Hanauer}.}
  \bibinfo{year}{2006}\natexlab{}.
\newblock \showarticletitle{EMERSE: the electronic medical record search
  engine}. In \bibinfo{booktitle}{\emph{AMIA annual symposium proceedings}},
  Vol.~\bibinfo{volume}{2006}. American Medical Informatics Association,
  \bibinfo{pages}{941}.
\newblock


\bibitem[Hannak et~al\mbox{.}(2013)]%
        {hannak2013measuring}
\bibfield{author}{\bibinfo{person}{Aniko Hannak}, \bibinfo{person}{Piotr
  Sapiezynski}, \bibinfo{person}{Arash~Molavi Kakhki},
  \bibinfo{person}{Balachander Krishnamurthy}, \bibinfo{person}{David Lazer},
  \bibinfo{person}{Alan Mislove}, {and} \bibinfo{person}{Christo Wilson}.}
  \bibinfo{year}{2013}\natexlab{}.
\newblock \showarticletitle{Measuring personalization of web search}. In
  \bibinfo{booktitle}{\emph{22nd International World Wide Web Conference, {WWW}
  '13, Rio de Janeiro, Brazil, May 13-17, 2013}}.
  \bibinfo{publisher}{International World Wide Web Conferences Steering
  Committee / {ACM}}, \bibinfo{pages}{527--538}.
\newblock
\urldef\tempurl%
\url{https://doi.org/10.1145/2488388.2488435}
\showDOI{\tempurl}


\bibitem[Harvey et~al\mbox{.}(2013)]%
        {harvey2013building}
\bibfield{author}{\bibinfo{person}{Morgan Harvey}, \bibinfo{person}{Fabio
  Crestani}, {and} \bibinfo{person}{Mark~James Carman}.}
  \bibinfo{year}{2013}\natexlab{}.
\newblock \showarticletitle{Building user profiles from topic models for
  personalised search}. In \bibinfo{booktitle}{\emph{22nd {ACM} International
  Conference on Information and Knowledge Management, CIKM'13, San Francisco,
  CA, USA, October 27 - November 1, 2013}}. \bibinfo{publisher}{{ACM}},
  \bibinfo{pages}{2309--2314}.
\newblock
\urldef\tempurl%
\url{https://doi.org/10.1145/2505515.2505642}
\showDOI{\tempurl}


\bibitem[Hewitt and Liang(2019)]%
        {hewitt:2019:emnlp:probingControlTasks}
\bibfield{author}{\bibinfo{person}{John Hewitt} {and} \bibinfo{person}{Percy
  Liang}.} \bibinfo{year}{2019}\natexlab{}.
\newblock \showarticletitle{Designing and Interpreting Probes with Control
  Tasks}. In \bibinfo{booktitle}{\emph{Proceedings of the 2019 Conference on
  Empirical Methods in Natural Language Processing and the 9th International
  Joint Conference on Natural Language Processing (EMNLP-IJCNLP)}}.
  \bibinfo{publisher}{Association for Computational Linguistics},
  \bibinfo{address}{Hong Kong, China}, \bibinfo{pages}{2733--2743}.
\newblock
\urldef\tempurl%
\url{https://doi.org/10.18653/v1/D19-1275}
\showDOI{\tempurl}


\bibitem[Hofst{\"{a}}tter et~al\mbox{.}(2021)]%
        {hoftaetter:2021:sigir:esm:etm}
\bibfield{author}{\bibinfo{person}{Sebastian Hofst{\"{a}}tter},
  \bibinfo{person}{Bhaskar Mitra}, \bibinfo{person}{Hamed Zamani},
  \bibinfo{person}{Nick Craswell}, {and} \bibinfo{person}{Allan Hanbury}.}
  \bibinfo{year}{2021}\natexlab{}.
\newblock \showarticletitle{Intra-Document Cascading: Learning to Select
  Passages for Neural Document Ranking}. In \bibinfo{booktitle}{\emph{{SIGIR}
  '21: The 44th International {ACM} {SIGIR} Conference on Research and
  Development in Information Retrieval, Virtual Event, Canada, July 11-15,
  2021}}. \bibinfo{publisher}{{ACM}}, \bibinfo{pages}{1349--1358}.
\newblock
\urldef\tempurl%
\url{https://doi.org/10.1145/3404835.3462889}
\showDOI{\tempurl}


\bibitem[Hofst{\"a}tter et~al\mbox{.}(2020)]%
        {hofstatter:2020:ecai:interpretablererank}
\bibfield{author}{\bibinfo{person}{Sebastian Hofst{\"a}tter},
  \bibinfo{person}{Markus Zlabinger}, {and} \bibinfo{person}{Allan Hanbury}.}
  \bibinfo{year}{2020}\natexlab{}.
\newblock \showarticletitle{Interpretable \& Time-Budget-Constrained
  Contextualization for Re-Ranking}.
\newblock In \bibinfo{booktitle}{\emph{ECAI 2020}}. \bibinfo{publisher}{IOS
  Press}, \bibinfo{pages}{513--520}.
\newblock


\bibitem[Holzmann and Anand(2016)]%
        {holzmann:www:2016:tempas}
\bibfield{author}{\bibinfo{person}{Helge Holzmann} {and}
  \bibinfo{person}{Avishek Anand}.} \bibinfo{year}{2016}\natexlab{}.
\newblock \showarticletitle{Tempas: Temporal archive search based on tags}. In
  \bibinfo{booktitle}{\emph{Proceedings of the 25th International Conference
  Companion on World Wide Web}}. \bibinfo{pages}{207--210}.
\newblock
\urldef\tempurl%
\url{https://doi.org/10.1145/2872518.2890555}
\showDOI{\tempurl}


\bibitem[Holzmann et~al\mbox{.}(2017a)]%
        {holzmann:webscience:2017:exploring}
\bibfield{author}{\bibinfo{person}{Helge Holzmann}, \bibinfo{person}{Wolfgang
  Nejdl}, {and} \bibinfo{person}{Avishek Anand}.}
  \bibinfo{year}{2017}\natexlab{a}.
\newblock \showarticletitle{Exploring web archives through temporal anchor
  texts}. In \bibinfo{booktitle}{\emph{Proceedings of the 2017 ACM on Web
  Science Conference}}. \bibinfo{pages}{289--298}.
\newblock


\bibitem[Holzmann et~al\mbox{.}(2017b)]%
        {holzmann:2017:websci:temporalanchor}
\bibfield{author}{\bibinfo{person}{H. Holzmann}, \bibinfo{person}{W. Nejdl},
  {and} \bibinfo{person}{A. Anand}.} \bibinfo{year}{2017}\natexlab{b}.
\newblock \showarticletitle{Exploring web archives through temporal anchor
  texts}. In \bibinfo{booktitle}{\emph{Proceedings of the 2017 ACM on Web
  Science Conference}}. \bibinfo{pages}{289--298}.
\newblock
\urldef\tempurl%
\url{https://doi.org/10.1145/3091478.3091500}
\showDOI{\tempurl}


\bibitem[Hooker et~al\mbox{.}(2019)]%
        {hooker:2019:roar-bechnmark}
\bibfield{author}{\bibinfo{person}{Sara Hooker}, \bibinfo{person}{Dumitru
  Erhan}, \bibinfo{person}{Pieter{-}Jan Kindermans}, {and}
  \bibinfo{person}{Been Kim}.} \bibinfo{year}{2019}\natexlab{}.
\newblock \showarticletitle{A Benchmark for Interpretability Methods in Deep
  Neural Networks}. In \bibinfo{booktitle}{\emph{Advances in Neural Information
  Processing Systems 32: Annual Conference on Neural Information Processing
  Systems 2019, NeurIPS 2019, December 8-14, 2019, Vancouver, BC, Canada}}.
  \bibinfo{pages}{9734--9745}.
\newblock
\urldef\tempurl%
\url{https://proceedings.neurips.cc/paper/2019/hash/fe4b8556000d0f0cae99daa5c5c5a410-Abstract.html}
\showURL{%
\tempurl}


\bibitem[Idahl et~al\mbox{.}(2021)]%
        {idahl:2021:utility}
\bibfield{author}{\bibinfo{person}{Maximilian Idahl}, \bibinfo{person}{Lijun
  Lyu}, \bibinfo{person}{Ujwal Gadiraju}, {and} \bibinfo{person}{Avishek
  Anand}.} \bibinfo{year}{2021}\natexlab{}.
\newblock \showarticletitle{Towards Benchmarking the Utility of Explanations
  for Model Debugging}. In \bibinfo{booktitle}{\emph{Proceedings of the First
  Workshop on Trustworthy Natural Language Processing}}.
  \bibinfo{publisher}{Association for Computational Linguistics},
  \bibinfo{address}{Online}, \bibinfo{pages}{68--73}.
\newblock
\urldef\tempurl%
\url{https://doi.org/10.18653/v1/2021.trustnlp-1.8}
\showDOI{\tempurl}


\bibitem[Jacovi and Goldberg(2021)]%
        {jacovi:2021:acl:socialattribution}
\bibfield{author}{\bibinfo{person}{Alon Jacovi} {and} \bibinfo{person}{Yoav
  Goldberg}.} \bibinfo{year}{2021}\natexlab{}.
\newblock \showarticletitle{Aligning Faithful Interpretations with their Social
  Attribution}.
\newblock \bibinfo{journal}{\emph{Transactions of the Association for
  Computational Linguistics}}  \bibinfo{volume}{9} (\bibinfo{year}{2021}),
  \bibinfo{pages}{294--310}.
\newblock
\urldef\tempurl%
\url{https://doi.org/10.1162/tacl_a_00367}
\showDOI{\tempurl}


\bibitem[Jiang et~al\mbox{.}(2021)]%
        {jiang:2021:blackboxnlp:bertbottleneck}
\bibfield{author}{\bibinfo{person}{Zhiying Jiang}, \bibinfo{person}{Raphael
  Tang}, \bibinfo{person}{Ji Xin}, {and} \bibinfo{person}{Jimmy Lin}.}
  \bibinfo{year}{2021}\natexlab{}.
\newblock \showarticletitle{How Does {BERT} Rerank Passages? An Attribution
  Analysis with Information Bottlenecks}. In
  \bibinfo{booktitle}{\emph{Proceedings of the Fourth BlackboxNLP Workshop on
  Analyzing and Interpreting Neural Networks for NLP}}.
  \bibinfo{publisher}{Association for Computational Linguistics},
  \bibinfo{address}{Punta Cana, Dominican Republic}, \bibinfo{pages}{496--509}.
\newblock
\urldef\tempurl%
\url{https://doi.org/10.18653/v1/2021.blackboxnlp-1.39}
\showDOI{\tempurl}


\bibitem[Joachims(2002)]%
        {joachims2002optimizingclick}
\bibfield{author}{\bibinfo{person}{Thorsten Joachims}.}
  \bibinfo{year}{2002}\natexlab{}.
\newblock \showarticletitle{Optimizing search engines using clickthrough data}.
  In \bibinfo{booktitle}{\emph{Proceedings of the eighth ACM SIGKDD
  international conference on Knowledge discovery and data mining}}. ACM,
  \bibinfo{pages}{133--142}.
\newblock


\bibitem[Khattab and Zaharia(2020)]%
        {khattab:2020:sigir:colbert}
\bibfield{author}{\bibinfo{person}{Omar Khattab} {and} \bibinfo{person}{Matei
  Zaharia}.} \bibinfo{year}{2020}\natexlab{}.
\newblock \showarticletitle{ColBERT: Efficient and Effective Passage Search via
  Contextualized Late Interaction over {BERT}}. In
  \bibinfo{booktitle}{\emph{Proceedings of the 43rd International {ACM} {SIGIR}
  conference on research and development in Information Retrieval, {SIGIR}
  2020, Virtual Event, China, July 25-30, 2020}}. \bibinfo{publisher}{{ACM}},
  \bibinfo{pages}{39--48}.
\newblock
\urldef\tempurl%
\url{https://doi.org/10.1145/3397271.3401075}
\showDOI{\tempurl}


\bibitem[Koh and Liang(2017)]%
        {Koh:2017:icml:influencefunction}
\bibfield{author}{\bibinfo{person}{Pang~Wei Koh} {and} \bibinfo{person}{Percy
  Liang}.} \bibinfo{year}{2017}\natexlab{}.
\newblock \showarticletitle{Understanding Black-box Predictions via Influence
  Functions}. In \bibinfo{booktitle}{\emph{Proceedings of the 34th
  International Conference on Machine Learning, {ICML} 2017, Sydney, NSW,
  Australia, 6-11 August 2017}} \emph{(\bibinfo{series}{Proceedings of Machine
  Learning Research}, Vol.~\bibinfo{volume}{70})}. \bibinfo{publisher}{{PMLR}},
  \bibinfo{pages}{1885--1894}.
\newblock
\urldef\tempurl%
\url{http://proceedings.mlr.press/v70/koh17a.html}
\showURL{%
\tempurl}


\bibitem[Kumar and Talukdar(2020)]%
        {kumar:2020:nile}
\bibfield{author}{\bibinfo{person}{Sawan Kumar} {and} \bibinfo{person}{Partha
  Talukdar}.} \bibinfo{year}{2020}\natexlab{}.
\newblock \showarticletitle{{NILE} : Natural Language Inference with Faithful
  Natural Language Explanations}. In \bibinfo{booktitle}{\emph{Proceedings of
  the 58th Annual Meeting of the Association for Computational Linguistics}}.
  \bibinfo{publisher}{Association for Computational Linguistics},
  \bibinfo{address}{Online}, \bibinfo{pages}{8730--8742}.
\newblock
\urldef\tempurl%
\url{https://doi.org/10.18653/v1/2020.acl-main.771}
\showDOI{\tempurl}


\bibitem[Lasri et~al\mbox{.}(2022)]%
        {lasri:2022:acl:probingusagegramnumber}
\bibfield{author}{\bibinfo{person}{Karim Lasri}, \bibinfo{person}{Tiago
  Pimentel}, \bibinfo{person}{Alessandro Lenci}, \bibinfo{person}{Thierry
  Poibeau}, {and} \bibinfo{person}{Ryan Cotterell}.}
  \bibinfo{year}{2022}\natexlab{}.
\newblock \showarticletitle{Probing for the Usage of Grammatical Number}. In
  \bibinfo{booktitle}{\emph{Proceedings of the 60th Annual Meeting of the
  Association for Computational Linguistics (Volume 1: Long Papers), {ACL}
  2022, Dublin, Ireland, May 22-27, 2022}}. \bibinfo{publisher}{Association for
  Computational Linguistics}, \bibinfo{pages}{8818--8831}.
\newblock
\urldef\tempurl%
\url{https://doi.org/10.18653/v1/2022.acl-long.603}
\showDOI{\tempurl}


\bibitem[Lavrenko and Croft(2001)]%
        {lavrenko:2001:rm3}
\bibfield{author}{\bibinfo{person}{Victor Lavrenko} {and}
  \bibinfo{person}{W.~Bruce Croft}.} \bibinfo{year}{2001}\natexlab{}.
\newblock \showarticletitle{Relevance Based Language Models}. In
  \bibinfo{booktitle}{\emph{Proceedings of the 24th Annual International ACM
  SIGIR Conference on Research and Development in Information Retrieval}} (New
  Orleans, Louisiana, USA) \emph{(\bibinfo{series}{SIGIR '01})}.
  \bibinfo{publisher}{Association for Computing Machinery},
  \bibinfo{address}{New York, NY, USA}, \bibinfo{pages}{120–127}.
\newblock
\showISBNx{1581133316}
\urldef\tempurl%
\url{https://doi.org/10.1145/383952.383972}
\showDOI{\tempurl}


\bibitem[Lehman et~al\mbox{.}(2019)]%
        {lehman:2019:naacl:inferringdataset}
\bibfield{author}{\bibinfo{person}{Eric Lehman}, \bibinfo{person}{Jay DeYoung},
  \bibinfo{person}{Regina Barzilay}, {and} \bibinfo{person}{Byron~C. Wallace}.}
  \bibinfo{year}{2019}\natexlab{}.
\newblock \showarticletitle{Inferring Which Medical Treatments Work from
  Reports of Clinical Trials}. In \bibinfo{booktitle}{\emph{Proceedings of the
  2019 Conference of the North {A}merican Chapter of the Association for
  Computational Linguistics: Human Language Technologies, Volume 1 (Long and
  Short Papers)}}. \bibinfo{publisher}{Association for Computational
  Linguistics}, \bibinfo{address}{Minneapolis, Minnesota},
  \bibinfo{pages}{3705--3717}.
\newblock
\urldef\tempurl%
\url{https://doi.org/10.18653/v1/N19-1371}
\showDOI{\tempurl}


\bibitem[Lei et~al\mbox{.}(2016)]%
        {lei:2016:emnlp:rationalizing}
\bibfield{author}{\bibinfo{person}{Tao Lei}, \bibinfo{person}{Regina Barzilay},
  {and} \bibinfo{person}{Tommi Jaakkola}.} \bibinfo{year}{2016}\natexlab{}.
\newblock \showarticletitle{Rationalizing Neural Predictions}. In
  \bibinfo{booktitle}{\emph{Proceedings of the 2016 Conference on Empirical
  Methods in Natural Language Processing}}. \bibinfo{publisher}{Association for
  Computational Linguistics}, \bibinfo{address}{Austin, Texas},
  \bibinfo{pages}{107--117}.
\newblock
\urldef\tempurl%
\url{https://doi.org/10.18653/v1/D16-1011}
\showDOI{\tempurl}


\bibitem[Leonhardt et~al\mbox{.}(2021)]%
        {leonhardt:2021:arxiv:hard:kuma:ranking}
\bibfield{author}{\bibinfo{person}{Jurek Leonhardt}, \bibinfo{person}{Koustav
  Rudra}, {and} \bibinfo{person}{Avishek Anand}.}
  \bibinfo{year}{2021}\natexlab{}.
\newblock \showarticletitle{Learnt Sparsity for Effective and Interpretable
  Document Ranking}.
\newblock \bibinfo{journal}{\emph{ArXiv preprint}}
  \bibinfo{volume}{abs/2106.12460} (\bibinfo{year}{2021}).
\newblock
\urldef\tempurl%
\url{https://arxiv.org/abs/2106.12460}
\showURL{%
\tempurl}


\bibitem[Lewis et~al\mbox{.}(2020)]%
        {lewis:2020:BART}
\bibfield{author}{\bibinfo{person}{Mike Lewis}, \bibinfo{person}{Yinhan Liu},
  \bibinfo{person}{Naman Goyal}, \bibinfo{person}{Marjan Ghazvininejad},
  \bibinfo{person}{Abdelrahman Mohamed}, \bibinfo{person}{Omer Levy},
  \bibinfo{person}{Veselin Stoyanov}, {and} \bibinfo{person}{Luke
  Zettlemoyer}.} \bibinfo{year}{2020}\natexlab{}.
\newblock \showarticletitle{{BART}: Denoising Sequence-to-Sequence Pre-training
  for Natural Language Generation, Translation, and Comprehension}. In
  \bibinfo{booktitle}{\emph{Proceedings of the 58th Annual Meeting of the
  Association for Computational Linguistics}}. \bibinfo{publisher}{Association
  for Computational Linguistics}, \bibinfo{address}{Online},
  \bibinfo{pages}{7871--7880}.
\newblock
\urldef\tempurl%
\url{https://doi.org/10.18653/v1/2020.acl-main.703}
\showDOI{\tempurl}


\bibitem[Lin(2004)]%
        {lin:2004:ROUGE}
\bibfield{author}{\bibinfo{person}{Chin-Yew Lin}.}
  \bibinfo{year}{2004}\natexlab{}.
\newblock \showarticletitle{{ROUGE}: A Package for Automatic Evaluation of
  Summaries}. In \bibinfo{booktitle}{\emph{Text Summarization Branches Out}}.
  \bibinfo{publisher}{Association for Computational Linguistics},
  \bibinfo{address}{Barcelona, Spain}, \bibinfo{pages}{74--81}.
\newblock
\urldef\tempurl%
\url{https://aclanthology.org/W04-1013}
\showURL{%
\tempurl}


\bibitem[Lin(2019)]%
        {lin:sigir:2019:recantation}
\bibfield{author}{\bibinfo{person}{Jimmy Lin}.}
  \bibinfo{year}{2019}\natexlab{}.
\newblock \showarticletitle{The neural hype, justified!: a recantation}.
\newblock \bibinfo{journal}{\emph{{SIGIR} Forum}} \bibinfo{volume}{53},
  \bibinfo{number}{2} (\bibinfo{year}{2019}), \bibinfo{pages}{88--93}.
\newblock
\urldef\tempurl%
\url{https://doi.org/10.1145/3458553.3458563}
\showDOI{\tempurl}


\bibitem[Liu et~al\mbox{.}(2019)]%
        {liu:2019:generative-explanation-framework}
\bibfield{author}{\bibinfo{person}{Hui Liu}, \bibinfo{person}{Qingyu Yin},
  {and} \bibinfo{person}{William~Yang Wang}.} \bibinfo{year}{2019}\natexlab{}.
\newblock \showarticletitle{Towards Explainable {NLP}: A Generative Explanation
  Framework for Text Classification}. In \bibinfo{booktitle}{\emph{Proceedings
  of the 57th Annual Meeting of the Association for Computational
  Linguistics}}. \bibinfo{publisher}{Association for Computational
  Linguistics}, \bibinfo{address}{Florence, Italy},
  \bibinfo{pages}{5570--5581}.
\newblock
\urldef\tempurl%
\url{https://doi.org/10.18653/v1/P19-1560}
\showDOI{\tempurl}


\bibitem[Lucchese et~al\mbox{.}(2022)]%
        {lucchese:2022:sigir:ilmart}
\bibfield{author}{\bibinfo{person}{Claudio Lucchese},
  \bibinfo{person}{Franco~Maria Nardini}, \bibinfo{person}{Salvatore Orlando},
  \bibinfo{person}{Raffaele Perego}, {and} \bibinfo{person}{Alberto Veneri}.}
  \bibinfo{year}{2022}\natexlab{}.
\newblock \showarticletitle{{ILMART:} Interpretable Ranking with Constrained
  LambdaMART}. In \bibinfo{booktitle}{\emph{{SIGIR} '22: The 45th International
  {ACM} {SIGIR} Conference on Research and Development in Information
  Retrieval, Madrid, Spain, July 11 - 15, 2022}}. \bibinfo{publisher}{{ACM}},
  \bibinfo{pages}{2255--2259}.
\newblock
\urldef\tempurl%
\url{https://doi.org/10.1145/3477495.3531840}
\showDOI{\tempurl}


\bibitem[Lundberg et~al\mbox{.}(2018)]%
        {lundberg:2018:treeshap}
\bibfield{author}{\bibinfo{person}{Scott~M. Lundberg},
  \bibinfo{person}{Gabriel~G. Erion}, {and} \bibinfo{person}{Su{-}In Lee}.}
  \bibinfo{year}{2018}\natexlab{}.
\newblock \showarticletitle{Consistent Individualized Feature Attribution for
  Tree Ensembles}.
\newblock \bibinfo{journal}{\emph{ArXiv preprint}}
  \bibinfo{volume}{abs/1802.03888} (\bibinfo{year}{2018}).
\newblock
\urldef\tempurl%
\url{https://arxiv.org/abs/1802.03888}
\showURL{%
\tempurl}


\bibitem[Lundberg and Lee(2017)]%
        {lundberg:2017:neurips:shap}
\bibfield{author}{\bibinfo{person}{Scott~M. Lundberg} {and}
  \bibinfo{person}{Su{-}In Lee}.} \bibinfo{year}{2017}\natexlab{}.
\newblock \showarticletitle{A Unified Approach to Interpreting Model
  Predictions}. In \bibinfo{booktitle}{\emph{Advances in Neural Information
  Processing Systems 30: Annual Conference on Neural Information Processing
  Systems 2017, December 4-9, 2017, Long Beach, CA, {USA}}}.
  \bibinfo{pages}{4765--4774}.
\newblock
\urldef\tempurl%
\url{https://proceedings.neurips.cc/paper/2017/hash/8a20a8621978632d76c43dfd28b67767-Abstract.html}
\showURL{%
\tempurl}


\bibitem[MacAvaney et~al\mbox{.}(2020)]%
        {macavaney:2020:arxiv:abnirml}
\bibfield{author}{\bibinfo{person}{Sean MacAvaney}, \bibinfo{person}{Sergey
  Feldman}, \bibinfo{person}{Nazli Goharian}, \bibinfo{person}{Doug Downey},
  {and} \bibinfo{person}{Arman Cohan}.} \bibinfo{year}{2020}\natexlab{}.
\newblock \showarticletitle{{{ABNIRML}}: {{Analyzing}} the {{Behavior}} of
  {{Neural IR Models}}}.
\newblock \bibinfo{journal}{\emph{ArXiv preprint}}
  \bibinfo{volume}{abs/2011.00696} (\bibinfo{year}{2020}).
\newblock
\urldef\tempurl%
\url{https://arxiv.org/abs/2011.00696}
\showURL{%
\tempurl}


\bibitem[Madsen et~al\mbox{.}(2021)]%
        {madsen:2021:eval-faithfulness}
\bibfield{author}{\bibinfo{person}{Andreas Madsen}, \bibinfo{person}{Nicholas
  Meade}, \bibinfo{person}{Vaibhav Adlakha}, {and} \bibinfo{person}{Siva
  Reddy}.} \bibinfo{year}{2021}\natexlab{}.
\newblock \showarticletitle{Evaluating the Faithfulness of Importance Measures
  in {NLP} by Recursively Masking Allegedly Important Tokens and Retraining}.
\newblock \bibinfo{journal}{\emph{ArXiv preprint}}
  \bibinfo{volume}{abs/2110.08412} (\bibinfo{year}{2021}).
\newblock
\urldef\tempurl%
\url{https://arxiv.org/abs/2110.08412}
\showURL{%
\tempurl}


\bibitem[McDonald et~al\mbox{.}(2018)]%
        {mcdonald:2018:P-DRMM}
\bibfield{author}{\bibinfo{person}{Ryan McDonald}, \bibinfo{person}{George
  Brokos}, {and} \bibinfo{person}{Ion Androutsopoulos}.}
  \bibinfo{year}{2018}\natexlab{}.
\newblock \showarticletitle{Deep Relevance Ranking Using Enhanced
  Document-Query Interactions}. In \bibinfo{booktitle}{\emph{Proceedings of the
  2018 Conference on Empirical Methods in Natural Language Processing}}.
  \bibinfo{publisher}{Association for Computational Linguistics},
  \bibinfo{address}{Brussels, Belgium}, \bibinfo{pages}{1849--1860}.
\newblock
\urldef\tempurl%
\url{https://doi.org/10.18653/v1/D18-1211}
\showDOI{\tempurl}


\bibitem[Miller(2019)]%
        {Miller:2019:insights-from-social-science}
\bibfield{author}{\bibinfo{person}{Tim Miller}.}
  \bibinfo{year}{2019}\natexlab{}.
\newblock \showarticletitle{Explanation in artificial intelligence: Insights
  from the social sciences}.
\newblock \bibinfo{journal}{\emph{Artif. Intell.}}  \bibinfo{volume}{267}
  (\bibinfo{year}{2019}), \bibinfo{pages}{1--38}.
\newblock
\urldef\tempurl%
\url{https://doi.org/10.1016/j.artint.2018.07.007}
\showDOI{\tempurl}


\bibitem[Molnar(2022)]%
        {molnar:2022:interpretable-ML-book}
\bibfield{author}{\bibinfo{person}{Christoph Molnar}.}
  \bibinfo{year}{2022}\natexlab{}.
\newblock \bibinfo{booktitle}{\emph{Interpretable Machine Learning}
  (\bibinfo{edition}{2} ed.)}.
\newblock
\urldef\tempurl%
\url{https://christophm.github.io/interpretable-ml-book}
\showURL{%
\tempurl}


\bibitem[Mowshowitz and Kawaguchi(2005)]%
        {mowshowitz:ipm:2005:measuring-ir-bias}
\bibfield{author}{\bibinfo{person}{Abbe Mowshowitz} {and}
  \bibinfo{person}{Akira Kawaguchi}.} \bibinfo{year}{2005}\natexlab{}.
\newblock \showarticletitle{Measuring search engine bias}.
\newblock \bibinfo{journal}{\emph{Inf. Process. Manag.}} \bibinfo{volume}{41},
  \bibinfo{number}{5} (\bibinfo{year}{2005}), \bibinfo{pages}{1193--1205}.
\newblock
\urldef\tempurl%
\url{https://doi.org/10.1016/j.ipm.2004.05.005}
\showDOI{\tempurl}


\bibitem[Mustafaraj et~al\mbox{.}(2020)]%
        {mustafaraj2020case}
\bibfield{author}{\bibinfo{person}{Eni Mustafaraj}, \bibinfo{person}{Emma
  Lurie}, {and} \bibinfo{person}{Claire Devine}.}
  \bibinfo{year}{2020}\natexlab{}.
\newblock \showarticletitle{The case for voter-centered audits of search
  engines during political elections}. In \bibinfo{booktitle}{\emph{Proceedings
  of the 2020 Conference on Fairness, Accountability, and Transparency}}.
  \bibinfo{pages}{559--569}.
\newblock


\bibitem[Nalisnick et~al\mbox{.}(2016)]%
        {nalisnick:2016:DESM}
\bibfield{author}{\bibinfo{person}{Eric Nalisnick}, \bibinfo{person}{Bhaskar
  Mitra}, \bibinfo{person}{Nick Craswell}, {and} \bibinfo{person}{Rich
  Caruana}.} \bibinfo{year}{2016}\natexlab{}.
\newblock \showarticletitle{Improving Document Ranking with Dual Word
  Embeddings}. In \bibinfo{booktitle}{\emph{Proceedings of the 25th
  International Conference Companion on World Wide Web}} (Montr\'{e}al,
  Qu\'{e}bec, Canada) \emph{(\bibinfo{series}{WWW '16 Companion})}.
  \bibinfo{publisher}{International World Wide Web Conferences Steering
  Committee}, \bibinfo{address}{Republic and Canton of Geneva, CHE},
  \bibinfo{pages}{83–84}.
\newblock
\showISBNx{9781450341448}
\urldef\tempurl%
\url{https://doi.org/10.1145/2872518.2889361}
\showDOI{\tempurl}


\bibitem[Nguyen et~al\mbox{.}(2016)]%
        {nguyen:2016:msmarco}
\bibfield{author}{\bibinfo{person}{Tri Nguyen}, \bibinfo{person}{Mir
  Rosenberg}, \bibinfo{person}{Xia Song}, \bibinfo{person}{Jianfeng Gao},
  \bibinfo{person}{Saurabh Tiwary}, \bibinfo{person}{Rangan Majumder}, {and}
  \bibinfo{person}{Li Deng}.} \bibinfo{year}{2016}\natexlab{}.
\newblock \showarticletitle{{MS} {MARCO:} {A} Human Generated MAchine Reading
  COmprehension Dataset}. In \bibinfo{booktitle}{\emph{Proceedings of the
  Workshop on Cognitive Computation: Integrating neural and symbolic approaches
  2016 co-located with the 30th Annual Conference on Neural Information
  Processing Systems {(NIPS} 2016), Barcelona, Spain, December 9, 2016}}
  \emph{(\bibinfo{series}{{CEUR} Workshop Proceedings},
  Vol.~\bibinfo{volume}{1773})},
  \bibfield{editor}{\bibinfo{person}{Tarek~Richard Besold},
  \bibinfo{person}{Antoine Bordes}, \bibinfo{person}{Artur~S. d'Avila Garcez},
  {and} \bibinfo{person}{Greg Wayne}} (Eds.). \bibinfo{publisher}{CEUR-WS.org}.
\newblock
\urldef\tempurl%
\url{http://ceur-ws.org/Vol-1773/CoCoNIPS\_2016\_paper9.pdf}
\showURL{%
\tempurl}


\bibitem[Papineni et~al\mbox{.}(2002)]%
        {Papineni:2002:BLEU}
\bibfield{author}{\bibinfo{person}{Kishore Papineni}, \bibinfo{person}{Salim
  Roukos}, \bibinfo{person}{Todd Ward}, {and} \bibinfo{person}{Wei-Jing Zhu}.}
  \bibinfo{year}{2002}\natexlab{}.
\newblock \showarticletitle{{B}leu: a Method for Automatic Evaluation of
  Machine Translation}. In \bibinfo{booktitle}{\emph{Proceedings of the 40th
  Annual Meeting of the Association for Computational Linguistics}}.
  \bibinfo{publisher}{Association for Computational Linguistics},
  \bibinfo{address}{Philadelphia, Pennsylvania, USA},
  \bibinfo{pages}{311--318}.
\newblock
\urldef\tempurl%
\url{https://doi.org/10.3115/1073083.1073135}
\showDOI{\tempurl}


\bibitem[Pariser(2011)]%
        {pariser2011filter}
\bibfield{author}{\bibinfo{person}{Eli Pariser}.}
  \bibinfo{year}{2011}\natexlab{}.
\newblock \bibinfo{booktitle}{\emph{The filter bubble: How the new personalized
  web is changing what we read and how we think}}.
\newblock \bibinfo{publisher}{Penguin}.
\newblock


\bibitem[Petroni et~al\mbox{.}(2021)]%
        {petroni:2021:naacl:kilt}
\bibfield{author}{\bibinfo{person}{Fabio Petroni}, \bibinfo{person}{Aleksandra
  Piktus}, \bibinfo{person}{Angela Fan}, \bibinfo{person}{Patrick Lewis},
  \bibinfo{person}{Majid Yazdani}, \bibinfo{person}{Nicola De~Cao},
  \bibinfo{person}{James Thorne}, \bibinfo{person}{Yacine Jernite},
  \bibinfo{person}{Vladimir Karpukhin}, \bibinfo{person}{Jean Maillard},
  \bibinfo{person}{Vassilis Plachouras}, \bibinfo{person}{Tim Rockt{\"a}schel},
  {and} \bibinfo{person}{Sebastian Riedel}.} \bibinfo{year}{2021}\natexlab{}.
\newblock \showarticletitle{{KILT}: a Benchmark for Knowledge Intensive
  Language Tasks}. In \bibinfo{booktitle}{\emph{Proceedings of the 2021
  Conference of the North American Chapter of the Association for Computational
  Linguistics: Human Language Technologies}}. \bibinfo{publisher}{Association
  for Computational Linguistics}, \bibinfo{address}{Online},
  \bibinfo{pages}{2523--2544}.
\newblock
\urldef\tempurl%
\url{https://doi.org/10.18653/v1/2021.naacl-main.200}
\showDOI{\tempurl}


\bibitem[Pimentel et~al\mbox{.}(2020)]%
        {pimentel:2020:emnlp:paretoprobing}
\bibfield{author}{\bibinfo{person}{Tiago Pimentel}, \bibinfo{person}{Naomi
  Saphra}, \bibinfo{person}{Adina Williams}, {and} \bibinfo{person}{Ryan
  Cotterell}.} \bibinfo{year}{2020}\natexlab{}.
\newblock \showarticletitle{{P}areto Probing: {T}rading Off Accuracy for
  Complexity}. In \bibinfo{booktitle}{\emph{Proceedings of the 2020 Conference
  on Empirical Methods in Natural Language Processing (EMNLP)}}.
  \bibinfo{publisher}{Association for Computational Linguistics},
  \bibinfo{address}{Online}, \bibinfo{pages}{3138--3153}.
\newblock
\urldef\tempurl%
\url{https://doi.org/10.18653/v1/2020.emnlp-main.254}
\showDOI{\tempurl}


\bibitem[Polley et~al\mbox{.}(2021)]%
        {polley:2021:exdocs}
\bibfield{author}{\bibinfo{person}{Sayantan Polley}, \bibinfo{person}{Atin
  Janki}, \bibinfo{person}{Juliane Thiel, Marcusand Hoebel-Mueller}, {and}
  \bibinfo{person}{Andreas Nuernberger}.} \bibinfo{year}{2021}\natexlab{}.
\newblock \showarticletitle{ExDocS: Evidence based Explainable Document
  Search}. In \bibinfo{booktitle}{\emph{ACM SIGIR Workshop on Causality in
  Search and Recommendation}}. ACM.
\newblock
\urldef\tempurl%
\url{https://csr21.github.io/polley-csr2021.pdf}
\showURL{%
\tempurl}


\bibitem[Ponte and Croft(1998)]%
        {pontecroft:1998:sigir:language}
\bibfield{author}{\bibinfo{person}{Jay~M Ponte} {and} \bibinfo{person}{W~Bruce
  Croft}.} \bibinfo{year}{1998}\natexlab{}.
\newblock \showarticletitle{A language modeling approach to information
  retrieval}. In \bibinfo{booktitle}{\emph{Proceedings of the 21st annual
  international ACM SIGIR conference on Research and development in information
  retrieval}}. ACM, \bibinfo{pages}{275--281}.
\newblock


\bibitem[Pratapa et~al\mbox{.}(2020)]%
        {pratapa:emnlp:2020:factchecking}
\bibfield{author}{\bibinfo{person}{Adithya Pratapa},
  \bibinfo{person}{Sai~Muralidhar Jayanthi}, {and} \bibinfo{person}{Kavya
  Nerella}.} \bibinfo{year}{2020}\natexlab{}.
\newblock \showarticletitle{Constrained fact verification for FEVER}. In
  \bibinfo{booktitle}{\emph{Proceedings of the 2020 Conference on Empirical
  Methods in Natural Language Processing (EMNLP)}}.
  \bibinfo{pages}{7826--7832}.
\newblock


\bibitem[Purpura et~al\mbox{.}(2021)]%
        {purpura:2021:ecir:neuralfs}
\bibfield{author}{\bibinfo{person}{Alberto Purpura}, \bibinfo{person}{Karolina
  Buchner}, \bibinfo{person}{Gianmaria Silvello}, {and}
  \bibinfo{person}{Gian~Antonio Susto}.} \bibinfo{year}{2021}\natexlab{}.
\newblock \showarticletitle{Neural feature selection for learning to rank}. In
  \bibinfo{booktitle}{\emph{European Conference on Information Retrieval}}.
  Springer, \bibinfo{pages}{342--349}.
\newblock


\bibitem[Qiao et~al\mbox{.}(2019)]%
        {qiao:2019:understanding-behavior-bert-ranking}
\bibfield{author}{\bibinfo{person}{Yifan Qiao}, \bibinfo{person}{Chenyan
  Xiong}, \bibinfo{person}{Zhenghao Liu}, {and} \bibinfo{person}{Zhiyuan Liu}.}
  \bibinfo{year}{2019}\natexlab{}.
\newblock \showarticletitle{Understanding the Behaviors of BERT in Ranking}.
\newblock \bibinfo{journal}{\emph{ArXiv preprint}}
  \bibinfo{volume}{abs/1904.07531} (\bibinfo{year}{2019}).
\newblock
\urldef\tempurl%
\url{https://arxiv.org/abs/1904.07531}
\showURL{%
\tempurl}


\bibitem[Qin and Liu(2013)]%
        {qin:2013:corr:letor4}
\bibfield{author}{\bibinfo{person}{Tao Qin} {and} \bibinfo{person}{Tie{-}Yan
  Liu}.} \bibinfo{year}{2013}\natexlab{}.
\newblock \showarticletitle{Introducing {LETOR} 4.0 Datasets}.
\newblock \bibinfo{journal}{\emph{CoRR}}  \bibinfo{volume}{abs/1306.2597}
  (\bibinfo{year}{2013}).
\newblock


\bibitem[Rahimi et~al\mbox{.}(2021)]%
        {rahimi:2021:arxiv:explainQuery}
\bibfield{author}{\bibinfo{person}{Razieh Rahimi}, \bibinfo{person}{Youngwoo
  Kim}, \bibinfo{person}{Hamed Zamani}, {and} \bibinfo{person}{James Allan}.}
  \bibinfo{year}{2021}\natexlab{}.
\newblock \showarticletitle{Explaining Documents' Relevance to Search Queries}.
\newblock \bibinfo{journal}{\emph{ArXiv preprint}}
  \bibinfo{volume}{abs/2111.01314} (\bibinfo{year}{2021}).
\newblock
\urldef\tempurl%
\url{https://arxiv.org/abs/2111.01314}
\showURL{%
\tempurl}


\bibitem[Rajani et~al\mbox{.}(2019)]%
        {rajani:2019:explain-yourself}
\bibfield{author}{\bibinfo{person}{Nazneen~Fatema Rajani},
  \bibinfo{person}{Bryan McCann}, \bibinfo{person}{Caiming Xiong}, {and}
  \bibinfo{person}{Richard Socher}.} \bibinfo{year}{2019}\natexlab{}.
\newblock \showarticletitle{Explain Yourself! Leveraging Language Models for
  Commonsense Reasoning}. In \bibinfo{booktitle}{\emph{Proceedings of the 57th
  Annual Meeting of the Association for Computational Linguistics}}.
  \bibinfo{publisher}{Association for Computational Linguistics},
  \bibinfo{address}{Florence, Italy}, \bibinfo{pages}{4932--4942}.
\newblock
\urldef\tempurl%
\url{https://doi.org/10.18653/v1/P19-1487}
\showDOI{\tempurl}


\bibitem[Raval and Verma(2020)]%
        {raval:2020:adv-attack-retrieval-models}
\bibfield{author}{\bibinfo{person}{Nisarg Raval} {and} \bibinfo{person}{Manisha
  Verma}.} \bibinfo{year}{2020}\natexlab{}.
\newblock \showarticletitle{One word at a time: adversarial attacks on
  retrieval models}.
\newblock \bibinfo{journal}{\emph{ArXiv preprint}}
  \bibinfo{volume}{abs/2008.02197} (\bibinfo{year}{2020}).
\newblock
\urldef\tempurl%
\url{https://arxiv.org/abs/2008.02197}
\showURL{%
\tempurl}


\bibitem[Rekabsaz and Schedl(2020)]%
        {rekabsaz:sigir:2020:ir-gender-bias}
\bibfield{author}{\bibinfo{person}{Navid Rekabsaz} {and}
  \bibinfo{person}{Markus Schedl}.} \bibinfo{year}{2020}\natexlab{}.
\newblock \showarticletitle{Do Neural Ranking Models Intensify Gender Bias?}.
  In \bibinfo{booktitle}{\emph{Proceedings of the 43rd International {ACM}
  {SIGIR} conference on research and development in Information Retrieval,
  {SIGIR} 2020, Virtual Event, China, July 25-30, 2020}}.
  \bibinfo{publisher}{{ACM}}, \bibinfo{pages}{2065--2068}.
\newblock
\urldef\tempurl%
\url{https://doi.org/10.1145/3397271.3401280}
\showDOI{\tempurl}


\bibitem[Rennings et~al\mbox{.}(2019)]%
        {rennings:2019:ecir:axiomDiagnoseNeural}
\bibfield{author}{\bibinfo{person}{Dani{\"{e}}l Rennings},
  \bibinfo{person}{Felipe Moraes}, {and} \bibinfo{person}{Claudia Hauff}.}
  \bibinfo{year}{2019}\natexlab{}.
\newblock \showarticletitle{An Axiomatic Approach to Diagnosing Neural {IR}
  Models}. In \bibinfo{booktitle}{\emph{Proceedings of {ECIR} 2019}}.
  \bibinfo{pages}{489--503}.
\newblock


\bibitem[Ribeiro et~al\mbox{.}(2016)]%
        {ribeiro:2016:kdd:lime}
\bibfield{author}{\bibinfo{person}{Marco~T{\'{u}}lio Ribeiro},
  \bibinfo{person}{Sameer Singh}, {and} \bibinfo{person}{Carlos Guestrin}.}
  \bibinfo{year}{2016}\natexlab{}.
\newblock \showarticletitle{"Why Should {I} Trust You?": Explaining the
  Predictions of Any Classifier}. In \bibinfo{booktitle}{\emph{Proceedings of
  the 22nd {ACM} {SIGKDD} International Conference on Knowledge Discovery and
  Data Mining, San Francisco, CA, USA, August 13-17, 2016}},
  \bibfield{editor}{\bibinfo{person}{Balaji Krishnapuram},
  \bibinfo{person}{Mohak Shah}, \bibinfo{person}{Alexander~J. Smola},
  \bibinfo{person}{Charu~C. Aggarwal}, \bibinfo{person}{Dou Shen}, {and}
  \bibinfo{person}{Rajeev Rastogi}} (Eds.). \bibinfo{publisher}{{ACM}},
  \bibinfo{pages}{1135--1144}.
\newblock
\urldef\tempurl%
\url{https://doi.org/10.1145/2939672.2939778}
\showDOI{\tempurl}


\bibitem[Rong et~al\mbox{.}(2022)]%
        {Rong:2022:eval-attribution-information-theory}
\bibfield{author}{\bibinfo{person}{Yao Rong}, \bibinfo{person}{Tobias Leemann},
  \bibinfo{person}{Vadim Borisov}, \bibinfo{person}{Gjergji Kasneci}, {and}
  \bibinfo{person}{Enkelejda Kasneci}.} \bibinfo{year}{2022}\natexlab{}.
\newblock \showarticletitle{A Consistent and Efficient Evaluation Strategy for
  Attribution Methods}. In \bibinfo{booktitle}{\emph{International Conference
  on Machine Learning, {ICML} 2022, 17-23 July 2022, Baltimore, Maryland,
  {USA}}} \emph{(\bibinfo{series}{Proceedings of Machine Learning Research},
  Vol.~\bibinfo{volume}{162})}. \bibinfo{publisher}{{PMLR}},
  \bibinfo{pages}{18770--18795}.
\newblock
\urldef\tempurl%
\url{https://proceedings.mlr.press/v162/rong22a.html}
\showURL{%
\tempurl}


\bibitem[Rosset et~al\mbox{.}(2019)]%
        {rosset:2019:sigir:axiomaticNeural}
\bibfield{author}{\bibinfo{person}{Corby Rosset}, \bibinfo{person}{Bhaskar
  Mitra}, \bibinfo{person}{Chenyan Xiong}, \bibinfo{person}{Nick Craswell},
  \bibinfo{person}{Xia Song}, {and} \bibinfo{person}{Saurabh Tiwary}.}
  \bibinfo{year}{2019}\natexlab{}.
\newblock \showarticletitle{An Axiomatic Approach to Regularizing Neural
  Ranking Models}. In \bibinfo{booktitle}{\emph{Proceedings of the 42nd
  International {ACM} {SIGIR} Conference on Research and Development in
  Information Retrieval, {SIGIR} 2019, Paris, France, July 21-25, 2019}}.
  \bibinfo{publisher}{{ACM}}, \bibinfo{pages}{981--984}.
\newblock
\urldef\tempurl%
\url{https://doi.org/10.1145/3331184.3331296}
\showDOI{\tempurl}


\bibitem[Roy and Anand(2021)]%
        {roy:2021:book:question}
\bibfield{author}{\bibinfo{person}{Rishiraj~Saha Roy} {and}
  \bibinfo{person}{Avishek Anand}.} \bibinfo{year}{2021}\natexlab{}.
\newblock \showarticletitle{Question Answering for the Curated Web: Tasks and
  Methods in QA over Knowledge Bases and Text Collections}.
\newblock \bibinfo{journal}{\emph{Synthesis Lectures onSynthesis Lectures on
  Information Concepts, Retrieval, and Services}} \bibinfo{volume}{13},
  \bibinfo{number}{4} (\bibinfo{year}{2021}), \bibinfo{pages}{1--194}.
\newblock
\urldef\tempurl%
\url{https://doi.org/10.1007/978-3-031-79512-1}
\showDOI{\tempurl}


\bibitem[Rudin(2019)]%
        {rudin:2019:nature:stopexplain}
\bibfield{author}{\bibinfo{person}{C. Rudin}.} \bibinfo{year}{2019}\natexlab{}.
\newblock \showarticletitle{Stop explaining black box machine learning models
  for high stakes decisions and use interpretable models instead}.
\newblock \bibinfo{journal}{\emph{Nature Machine Intelligence}}
  \bibinfo{volume}{1}, \bibinfo{number}{5} (\bibinfo{year}{2019}),
  \bibinfo{pages}{206}.
\newblock


\bibitem[Samek et~al\mbox{.}(2019)]%
        {wojciech:2019:explainable-ai-book}
\bibfield{editor}{\bibinfo{person}{Wojciech Samek},
  \bibinfo{person}{Gr{\'{e}}goire Montavon}, \bibinfo{person}{Andrea Vedaldi},
  \bibinfo{person}{Lars~Kai Hansen}, {and} \bibinfo{person}{Klaus{-}Robert
  M{\"{u}}ller}} (Eds.). \bibinfo{year}{2019}\natexlab{}.
\newblock \bibinfo{booktitle}{\emph{Explainable {AI:} Interpreting, Explaining
  and Visualizing Deep Learning}}. \bibinfo{series}{Lecture Notes in Computer
  Science}, Vol.~\bibinfo{volume}{11700}.
\newblock \bibinfo{publisher}{Springer}.
\newblock
\showISBNx{978-3-030-28953-9}
\urldef\tempurl%
\url{https://doi.org/10.1007/978-3-030-28954-6}
\showDOI{\tempurl}


\bibitem[Sarker et~al\mbox{.}(2021)]%
        {sarker:2021:AIC:neurosymbolic}
\bibfield{author}{\bibinfo{person}{Md~Kamruzzaman Sarker}, \bibinfo{person}{Lu
  Zhou}, \bibinfo{person}{Aaron Eberhart}, {and} \bibinfo{person}{Pascal
  Hitzler}.} \bibinfo{year}{2021}\natexlab{}.
\newblock \showarticletitle{Neuro-Symbolic Artificial Intelligence}.
\newblock \bibinfo{journal}{\emph{AI Commun.}} \bibinfo{volume}{34},
  \bibinfo{number}{3} (\bibinfo{year}{2021}), \bibinfo{pages}{197–209}.
\newblock
\showISSN{0921-7126}
\urldef\tempurl%
\url{https://doi.org/10.3233/AIC-210084}
\showDOI{\tempurl}


\bibitem[Sen et~al\mbox{.}(2020)]%
        {sen:2020:explain-document-scores-within-and-across-ranking-models}
\bibfield{author}{\bibinfo{person}{Procheta Sen}, \bibinfo{person}{Debasis
  Ganguly}, \bibinfo{person}{Manisha Verma}, {and} \bibinfo{person}{Gareth
  J.~F. Jones}.} \bibinfo{year}{2020}\natexlab{}.
\newblock \showarticletitle{The Curious Case of {IR} Explainability: Explaining
  Document Scores within and across Ranking Models}. In
  \bibinfo{booktitle}{\emph{Proceedings of the 43rd International {ACM} {SIGIR}
  conference on research and development in Information Retrieval, {SIGIR}
  2020, Virtual Event, China, July 25-30, 2020}}. \bibinfo{publisher}{{ACM}},
  \bibinfo{pages}{2069--2072}.
\newblock
\urldef\tempurl%
\url{https://doi.org/10.1145/3397271.3401286}
\showDOI{\tempurl}


\bibitem[Shrikumar et~al\mbox{.}(2017)]%
        {shrikumar:2017:icml:deeplift}
\bibfield{author}{\bibinfo{person}{Avanti Shrikumar}, \bibinfo{person}{Peyton
  Greenside}, {and} \bibinfo{person}{Anshul Kundaje}.}
  \bibinfo{year}{2017}\natexlab{}.
\newblock \showarticletitle{Learning Important Features Through Propagating
  Activation Differences}. In \bibinfo{booktitle}{\emph{Proceedings of the 34th
  International Conference on Machine Learning, {ICML} 2017, Sydney, NSW,
  Australia, 6-11 August 2017}} \emph{(\bibinfo{series}{Proceedings of Machine
  Learning Research}, Vol.~\bibinfo{volume}{70})},
  \bibfield{editor}{\bibinfo{person}{Doina Precup} {and}
  \bibinfo{person}{Yee~Whye Teh}} (Eds.). \bibinfo{publisher}{{PMLR}},
  \bibinfo{pages}{3145--3153}.
\newblock
\urldef\tempurl%
\url{http://proceedings.mlr.press/v70/shrikumar17a.html}
\showURL{%
\tempurl}


\bibitem[Singh and Anand(2018)]%
        {singh:2018:ears:posthocSecondary}
\bibfield{author}{\bibinfo{person}{Jaspreet Singh} {and}
  \bibinfo{person}{Avishek Anand}.} \bibinfo{year}{2018}\natexlab{}.
\newblock \showarticletitle{Posthoc Interpretability of Learning to Rank Models
  using Secondary Training Data}. In \bibinfo{booktitle}{\emph{Workshop on
  ExplainAble Recommendation and Search (EARS 2018) at SIGIR 2018}}.
\newblock
\urldef\tempurl%
\url{https://ears2018.github.io/ears18-singh.pdf}
\showURL{%
\tempurl}


\bibitem[Singh and Anand(2019)]%
        {singh:2019:wsdm:exs}
\bibfield{author}{\bibinfo{person}{Jaspreet Singh} {and}
  \bibinfo{person}{Avishek Anand}.} \bibinfo{year}{2019}\natexlab{}.
\newblock \showarticletitle{{EXS:} Explainable Search Using Local Model
  Agnostic Interpretability}. In \bibinfo{booktitle}{\emph{Proceedings of the
  Twelfth {ACM} International Conference on Web Search and Data Mining, {WSDM}
  2019, Melbourne, VIC, Australia, February 11-15, 2019}},
  \bibfield{editor}{\bibinfo{person}{J.~Shane Culpepper},
  \bibinfo{person}{Alistair Moffat}, \bibinfo{person}{Paul~N. Bennett}, {and}
  \bibinfo{person}{Kristina Lerman}} (Eds.). \bibinfo{publisher}{{ACM}},
  \bibinfo{pages}{770--773}.
\newblock
\urldef\tempurl%
\url{https://doi.org/10.1145/3289600.3290620}
\showDOI{\tempurl}


\bibitem[Singh and Anand(2020)]%
        {singh:2020:fat:intentmodel}
\bibfield{author}{\bibinfo{person}{Jaspreet Singh} {and}
  \bibinfo{person}{Avishek Anand}.} \bibinfo{year}{2020}\natexlab{}.
\newblock \showarticletitle{Model agnostic interpretability of rankers via
  intent modelling}. In \bibinfo{booktitle}{\emph{Proceedings of the 2020
  Conference on Fairness, Accountability, and Transparency}}.
  \bibinfo{pages}{618--628}.
\newblock
\urldef\tempurl%
\url{https://doi.org/10.1145/3351095.3375234}
\showDOI{\tempurl}


\bibitem[Singh et~al\mbox{.}(2016a)]%
        {singh:cikm:2016:entity}
\bibfield{author}{\bibinfo{person}{Jaspreet Singh}, \bibinfo{person}{Johannes
  Hoffart}, {and} \bibinfo{person}{Avishek Anand}.}
  \bibinfo{year}{2016}\natexlab{a}.
\newblock \showarticletitle{Discovering entities with just a little help from
  you}. In \bibinfo{booktitle}{\emph{Proceedings of the 25th ACM International
  on Conference on Information and Knowledge Management}}.
  \bibinfo{pages}{1331--1340}.
\newblock
\urldef\tempurl%
\url{https://doi.org/10.1145/2983323.2983798}
\showDOI{\tempurl}


\bibitem[Singh et~al\mbox{.}(2021)]%
        {singh:2021:ValidSetLtR}
\bibfield{author}{\bibinfo{person}{Jaspreet Singh}, \bibinfo{person}{Megha
  Khosla}, \bibinfo{person}{Wang Zhenye}, {and} \bibinfo{person}{Avishek
  Anand}.} \bibinfo{year}{2021}\natexlab{}.
\newblock \showarticletitle{Extracting per Query Valid Explanations for
  Blackbox Learning-to-Rank Models}. In \bibinfo{booktitle}{\emph{Proceedings
  of the 2021 ACM SIGIR International Conference on Theory of Information
  Retrieval}} (Virtual Event, Canada) \emph{(\bibinfo{series}{ICTIR '21})}.
  \bibinfo{publisher}{Association for Computing Machinery},
  \bibinfo{address}{New York, NY, USA}, \bibinfo{pages}{203–210}.
\newblock
\showISBNx{9781450386111}
\urldef\tempurl%
\url{https://doi.org/10.1145/3471158.3472241}
\showDOI{\tempurl}


\bibitem[Singh et~al\mbox{.}(2016b)]%
        {singh:sigir:2016:expedition}
\bibfield{author}{\bibinfo{person}{Jaspreet Singh}, \bibinfo{person}{Wolfgang
  Nejdl}, {and} \bibinfo{person}{Avishek Anand}.}
  \bibinfo{year}{2016}\natexlab{b}.
\newblock \showarticletitle{Expedition: a time-aware exploratory search system
  designed for scholars}. In \bibinfo{booktitle}{\emph{Proceedings of the 39th
  International ACM SIGIR conference on Research and Development in Information
  Retrieval}}. \bibinfo{pages}{1105--1108}.
\newblock
\urldef\tempurl%
\url{https://doi.org/10.1145/2911451.2911465}
\showDOI{\tempurl}


\bibitem[S{\o}gaard(2021)]%
        {Sogaard:2021:explainable-nlp}
\bibfield{author}{\bibinfo{person}{Anders S{\o}gaard}.}
  \bibinfo{year}{2021}\natexlab{}.
\newblock \bibinfo{booktitle}{\emph{Explainable Natural Language Processing}}.
\newblock \bibinfo{publisher}{Morgan {\&} Claypool Publishers}.
\newblock
\urldef\tempurl%
\url{https://doi.org/10.2200/S01118ED1V01Y202107HLT051}
\showDOI{\tempurl}


\bibitem[Springenberg et~al\mbox{.}(2014)]%
        {springenberg:2014:guidedBP}
\bibfield{author}{\bibinfo{person}{Jost~Tobias Springenberg},
  \bibinfo{person}{Alexey Dosovitskiy}, \bibinfo{person}{Thomas Brox}, {and}
  \bibinfo{person}{Martin Riedmiller}.} \bibinfo{year}{2014}\natexlab{}.
\newblock \showarticletitle{Striving for simplicity: The all convolutional
  net}.
\newblock \bibinfo{journal}{\emph{arXiv preprint arXiv:1412.6806}}
  (\bibinfo{year}{2014}).
\newblock


\bibitem[Sturmfels et~al\mbox{.}(2020)]%
        {sturmfels:2020:baselines}
\bibfield{author}{\bibinfo{person}{Pascal Sturmfels}, \bibinfo{person}{Scott
  Lundberg}, {and} \bibinfo{person}{Su-In Lee}.}
  \bibinfo{year}{2020}\natexlab{}.
\newblock \showarticletitle{Visualizing the Impact of Feature Attribution
  Baselines}.
\newblock \bibinfo{journal}{\emph{Distill}} (\bibinfo{year}{2020}).
\newblock
\urldef\tempurl%
\url{https://doi.org/10.23915/distill.00022}
\showDOI{\tempurl}
\newblock
\shownote{https://distill.pub/2020/attribution-baselines}.


\bibitem[Sundararajan et~al\mbox{.}(2017)]%
        {sundararajan:2017:integratedgradients}
\bibfield{author}{\bibinfo{person}{Mukund Sundararajan}, \bibinfo{person}{Ankur
  Taly}, {and} \bibinfo{person}{Qiqi Yan}.} \bibinfo{year}{2017}\natexlab{}.
\newblock \showarticletitle{Axiomatic Attribution for Deep Networks}. In
  \bibinfo{booktitle}{\emph{Proceedings of the 34th International Conference on
  Machine Learning, {ICML} 2017, Sydney, NSW, Australia, 6-11 August 2017}}
  \emph{(\bibinfo{series}{Proceedings of Machine Learning Research},
  Vol.~\bibinfo{volume}{70})}, \bibfield{editor}{\bibinfo{person}{Doina Precup}
  {and} \bibinfo{person}{Yee~Whye Teh}} (Eds.). \bibinfo{publisher}{{PMLR}},
  \bibinfo{pages}{3319--3328}.
\newblock
\urldef\tempurl%
\url{http://proceedings.mlr.press/v70/sundararajan17a.html}
\showURL{%
\tempurl}


\bibitem[Tenney et~al\mbox{.}(2019a)]%
        {tenney:2019:acl:bert}
\bibfield{author}{\bibinfo{person}{Ian Tenney}, \bibinfo{person}{Dipanjan Das},
  {and} \bibinfo{person}{Ellie Pavlick}.} \bibinfo{year}{2019}\natexlab{a}.
\newblock \showarticletitle{{BERT} Rediscovers the Classical {NLP} Pipeline}.
  In \bibinfo{booktitle}{\emph{Proceedings of the 57th Annual Meeting of the
  Association for Computational Linguistics}}. \bibinfo{publisher}{Association
  for Computational Linguistics}, \bibinfo{address}{Florence, Italy},
  \bibinfo{pages}{4593--4601}.
\newblock
\urldef\tempurl%
\url{https://doi.org/10.18653/v1/P19-1452}
\showDOI{\tempurl}


\bibitem[Tenney et~al\mbox{.}(2019b)]%
        {tenney:2019:iclr:you}
\bibfield{author}{\bibinfo{person}{Ian Tenney}, \bibinfo{person}{Patrick Xia},
  \bibinfo{person}{Berlin Chen}, \bibinfo{person}{Alex Wang},
  \bibinfo{person}{Adam Poliak}, \bibinfo{person}{R.~Thomas McCoy},
  \bibinfo{person}{Najoung Kim}, \bibinfo{person}{Benjamin~Van Durme},
  \bibinfo{person}{Samuel~R. Bowman}, \bibinfo{person}{Dipanjan Das}, {and}
  \bibinfo{person}{Ellie Pavlick}.} \bibinfo{year}{2019}\natexlab{b}.
\newblock \showarticletitle{What do you learn from context? Probing for
  sentence structure in contextualized word representations}. In
  \bibinfo{booktitle}{\emph{7th International Conference on Learning
  Representations, {ICLR} 2019, New Orleans, LA, USA, May 6-9, 2019}}.
  \bibinfo{publisher}{OpenReview.net}.
\newblock
\urldef\tempurl%
\url{https://openreview.net/forum?id=SJzSgnRcKX}
\showURL{%
\tempurl}


\bibitem[Tomsett et~al\mbox{.}(2020)]%
        {tomsett:2020:saliency-metrics}
\bibfield{author}{\bibinfo{person}{Richard Tomsett}, \bibinfo{person}{Dan
  Harborne}, \bibinfo{person}{Supriyo Chakraborty}, \bibinfo{person}{Prudhvi
  Gurram}, {and} \bibinfo{person}{Alun~D. Preece}.}
  \bibinfo{year}{2020}\natexlab{}.
\newblock \showarticletitle{Sanity Checks for Saliency Metrics}. In
  \bibinfo{booktitle}{\emph{The Thirty-Fourth {AAAI} Conference on Artificial
  Intelligence, {AAAI} 2020, February 7-12, 2020}}. \bibinfo{publisher}{{AAAI}
  Press}, \bibinfo{pages}{6021--6029}.
\newblock
\urldef\tempurl%
\url{https://aaai.org/ojs/index.php/AAAI/article/view/6064}
\showURL{%
\tempurl}


\bibitem[van Aken et~al\mbox{.}(2019)]%
        {vanaken:2019:cikm:howdoesbertQA}
\bibfield{author}{\bibinfo{person}{Betty van Aken}, \bibinfo{person}{Benjamin
  Winter}, \bibinfo{person}{Alexander L{\"{o}}ser}, {and}
  \bibinfo{person}{Felix~A. Gers}.} \bibinfo{year}{2019}\natexlab{}.
\newblock \showarticletitle{How Does {BERT} Answer Questions?: {A} Layer-Wise
  Analysis of Transformer Representations}. In
  \bibinfo{booktitle}{\emph{Proceedings of the 28th {ACM} International
  Conference on Information and Knowledge Management, {CIKM} 2019, Beijing,
  China, November 3-7, 2019}}. \bibinfo{publisher}{{ACM}},
  \bibinfo{pages}{1823--1832}.
\newblock
\urldef\tempurl%
\url{https://doi.org/10.1145/3357384.3358028}
\showDOI{\tempurl}


\bibitem[Verma and Ganguly(2019)]%
        {verma:2019:sigir:lirme}
\bibfield{author}{\bibinfo{person}{Manisha Verma} {and}
  \bibinfo{person}{Debasis Ganguly}.} \bibinfo{year}{2019}\natexlab{}.
\newblock \showarticletitle{{LIRME:} Locally Interpretable Ranking Model
  Explanation}. In \bibinfo{booktitle}{\emph{Proceedings of the 42nd
  International {ACM} {SIGIR} Conference on Research and Development in
  Information Retrieval, {SIGIR} 2019, Paris, France, July 21-25, 2019}},
  \bibfield{editor}{\bibinfo{person}{Benjamin Piwowarski}, \bibinfo{person}{Max
  Chevalier}, \bibinfo{person}{{\'{E}}ric Gaussier}, \bibinfo{person}{Yoelle
  Maarek}, \bibinfo{person}{Jian{-}Yun Nie}, {and} \bibinfo{person}{Falk
  Scholer}} (Eds.). \bibinfo{publisher}{{ACM}}, \bibinfo{pages}{1281--1284}.
\newblock
\urldef\tempurl%
\url{https://doi.org/10.1145/3331184.3331377}
\showDOI{\tempurl}


\bibitem[Vig(2019)]%
        {vig:2019:bertviz}
\bibfield{author}{\bibinfo{person}{Jesse Vig}.}
  \bibinfo{year}{2019}\natexlab{}.
\newblock \showarticletitle{A Multiscale Visualization of Attention in the
  Transformer Model}. In \bibinfo{booktitle}{\emph{Proceedings of the 57th
  Annual Meeting of the Association for Computational Linguistics: System
  Demonstrations}}. \bibinfo{publisher}{Association for Computational
  Linguistics}, \bibinfo{address}{Florence, Italy}, \bibinfo{pages}{37--42}.
\newblock
\urldef\tempurl%
\url{https://doi.org/10.18653/v1/P19-3007}
\showDOI{\tempurl}


\bibitem[Voita and Titov(2020)]%
        {voita:2020:emnlp:ProbingMDL}
\bibfield{author}{\bibinfo{person}{Elena Voita} {and} \bibinfo{person}{Ivan
  Titov}.} \bibinfo{year}{2020}\natexlab{}.
\newblock \showarticletitle{Information-Theoretic Probing with Minimum
  Description Length}. In \bibinfo{booktitle}{\emph{Proceedings of the 2020
  Conference on Empirical Methods in Natural Language Processing (EMNLP)}}.
  \bibinfo{publisher}{Association for Computational Linguistics},
  \bibinfo{address}{Online}, \bibinfo{pages}{183--196}.
\newblock
\urldef\tempurl%
\url{https://doi.org/10.18653/v1/2020.emnlp-main.14}
\showDOI{\tempurl}


\bibitem[V{\"{o}}lske et~al\mbox{.}(2021)]%
        {voelske:2021:ictir:axiomaticExplanations4NeuralRanking}
\bibfield{author}{\bibinfo{person}{Michael V{\"{o}}lske},
  \bibinfo{person}{Alexander Bondarenko}, \bibinfo{person}{Maik Fr{\"{o}}be},
  \bibinfo{person}{Benno Stein}, \bibinfo{person}{Jaspreet Singh},
  \bibinfo{person}{Matthias Hagen}, {and} \bibinfo{person}{Avishek Anand}.}
  \bibinfo{year}{2021}\natexlab{}.
\newblock \showarticletitle{Towards Axiomatic Explanations for Neural Ranking
  Models}. In \bibinfo{booktitle}{\emph{{ICTIR} '21: The 2021 {ACM} {SIGIR}
  International Conference on the Theory of Information Retrieval, Virtual
  Event, Canada, July 11, 2021}}, \bibfield{editor}{\bibinfo{person}{Faegheh
  Hasibi}, \bibinfo{person}{Yi~Fang}, {and} \bibinfo{person}{Akiko Aizawa}}
  (Eds.). \bibinfo{publisher}{{ACM}}, \bibinfo{pages}{13--22}.
\newblock
\urldef\tempurl%
\url{https://doi.org/10.1145/3471158.3472256}
\showDOI{\tempurl}


\bibitem[Voorhees(2006)]%
        {voorhees:2006:trec:robust}
\bibfield{author}{\bibinfo{person}{Ellen~M Voorhees}.}
  \bibinfo{year}{2006}\natexlab{}.
\newblock \showarticletitle{The TREC 2005 robust track}. In
  \bibinfo{booktitle}{\emph{ACM SIGIR Forum}}, Vol.~\bibinfo{volume}{40}. ACM
  New York, NY, USA, \bibinfo{pages}{41--48}.
\newblock


\bibitem[Wallace et~al\mbox{.}(2019)]%
        {wallace:2019:emnlp:universal}
\bibfield{author}{\bibinfo{person}{Eric Wallace}, \bibinfo{person}{Shi Feng},
  \bibinfo{person}{Nikhil Kandpal}, \bibinfo{person}{Matt Gardner}, {and}
  \bibinfo{person}{Sameer Singh}.} \bibinfo{year}{2019}\natexlab{}.
\newblock \showarticletitle{Universal Adversarial Triggers for Attacking and
  Analyzing {NLP}}. In \bibinfo{booktitle}{\emph{Proceedings of the 2019
  Conference on Empirical Methods in Natural Language Processing and the 9th
  International Joint Conference on Natural Language Processing
  (EMNLP-IJCNLP)}}. \bibinfo{publisher}{Association for Computational
  Linguistics}, \bibinfo{address}{Hong Kong, China},
  \bibinfo{pages}{2153--2162}.
\newblock
\urldef\tempurl%
\url{https://doi.org/10.18653/v1/D19-1221}
\showDOI{\tempurl}


\bibitem[Wallat et~al\mbox{.}(2020)]%
        {DBLP:conf/blackboxnlp/bertnesia}
\bibfield{author}{\bibinfo{person}{Jonas Wallat}, \bibinfo{person}{Jaspreet
  Singh}, {and} \bibinfo{person}{Avishek Anand}.}
  \bibinfo{year}{2020}\natexlab{}.
\newblock \showarticletitle{{BERT}nesia: Investigating the capture and
  forgetting of knowledge in {BERT}}. In \bibinfo{booktitle}{\emph{Proceedings
  of the Third BlackboxNLP Workshop on Analyzing and Interpreting Neural
  Networks for NLP}}. \bibinfo{publisher}{Association for Computational
  Linguistics}, \bibinfo{address}{Online}, \bibinfo{pages}{174--183}.
\newblock
\urldef\tempurl%
\url{https://doi.org/10.18653/v1/2020.blackboxnlp-1.17}
\showDOI{\tempurl}


\bibitem[Wang et~al\mbox{.}(2022)]%
        {wang:2022:ictir:bert}
\bibfield{author}{\bibinfo{person}{Yumeng Wang}, \bibinfo{person}{Lijun Lyu},
  {and} \bibinfo{person}{Avishek Anand}.} \bibinfo{year}{2022}\natexlab{}.
\newblock \showarticletitle{BERT Rankers are Brittle: A Study using Adversarial
  Document Perturbations}. In \bibinfo{booktitle}{\emph{Proceedings of the 2022
  ACM SIGIR International Conference on Theory of Information Retrieval}}.
  \bibinfo{pages}{115--120}.
\newblock


\bibitem[Wiegreffe and Marasovic(2021)]%
        {wiegreffe:2021:teachmetoexplain}
\bibfield{author}{\bibinfo{person}{Sarah Wiegreffe} {and} \bibinfo{person}{Ana
  Marasovic}.} \bibinfo{year}{2021}\natexlab{}.
\newblock \showarticletitle{Teach Me to Explain: A Review of Datasets for
  Explainable Natural Language Processing}. In
  \bibinfo{booktitle}{\emph{Proceedings of the Neural Information Processing
  Systems Track on Datasets and Benchmarks}},
  \bibfield{editor}{\bibinfo{person}{J.~Vanschoren} {and}
  \bibinfo{person}{S.~Yeung}} (Eds.), Vol.~\bibinfo{volume}{1}.
\newblock
\urldef\tempurl%
\url{https://datasets-benchmarks-proceedings.neurips.cc/paper/2021/file/698d51a19d8a121ce581499d7b701668-Paper-round1.pdf}
\showURL{%
\tempurl}


\bibitem[Wojtas and Chen(2020)]%
        {wojtas:2020:nips:featureranking}
\bibfield{author}{\bibinfo{person}{Maksymilian Wojtas} {and}
  \bibinfo{person}{Ke Chen}.} \bibinfo{year}{2020}\natexlab{}.
\newblock \showarticletitle{Feature Importance Ranking for Deep Learning}. In
  \bibinfo{booktitle}{\emph{Advances in Neural Information Processing
  Systems}}, Vol.~\bibinfo{volume}{33}. \bibinfo{publisher}{Curran Associates,
  Inc.}, \bibinfo{pages}{5105--5114}.
\newblock
\urldef\tempurl%
\url{https://proceedings.neurips.cc/paper/2020/file/36ac8e558ac7690b6f44e2cb5ef93322-Paper.pdf}
\showURL{%
\tempurl}


\bibitem[Wu et~al\mbox{.}(2022)]%
        {wu:2022:arxiv:prada}
\bibfield{author}{\bibinfo{person}{Chen Wu}, \bibinfo{person}{Ruqing Zhang},
  \bibinfo{person}{Jiafeng Guo}, \bibinfo{person}{Maarten de Rijke},
  \bibinfo{person}{Yixing Fan}, {and} \bibinfo{person}{Xueqi Cheng}.}
  \bibinfo{year}{2022}\natexlab{}.
\newblock \showarticletitle{PRADA: Practical Black-Box Adversarial Attacks
  against Neural Ranking Models}.
\newblock \bibinfo{journal}{\emph{ArXiv preprint}}
  \bibinfo{volume}{abs/2204.01321} (\bibinfo{year}{2022}).
\newblock
\urldef\tempurl%
\url{https://arxiv.org/abs/2204.01321}
\showURL{%
\tempurl}


\bibitem[Xiong et~al\mbox{.}(2017)]%
        {xiong:2017:sigir:kernelpooling}
\bibfield{author}{\bibinfo{person}{Chenyan Xiong}, \bibinfo{person}{Zhuyun
  Dai}, \bibinfo{person}{Jamie Callan}, \bibinfo{person}{Zhiyuan Liu}, {and}
  \bibinfo{person}{Russell Power}.} \bibinfo{year}{2017}\natexlab{}.
\newblock \showarticletitle{End-to-End Neural Ad-hoc Ranking with Kernel
  Pooling}. In \bibinfo{booktitle}{\emph{Proceedings of the 40th International
  {ACM} {SIGIR} Conference on Research and Development in Information
  Retrieval, Shinjuku, Tokyo, Japan, August 7-11, 2017}}.
  \bibinfo{publisher}{{ACM}}, \bibinfo{pages}{55--64}.
\newblock
\urldef\tempurl%
\url{https://doi.org/10.1145/3077136.3080809}
\showDOI{\tempurl}


\bibitem[Yang and Kim(2019)]%
        {yang:2019:bam-benchmark}
\bibfield{author}{\bibinfo{person}{Mengjiao Yang} {and} \bibinfo{person}{Been
  Kim}.} \bibinfo{year}{2019}\natexlab{}.
\newblock \showarticletitle{{Benchmarking Attribution Methods with Relative
  Feature Importance}}.
\newblock \bibinfo{journal}{\emph{CoRR}}  \bibinfo{volume}{abs/1907.09701}
  (\bibinfo{year}{2019}).
\newblock


\bibitem[Yu et~al\mbox{.}(2022a)]%
        {yu:2022:sigir:generateListExplanation}
\bibfield{author}{\bibinfo{person}{Puxuan Yu}, \bibinfo{person}{Razieh Rahimi},
  {and} \bibinfo{person}{James Allan}.} \bibinfo{year}{2022}\natexlab{a}.
\newblock \showarticletitle{Towards Explainable Search Results: {A} Listwise
  Explanation Generator}. In \bibinfo{booktitle}{\emph{{SIGIR} '22: The 45th
  International {ACM} {SIGIR} Conference on Research and Development in
  Information Retrieval, Madrid, Spain, July 11 - 15, 2022}}.
  \bibinfo{publisher}{{ACM}}, \bibinfo{pages}{669--680}.
\newblock
\urldef\tempurl%
\url{https://doi.org/10.1145/3477495.3532067}
\showDOI{\tempurl}


\bibitem[Yu et~al\mbox{.}(2022b)]%
        {Yu:2022:sigir:iot-match}
\bibfield{author}{\bibinfo{person}{Weijie Yu}, \bibinfo{person}{Zhongxiang
  Sun}, \bibinfo{person}{Jun Xu}, \bibinfo{person}{Zhenhua Dong},
  \bibinfo{person}{Xu Chen}, \bibinfo{person}{Hongteng Xu}, {and}
  \bibinfo{person}{Ji{-}Rong Wen}.} \bibinfo{year}{2022}\natexlab{b}.
\newblock \showarticletitle{Explainable Legal Case Matching via Inverse Optimal
  Transport-based Rationale Extraction}. In \bibinfo{booktitle}{\emph{{SIGIR}
  '22: The 45th International {ACM} {SIGIR} Conference on Research and
  Development in Information Retrieval, Madrid, Spain, July 11 - 15, 2022}}.
  \bibinfo{publisher}{{ACM}}, \bibinfo{pages}{657--668}.
\newblock
\urldef\tempurl%
\url{https://doi.org/10.1145/3477495.3531974}
\showDOI{\tempurl}


\bibitem[Zamani et~al\mbox{.}(2020)]%
        {Zamani:2020:MIMICS}
\bibfield{author}{\bibinfo{person}{Hamed Zamani}, \bibinfo{person}{Gord Lueck},
  \bibinfo{person}{Everest Chen}, \bibinfo{person}{Rodolfo Quispe},
  \bibinfo{person}{Flint Luu}, {and} \bibinfo{person}{Nick Craswell}.}
  \bibinfo{year}{2020}\natexlab{}.
\newblock \showarticletitle{{MIMICS:} {A} Large-Scale Data Collection for
  Search Clarification}. In \bibinfo{booktitle}{\emph{{CIKM} '20: The 29th
  {ACM} International Conference on Information and Knowledge Management,
  Virtual Event, Ireland, October 19-23, 2020}},
  \bibfield{editor}{\bibinfo{person}{Mathieu d'Aquin}, \bibinfo{person}{Stefan
  Dietze}, \bibinfo{person}{Claudia Hauff}, \bibinfo{person}{Edward Curry},
  {and} \bibinfo{person}{Philippe Cudr{\'{e}}{-}Mauroux}} (Eds.).
  \bibinfo{publisher}{{ACM}}, \bibinfo{pages}{3189--3196}.
\newblock
\urldef\tempurl%
\url{https://doi.org/10.1145/3340531.3412772}
\showDOI{\tempurl}


\bibitem[Zhan et~al\mbox{.}(2020)]%
        {zhan:2020:sigir:AnalysisBERTDocRanking}
\bibfield{author}{\bibinfo{person}{Jingtao Zhan}, \bibinfo{person}{Jiaxin Mao},
  \bibinfo{person}{Yiqun Liu}, \bibinfo{person}{Min Zhang}, {and}
  \bibinfo{person}{Shaoping Ma}.} \bibinfo{year}{2020}\natexlab{}.
\newblock \showarticletitle{An Analysis of {BERT} in Document Ranking}. In
  \bibinfo{booktitle}{\emph{Proceedings of the 43rd International {ACM} {SIGIR}
  conference on research and development in Information Retrieval, {SIGIR}
  2020, Virtual Event, China, July 25-30, 2020}}. \bibinfo{publisher}{{ACM}},
  \bibinfo{pages}{1941--1944}.
\newblock
\urldef\tempurl%
\url{https://doi.org/10.1145/3397271.3401325}
\showDOI{\tempurl}


\bibitem[Zhang and Bowman(2018)]%
        {zhang:2018:bbnlp:ProbingBaselines}
\bibfield{author}{\bibinfo{person}{Kelly Zhang} {and} \bibinfo{person}{Samuel
  Bowman}.} \bibinfo{year}{2018}\natexlab{}.
\newblock \showarticletitle{Language Modeling Teaches You More than Translation
  Does: Lessons Learned Through Auxiliary Syntactic Task Analysis}. In
  \bibinfo{booktitle}{\emph{Proceedings of the 2018 {EMNLP} Workshop
  {B}lackbox{NLP}: Analyzing and Interpreting Neural Networks for {NLP}}}.
  \bibinfo{publisher}{Association for Computational Linguistics},
  \bibinfo{address}{Brussels, Belgium}, \bibinfo{pages}{359--361}.
\newblock
\urldef\tempurl%
\url{https://doi.org/10.18653/v1/W18-5448}
\showDOI{\tempurl}


\bibitem[Zhang et~al\mbox{.}(2020a)]%
        {zhang:2020:cikm:intentGeneration}
\bibfield{author}{\bibinfo{person}{Ruqing Zhang}, \bibinfo{person}{Jiafeng
  Guo}, \bibinfo{person}{Yixing Fan}, \bibinfo{person}{Yanyan Lan}, {and}
  \bibinfo{person}{Xueqi Cheng}.} \bibinfo{year}{2020}\natexlab{a}.
\newblock \showarticletitle{Query Understanding via Intent Description
  Generation}. In \bibinfo{booktitle}{\emph{{CIKM} '20: The 29th {ACM}
  International Conference on Information and Knowledge Management, Virtual
  Event, Ireland, October 19-23, 2020}},
  \bibfield{editor}{\bibinfo{person}{Mathieu d'Aquin}, \bibinfo{person}{Stefan
  Dietze}, \bibinfo{person}{Claudia Hauff}, \bibinfo{person}{Edward Curry},
  {and} \bibinfo{person}{Philippe Cudr{\'{e}}{-}Mauroux}} (Eds.).
  \bibinfo{publisher}{{ACM}}, \bibinfo{pages}{1823--1832}.
\newblock
\urldef\tempurl%
\url{https://doi.org/10.1145/3340531.3411999}
\showDOI{\tempurl}


\bibitem[Zhang et~al\mbox{.}(2020b)]%
        {Zhang:2020:BERTScore}
\bibfield{author}{\bibinfo{person}{Tianyi Zhang}, \bibinfo{person}{Varsha
  Kishore}, \bibinfo{person}{Felix Wu}, \bibinfo{person}{Kilian~Q. Weinberger},
  {and} \bibinfo{person}{Yoav Artzi}.} \bibinfo{year}{2020}\natexlab{b}.
\newblock \showarticletitle{BERTScore: Evaluating Text Generation with {BERT}}.
  In \bibinfo{booktitle}{\emph{8th International Conference on Learning
  Representations, {ICLR} 2020, Addis Ababa, Ethiopia, April 26-30, 2020}}.
  \bibinfo{publisher}{OpenReview.net}.
\newblock
\urldef\tempurl%
\url{https://openreview.net/forum?id=SkeHuCVFDr}
\showURL{%
\tempurl}


\bibitem[Zhang and Chen(2020)]%
        {zhang:2020:explainable-recommendation-book}
\bibfield{author}{\bibinfo{person}{Yongfeng Zhang} {and} \bibinfo{person}{Xu
  Chen}.} \bibinfo{year}{2020}\natexlab{}.
\newblock \showarticletitle{Explainable Recommendation: {A} Survey and New
  Perspectives}.
\newblock \bibinfo{journal}{\emph{Found. Trends Inf. Retr.}}
  \bibinfo{volume}{14}, \bibinfo{number}{1} (\bibinfo{year}{2020}),
  \bibinfo{pages}{1--101}.
\newblock
\urldef\tempurl%
\url{https://doi.org/10.1561/1500000066}
\showDOI{\tempurl}


\bibitem[Zhang et~al\mbox{.}(2021a)]%
        {zhang:2021:wsdm:expred}
\bibfield{author}{\bibinfo{person}{Zijian Zhang}, \bibinfo{person}{Koustav
  Rudra}, {and} \bibinfo{person}{Avishek Anand}.}
  \bibinfo{year}{2021}\natexlab{a}.
\newblock \showarticletitle{Explain and Predict, and then Predict Again}. In
  \bibinfo{booktitle}{\emph{{WSDM} '21, The Fourteenth {ACM} International
  Conference on Web Search and Data Mining, Virtual Event, Israel, March 8-12,
  2021}}. \bibinfo{publisher}{{ACM}}, \bibinfo{pages}{418--426}.
\newblock
\urldef\tempurl%
\url{https://doi.org/10.1145/3437963.3441758}
\showDOI{\tempurl}


\bibitem[Zhang et~al\mbox{.}(2021b)]%
        {zhang:cikm:2021:faxplainac}
\bibfield{author}{\bibinfo{person}{Zijian Zhang}, \bibinfo{person}{Koustav
  Rudra}, {and} \bibinfo{person}{Avishek Anand}.}
  \bibinfo{year}{2021}\natexlab{b}.
\newblock \showarticletitle{FaxPlainAC: A Fact-Checking Tool Based on
  EXPLAINable Models with HumAn Correction in the Loop}. In
  \bibinfo{booktitle}{\emph{Proceedings of the 30th ACM International
  Conference on Information \& Knowledge Management}}.
  \bibinfo{pages}{4823--4827}.
\newblock
\urldef\tempurl%
\url{https://doi.org/10.1145/3459637.3481985}
\showDOI{\tempurl}


\bibitem[Zheng and Fang(2010)]%
        {zheng:2010:ecir:queryTermWeight}
\bibfield{author}{\bibinfo{person}{Wei Zheng} {and} \bibinfo{person}{Hui
  Fang}.} \bibinfo{year}{2010}\natexlab{}.
\newblock \showarticletitle{Query Aspect Based Term Weighting Regularization in
  Information Retrieval}. In \bibinfo{booktitle}{\emph{Proceedings of {ECIR}
  2010}}. \bibinfo{pages}{344--356}.
\newblock


\bibitem[Zhuang et~al\mbox{.}(2021)]%
        {zhuang:2021:wsdm:gam}
\bibfield{author}{\bibinfo{person}{Honglei Zhuang}, \bibinfo{person}{Xuanhui
  Wang}, \bibinfo{person}{Michael Bendersky}, \bibinfo{person}{Alexander
  Grushetsky}, \bibinfo{person}{Yonghui Wu}, \bibinfo{person}{Petr Mitrichev},
  \bibinfo{person}{Ethan Sterling}, \bibinfo{person}{Nathan Bell},
  \bibinfo{person}{Walker Ravina}, {and} \bibinfo{person}{Hai Qian}.}
  \bibinfo{year}{2021}\natexlab{}.
\newblock \showarticletitle{Interpretable Ranking with Generalized Additive
  Models}. In \bibinfo{booktitle}{\emph{{WSDM} '21, The Fourteenth {ACM}
  International Conference on Web Search and Data Mining, Virtual Event,
  Israel, March 8-12, 2021}}. \bibinfo{publisher}{{ACM}},
  \bibinfo{pages}{499--507}.
\newblock
\urldef\tempurl%
\url{https://doi.org/10.1145/3437963.3441796}
\showDOI{\tempurl}


\end{thebibliography}

\appendix


\end{document}